\documentclass[twocolumn]{aastex63}
\usepackage{placeins}
\usepackage{textcomp, gensymb}
\usepackage{hyperref}
\usepackage{booktabs} 
\usepackage{rotating}
\usepackage{multirow}
\usepackage[utf8]{inputenc}

\definecolor{purp}{RGB}{153, 51, 255}

\newcommand\z{\textit{z}}
\newcommand\HST{\textit{HST}}
\newcommand\jwst{\textit{JWST}}
\newcommand\nancy{\textit{Roman}}
\newcommand\Euclid{\textit{Euclid}}
\renewcommand{\textbf}[1]{#1}
\newcommand{\revtwo}[1]{{\bfseries #1}}
\renewcommand{\revtwo}[1]{#1}
\shorttitle{Galaxy Merger Identification at $z\sim1$}
\shortauthors{Schechter et al.}

\graphicspath{{./}{figures/}}

\begin{document}

\title{Beyond the Brightest: A Deep Learning Approach to Identifying Major and Minor Galaxy Mergers in CANDELS at $z \sim 1$}
 
\correspondingauthor{Aimee Schechter}
\email{aimee.schechter@colorado.edu}

\author{Aimee L. Schechter}
\affiliation{Department of Astrophysical and Planetary Sciences, University of Colorado, Boulder, CO 80309, USA}
\author{Aleksandra \'Ciprijanovi\'c}
\affiliation{Fermi National Accelerator Laboratory, Batavia, IL 60510, USA}
\affiliation{Department of Astronomy and Astrophysics, University of Chicago, Chicago, IL 60637, USA}
\affiliation{NSF-Simons AI Institute for the Sky (SkAI), Chicago, IL 60611, USA}
\author{Xuejian Shen}
\affiliation{TAPIR, California Institute of Technology, Pasadena, CA 91106, USA}
\affiliation{Kavli Institute for Astrophysics and Space Research, Massachusetts Institute of Technology, Cambridge, MA 02139, USA}
\author{Rebecca Nevin}
\affiliation{Fermi National Accelerator Laboratory, Batavia, IL 60510, USA}
\author{Julia M. Comerford}
\affiliation{Department of Astrophysical and Planetary Sciences, University of Colorado, Boulder, CO 80309, USA}
\author{Aaron Stemo}
\affiliation{Vanderbilt University, Department of Physics \& Astronomy, 6301 Stevenson Center, Nashville, TN 37235, USA}
\author{Laura Blecha}
\affiliation{Department of Physics, University of Florida, Gainesville, FL 32611, USA}
\author{Austin Fraley}
\affiliation{Department of Physics, University of Florida, Gainesville, FL 32611, USA}

\begin{abstract}

Galaxy mergers play an important role in galaxy evolution. Therefore, accurate merger identifications are paramount for achieving a complete understanding of how galaxies evolve.
As we enter the era of \textbf{large, deep, high-resolution} imaging surveys, we can observe mergers extending to even lower masses and higher redshifts. Despite low-mass galaxies being more common, many previous merger identification methods were calibrated for high-mass galaxies which are easier to identify. 
To prepare for upcoming surveys, we train a convolutional neural network (CNN) using mock \HST{} CANDELS images at $z\sim1$ created from the IllustrisTNG50 cosmological simulation. We successfully identify galaxy mergers between a wide range of galaxies ($10^8M_\odot < M_\star < 10^{12.5}M_\odot$, and $q\geq1:10$), achieving overall accuracy, purity, and completeness of \textbf{$\sim65\%$}.
We show, for the first time, that a CNN trained on this diverse set of galaxies is capable of identifying major mergers, \textbf{especially at early stages (74\% accuracy)}, similar to that of networks trained at lower redshifts and/or higher masses (with accuracies between $66-80\%$). 
We discuss the inherent limits of galaxy merger identification due to orientation angle, finding \textbf{98\% of mergers are correctly identified from at least one angle, and 61\% from the majority of angles.}
\textbf{We additionally} explore the confounding variables, such as star formation, to consider when applying to real data.
This network enables the exploration of the impact of previously overlooked mergers of high mass ratio and low stellar masses on galaxy evolution in CANDELS, and can be expanded to surveys from \jwst{}, Rubin, \nancy{}, and \textit{Euclid}.

\end{abstract}


\section{Introduction} \label{sec:intro}
Galaxy mergers are one of the main avenues for galaxies to evolve from clumpy, high-redshift galaxies into the organized structures we see in the local universe, both with large morphological changes due to major mergers and building stellar mass through minor mergers \citep[e.g.][]{toomre_galactic_1972, mihos_gasdynamics_1996, buitrago_early-type_2013, martin_role_2018}. 
Identifying their role in galaxy evolution is a key task for observational astronomy, especially in the era of high redshift science with \jwst~\citep{jwst}, and large astronomical surveys and telescopes such as the Vera C. Rubin Observatory's (Rubin) Legacy Survey of Space and Time~\citep[LSST;][]{ivezic_lsst_2019}, Nancy Grace Roman Space Telescope~\citep[\nancy{};][]{roman}, and \Euclid{}~\citep{euclid}.

Though the merging process influences galaxy structure, mergers can prove tricky to identify.
A galaxy merger can take hundreds of Myr to a few Gyr, depending on factors such as mass ratio and orbital parameters \citep{lotz_galaxy_2008}.
Merger stage can also complicate merger identification, as the galaxies do not experience each stage for an equal amount of time, and thus some stages are more common than others.
There is a range of morphologies associated with each merger stage, and thus the method used influences which mergers are found and which are missed, as certain methods are more sensitive to certain morphologies \citep{lotz_galaxy_2008}.

Mergers experiencing their first pass (early stage) are easier to identify than mergers experiencing the coalescence of the nuclei (late stage). 
The human eye is historically trustworthy at identifying low-redshift, early stage mergers, especially when multiple people visually inspect each image (e.g., \citealt{darg_galaxy_2010}). 
Visual (i.e., ``by-eye", performed by scientists) classification efforts have enabled huge amounts of science on the role of mergers, but they have some drawbacks.
First, they require many human hours. 
Some of this has been outsourced to citizen scientists with projects like Galaxy Zoo \citep{lintott_galaxy_2008}.
Second, the human eye may struggle to capture a range of galaxy mergers.
Visual classification studies are biased to early stage, major (merger mass ratio $q\geq 1:4$) mergers, due to these systems having obvious morphological disturbances and/or signatures of multiple galaxies, such as two stellar bulges.
Late stage, minor mergers (mass ratio $q<1:4$) may be missed since they can be mistaken for isolated galaxies, not highly disturbed, or even just visually overlapping galaxies. 
\textbf{Third, visual classification catalogs inherently lack reproducibility, as individual observers may prioritize different morphological features.}

These biases in visual classifications lead to the use of quantitative imaging predictors such as the concentration (C), asymmetry (A), and smoothness (clumpiness) (S) i.e., CAS parameters \citep{conselice_relationship_2003} and Gini-M$_{20}$ \citep{lotz_new_2004}. 
These non-parametric morphology measurements are all based on various ways of measuring distributions of light in images, often separating quiescent galaxies from star-forming galaxies, and more defined morphologies from disturbed systems.
Separating those types of morphologies is extremely useful in identifying mergers.
The merger stage and image quality affect which method correctly identifies more mergers.
Morphological asymmetry is more accurate for mergers in early stages, and Gini-M$_{20}$ is more successful with mergers near the end of the process \citep{nevin_accurate_2019, wilkinson_limitations_2024}, \textbf{since asymmetrical features fade with time \citep{pawlik_shape_2016}}.
However, all of these methods are calibrated on lower-\z\ galaxies and may need to be adjusted for high redshift galaxies.

Galaxies and galaxy mergers at higher redshifts, $z$, \textbf{may} look different than those in our local universe (\citealt{conselice_evolution_2014} and references therein). 
\textbf{For example, galaxies at $z > 1$ contain more gas and dust, and thus, when viewed through optical filters probing the rest-frame ultraviolet, can appear clumpier compared to low-redshift galaxies. 
However, recent observations in rest-frame optical and infrared with the {\jwst} show that the Hubble Sequence is largely in place by $z\sim3$ (maybe even earlier), especially for massive galaxies \citep{ferreira_jwst_2023, lee_morphology_2024, huertas-company_galaxy_2024, huertas-company_cosmos-web_2025}.}
The physical morphology of high-redshift galaxies \textbf{observed at optical wavelengths} could make it harder to identify mergers visually, since the clumpy, uneven structure could make these galaxies appear to be merging even when they are actually isolated. 
These galaxies are also often smaller than their present-day counterparts. 
Therefore, at optical wavelengths, how we identify low-redshift mergers may need to be different from how we identify high-redshift mergers.
There have been by-eye classification efforts out to $z \sim 7$
\citep{kartaltepe_candels_2015,simmons_galaxy_2017, willett_galaxy_2017,smethurst_galaxy_2025}, though not all by-eye classifications can identify mergers.

\textbf{Photometric close pair studies can identify early stage mergers by finding galaxies close together on the sky and in photometric redshifts \citep[e.g.,][]{lopez-sanjuan_spectro-photometric_2010, lopez-sanjuan_alhambra_2015, barrows_census_2023}.
However, these studies can have high contamination from nonmergers as photometric redshifts can be unreliable.
When available, spectra can help weed out interloping nonmergers by finding pairs of galaxies within a given rest-frame velocity difference and projected separation \citep[e.g.,][]{ellison_galaxy_2008, mantha_major_2018, duncan_observational_2019, desmons_galaxy_2023}. 
Though close pairs discovery via spectroscopic observations benefits from having no human bias, it is still limited by spectral resolution, redshift of the galaxies, and fiber collision for pairs at very small separation and close to coalescence.
Additionally, just as in imaging, there may be interlopers that appear within the line-of-sight and pass the velocity difference and projected separation cuts but are not true mergers.
However, the biggest drawback is that spectra are time consuming and expensive compared to photometry.
}

Convolutional Neural Networks (CNNs) are a type of neural network for working with imaging data. 
Their ability to extract useful features from images (even those that might not be obvious to the human eye) makes them a promising tool with which to identify galaxy mergers. 
Many studies have successfully shown that CNNs outperform both by-eye identification and Gini-M$_{20}$/CAS at $z < 1$ \citep{bickley_convolutional_2021, ackermann_using_2018}. 
By using machine learning rather than visual, by-eye identification, we avoid biases humans may have of identifying only more obvious mergers, such as a merger between two large spiral galaxies. 
Additionally, CNNs, \textbf{once trained,} are much faster than human identification, which is an increasingly important consideration as we prepare for large observational survey telescopes. 

When considering CNNs as a merger identification tool, it is important to take into account that CNNs, as any other machine learning model, can only learn from the information contained in the training set.
Therefore, which mergers astronomers decide to include in their training set dictates which masses, mass ratios, and merger stages the CNN will be most likely to find.
CNNs have been commonly trained on samples of galaxies that are identified by eye, both by citizen scientists and by experts. 
Many papers have used the Galaxy Zoo data to train or test the performance of their network to identify galaxy morphologies (e.g., \citealt{dieleman_rotation-invariant_2015, dominguez_sanchez_improving_2018, cavanagh_morphological_2021}). 
However, \cite{bickley_convolutional_2021} showed that compared to CNNs, visual inspection can often lead to many different classifications of any given galaxy depending on the training of the individual performing the classification. 
Therefore, it is becoming more common to train networks on mock images of simulated galaxies because this ensures that the mergers/nonmergers used in the training set are known a-priori. 

Mock images from the IllustrisTNG cosmological simulation suite \citep{marinacci_first_2018, naiman_first_2018,nelson_first_2018, pillepich_first_2018, springel_first_2018} are a common choice for a training set due to TNG's wide range of galaxy morphologies and match of observational galaxy properties at $z = 0$. 
Many papers choose to make mock images from TNG100-1 or TNG300-1, with the $\sim100$Mpc per side and $\sim300$Mpc per side boxes respectively, and use neural networks to classify mergers
\citep{bickley_convolutional_2021, bottrell_combined_2022,ferreira_simulation-driven_2022, avirett-mackenzie_post-merger_2024, margalef-bentabol_galaxy_2024,ferreira_galaxy_2024, ferreira_galaxy_2026}.
The highest spatial resolution run of TNG, TNG50 \citep[$\sim50$Mpc per side box;][]{nelson_first_2019, pillepich_first_2019}, was used as the training set in \citet{omori_galaxy_2023}. 
The TNG100 and TNG300 boxes contain significantly more volume than the TNG50 box, and thus enable much larger training sets of mergers and nonmergers. 
These works show that machine learning combined with TNG can be successful at identifying mergers at both early and late stages at lower redshifts $z < 1$ and higher masses $M_\star\gtrsim10^9 M_\odot$ for a variety of mocked imaging surveys. 
\citet{pearson_identifying_2019} also showed successful ML merger identification with with the EAGLE simulation \citep{crain_eagle_2015, schaye_eagle_2015}, \citet{dominguez_sanchez_identification_2023} with the Horizon-AGN simulation \citep{kaviraj_horizon-agn_2017}, \textbf{and \citet{la_marca_dust_2024} combining TNG100, TNG300, and Horizon-AGN.}

Along with the use of neural networks, recent years have witnessed an explosion in the field of eXplainable aritificial intelligence (XAI), which seeks to promote interpretability in machine learning tools (see \citealt{arrieta_explainable_2019} for a review). 
Pixel attribution methods such as saliency maps \citep{simonyan_deep_2013} and Gradient-Weighted Class Activation Mapping \citep{selvaraju_grad-cam_2020} highlight specific pixels and regions of an input image that the CNN relied on for its final classification.
Additionally, inspecting feature maps from hidden layers can also provide physical intuition into why the network makes the decisions it does.
This makes CNNs less of a black box and more useful for astronomical purposes. 
\citet{ciprijanovic_deepmerge_2020} identified merging galaxies at $z = 2$ in the original Illustris~\citep{vogelsberger_introducing_2014} simulation with a CNN for the first time.
They used XAI to show that when identifying mergers, their CNN focused on larger areas of the image that contained faint substructures of a galaxy, but when identifying nonmergers, the influential pixels were in a more compact area.
This provided insight into what physical processes the galaxy images contained that the network observed to make a decision between merger and nonmerger.

Using a galaxy merger sample from IllustrisTNG50, we create mock galaxy images from the \textit{HST} \textbf{Cosmic Assembly Near-IR Deep Extragalactic Legacy Survey}~\citep[CANDELS;][]{koekemoer_candels_2011, grogin_candels_2011}.
This survey has some overlapping wavelength coverage with \nancy, but since the data are already public, it has the benefit of creating mock images with real backgrounds instead of simulated background noise.
We want to study mergers around cosmic noon (1$ < z < $3), the peak of morphological evolution, for which CANDELS provides a large sample of galaxies. 
CANDELS is $>90\%$ complete at $z<$3 \citep{guo_candels_2013, mantha_major_2018}.

In this paper, we use extremely realistic, fully radiatively transferred mock \HST{} images from IllustrisTNG50 to classify mergers in optical and infrared filters at $1 \leq z \leq 1.5$. 
We include galaxies down to $M_\star = 10^8 M_\odot$, while still including mass ratios down to 1:10 and multiple merger stages.
We aim to create a trustworthy classifier to find mergers at high-$z$ and lower stellar masses with \jwst{}, Rubin, \textit{Roman}, and \Euclid{} with the help of a high resolution simulation and XAI interpretation.

Our merger selection from TNG and mock image process is discussed in Section~\ref{sec:data}, the CNN architecture, performance evaluation metrics and model interpretability are discussed in Section~\ref{sec:methods}, the performance of our CNN is discussed in Section~\ref{sec:results}, and a discussion of the limitations of our sample and model is in Section~\ref{sec:discussion}.
We use the Planck cosmological parameters \citep{planck_collaboration_planck_2016}, the same ones used by IllustrisTNG: a matter density $\Omega_m = \Omega_{dm} + \Omega_b =
0.3089$, baryonic density $\Omega_b = 0.0486$, cosmological constant
$\Omega_{\Lambda}= 0.6911$, Hubble constant $H_0 = 100\, h$ km s$^{-1}$ Mpc$^{-1}$ with
$h = 0.6774$, normalization $\sigma_8 = 0.8159$ and spectral index
$n_s = 0.9667$.

\section{Data} \label{sec:data}

To quantify the role of mergers in processes at cosmic noon ($1 < z < 3$) such as galaxy morphological evolution, cosmic star formation, and black hole activity, a reliable classifier that can separate mergers from nonmergers at low masses, including among peculiar (not spiral or elliptical) galaxies is necessary.
Therefore, we aim to build a CNN that is able to identify both major and minor mergers, mergers at early and late stages, and mergers between galaxies with $M_\star > 10^8M_\odot$ at $z\sim1$. 

We create a dataset of mock \HST{} imaging with galaxies in the $1\leq z \leq 1.5$ redshift range.
This redshift range is the highest redshift where, \textbf{by \HST{} morphology,} spiral and elliptical galaxies are each about as common as peculiar galaxies~\citep[which would include mergers;][]{buitrago_early-type_2013}. 
Above this redshift, the universe has a different makeup than it does locally, with peculiar galaxies being much more prevalent, especially among low mass galaxies \citep[e.g.,][]{ferreira_jwst_2023}.
Additionally, we know minor mergers are an important avenue for galaxies to build up stellar mass over cosmic time, and that low mass galaxies are more common at higher redshifts \citep{martin_role_2018}.

The \HST{} CANDELS survey is the basis for our mock images due to its high spatial resolution, coverage around the peak of cosmic star formation, and, importantly, existing catalogs with which to compare our results and build additional trust in our model before applying to newer surveys.
In order to train a successful network we need a highly realistic training set that includes galaxy mergers with these mass ratios, merger stages, and stellar masses.
We use galaxies from the TNG50 cosmological simulation as our training set.
We want to use high-resolution simulations and images like those from TNG50 (with a median spatial resolution of $\sim$0.1kpc, which is comparable to or better than the resolution of CANDELS images at $z > 0.2$) for the merger identification process to be sure the CNN can separate small, clumpy, isolated galaxies from merging galaxies.
Using mock images from TNG50 will also provide a strong training set since we will know the true classification of each galaxy, allowing our tool to be better trained at identifying mergers close to cosmic noon. 
We create mock images in three \HST{} filters (F814W, ~F160W, and ~F606W) for a three-channel CNN.
This section describes the merger selection process from the simulation, and how we go from the simulated galaxies to fully realistic mock images.

\subsection{Cosmological Simulation and Galaxy Selection}\label{sec:sim}

IllustrisTNG  is a suite of cosmological magnetohydrodynamical simulations spanning 0 $<$ \textit{z} $<$ 127 with improved physics from the original Illustris simulation \citep{vogelsberger_introducing_2014}. 
It comes in three box sizes, the largest being TNG300 at roughly 300Mpc/side and the smallest, highest resolution run, TNG50, at roughly 50Mpc/side \citep{nelson_first_2019, pillepich_first_2019}. 
All volumes have both ``dark matter only" and ``baryonic physics" runs. 
The IllustrisTNG model includes physical processes such as gas cooling (primordial and metal-line), star formation, supermassive black hole formation and mergers, chemical enrichment from supernovae, stellar and black hole feedback, and cosmic magnetic field evolution, all of which impact galaxy morphologies and evolution. 
This self-consistent simulation enables us to investigate how morphologies change over time and how this may affect merger identification techniques at different redshifts.

IllustrisTNG has ``full" and ``mini" snapshots. 
Both snapshots trace all of the same subhalos through cosmic time and include the full 50 co-moving Mpc/side box. 
A ``full" snapshot contains all of the physics that TNG calculates, while the mini snapshots do not calculate physics for all particle fields. 
Since we aim to run radiative transfer, we only can only create images from snapshots that are ``full", which are necessary to make the mock images described in Section \ref{sec:methodmocks}. 
The redshift bin centers of $z = 1$ and $z = 1.5$ are both ``full" snapshots.
Notable examples of outputs that are only in the ``full" snapshots are: magnetic fields, neutral hydrogen density, dark matter density, and stellar metallicity.

The {\sc SubLink} assembly history traces a subhalo back in time and connects it to subhalos in previous snapshots (higher redshifts), that it evolved from.
For a given snapshots, we identify all subhalos with a stellar mass greater than 1000 times the baryonic mass resolution of TNG50, which is $8.5\times10^4M_\odot$. 
We trace the subhalos across cosmic time using the merger trees and merger definitions from \cite{rodriguez-gomez_merger_2015}, which link the identities of a given subhalo to its progenitor and descendant subhalos at the previous and following snapshot, respectively.
We define a merger \textbf{as in \citet{schechter_enhanced_2025}: to be considered a merger, a subhalo needs to have at least two progenitors at the previous snapshot in time, and both progenitors must share a descendant history with the given merged subhalo.}
We additionally require that the stellar mass ratio of the first and next progenitor be greater than 0.1 to include minor and major mergers \textbf{(as mentioned before, major mergers have merger mass ratio $q\geq 1:4$, while minor mergers have mass ratio $1:10 \leq q < 1:4$).}

We select mergers from TNG50 in two redshift bins centered at \z\ = 1 and \z\ = 1.5, \textbf{which are both ``full" snapshots}. 
To identify mergers (and nonmergers) for each redshift bin, we begin by identifying all snapshots that are within 250 Myr of the bin center, \textbf{giving a total merger window of 500Myr around the central snapshot (spanning three snapshots total-- one ``full" snapshot in the middle of the bin, and a ``mini" snapshot on either side)}. 
There are a few hundred independent mergers in each redshift bin. 
We then find mass-matched nonmerging galaxies by adapting the matching scheme from \citet{bickley_convolutional_2021}, as described in \citet{schechter_enhanced_2025}.
Each merging galaxy is matched with a nonmerging galaxy in the same snapshot by searching for a galaxy within a factor of $e^{0.1}$ and expanding the threshold by an exponential factor of 1.5 if a match is not found.
\textbf{Each matched pair is unique;} no nonmergers can be matched to multiple merging galaxies. 
Additionally, a nonmerger cannot be selected if it has gone through a merger within the past 2Gyr, \textbf{to avoid potential contamination from slowly coalescing mergers.
We are not able to impose a cut on nonmergers merging in the future, beyond the three snapshots included in the redshift bin, due to the small box size of TNG50; imposing this cut leaves too few nonmergers with a similar mass to choose from, and galaxies with dissimilar stellar masses end up as matched pairs.}

After following this procedure, relative to the central, ``full" snapshot, we have galaxies that merged at the previous snapshot back in time (a ``mini" snapshot), galaxies that merge at the center ``full" snapshot, and we have galaxies that will merge at the next snapshot forward in time (a ``mini" snapshot).
\textbf{In other words the merger can happen any time in the merger window (500Myr around the central snapshot). 
We then trace all mergers} and take their image from the ``full" snapshot bin center instead of the ``mini" snapshot in which they may truly merge.
That provides different merger stages, as we are imaging some galaxies pre-merging and post-merging. 
They may be true mergers one snapshot forward or backward in time respectively, but are imaged at these pre or post phases in ``full", central snapshot. 
For our analysis, we consider the pre-merging galaxies our ``pre-coalescence" or ``early stage" galaxy merger sample since the merger trees still identify two subhalos, and both the true merger galaxies in the central snapshot and the post-merging galaxies as ``post-coalescence" or ``late stage" mergers as the merger trees only identify one subhalo in both of these stages.
In our case, because we are using theoretical definitions from the TNG merger trees, these may not match exactly to observational merger stages such as ``first passage" or ``nearing coalescence".

\begin{figure*}
    \centering
    \includegraphics[width=\linewidth]{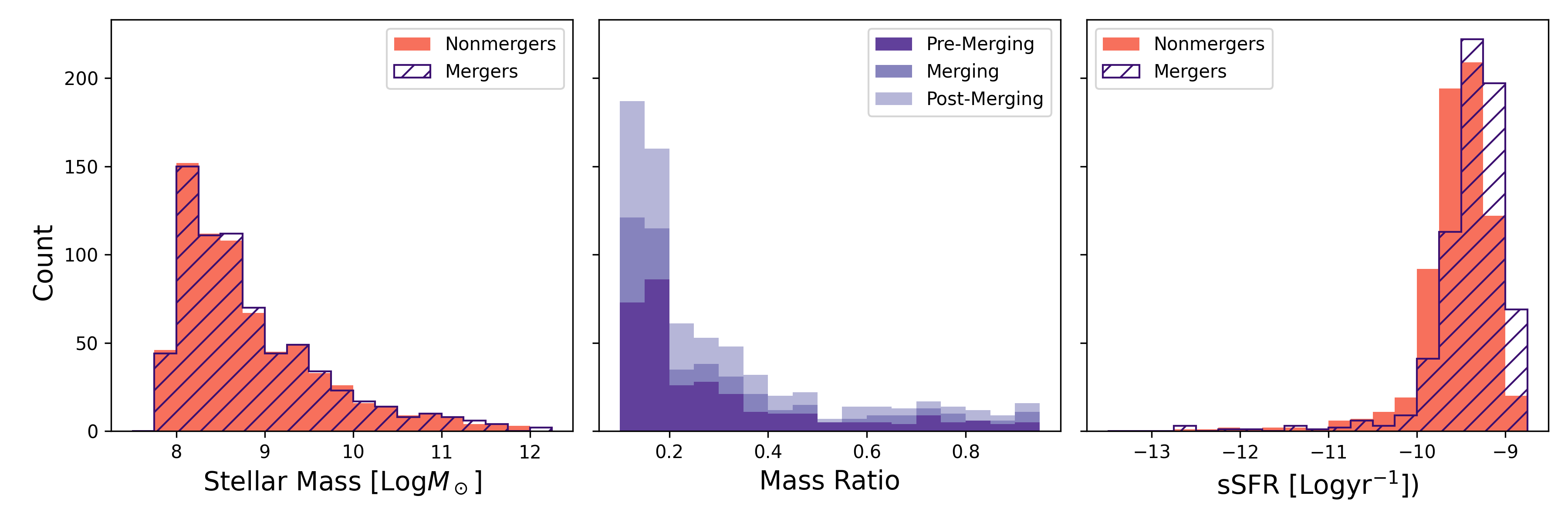}
    \caption{\emph{Left:} Distributions of merging and nonmerging galaxies' stellar masses. \emph{Center:} Stacked histogram of merger mass ratios in the merging sample color coded by merger stage as defined in Section \ref{sec:sim}. \emph{Right:} Distributions of sSFR for mergers and nonmergers.}
    \label{fig:demographics}
\end{figure*}

The resultant sample contains 260 mergers and 260 mass-matched nonmergers at $z = 1$ and 446 mergers and 446 mass-matched nonmergers at $z = 1.5$.
We combine the two redshift bins into one dataset, and the distributions of stellar mass (left figure), merger stage, and mass ratio (middle figure) are seen in Figure \ref{fig:demographics}.
The stellar mass ranges from $10^{7.9} -  10^{12.5}M_\odot$. 
The vast majority of our galaxies have lower masses ($M_\star < 10^{9.5}M_\odot$), which could make the classification task more challenging for the CNN. 
However, we want to utilize the lower stellar mass galaxies as well to get a full picture of the role mergers overall play in galaxy evolution and cosmic star formation, which the high resolution of TNG50 enables us to do.
All merger classes (pre-coalescence, merger, and post-coalescence) are combined into one class of ``merger" for the CNN. 
The middle plot on Figure \ref{fig:demographics} is a stacked histogram, so the overall distribution is for the entire merger class. 
The shading displays how the stages are broken down within the sample.
The specific star formation rates (sSFRs) are shown in Figure \ref{fig:demographics} (right plot), showing that the merging distriubtion peaks at a slightly higher sSFR than the nonmerging distribution. 
We show sSFR instead of SFR to remove the mass dependence of the star formation main sequence. 
While our sample is mass matched, it is not matched in SFR, due to the small box size.
Matching in both SFR and mass required expanding the mass match threshold many times, so we elect to use a close mass match and no SFR match.

\subsection{Realistic Mock Images} \label{sec:methodmocks}

We recognize that higher spatial resolution and deeper imaging at $z \sim 1$ exists with \jwst{}. 
We choose to use CANDELS in this work in order to compare our upcoming merger catalog (which will be the focus of our future work) to existing merger catalogs created by non-ML methods to build trust in our CNN. 
It also gives us a baseline at optical wavelengths to compare to when applying to Rubin, \nancy{}, \textbf{and \textit{Euclid}}.
Lastly, in order to not just identify mergers but draw statistical conclusions about their role in galaxy evolution, we require a large sample of observed galaxies, ideally with known stellar masses and star formation rates.
CANDELS provides a large field \textbf{(0.2deg$^2$)}, which \jwst{} does not, with existing value-added catalogs. 

To train the CNN on IllustrisTNG50 images for this goal, we must first make them as similar to real \textit{HST} images as possible. 
\cite{bottrell_deep_2019} investigated how important realistic training images are to the final classifications of a CNN while using one to not only identify mergers but also predict the merger stage. 
They found that as long as a network is exposed to background noise, other sources, and the spatial resolution of the telescope in training, it is able to predict mergers accurately when given real data. 
Notably, this paper discovered that realistic environments are more important to accurate predictions than radiative transfer in making mock images to train a CNN. 
\cite{nevin_accurate_2019} showed detailed steps to create SDSS mock images to use for machine learning and quantitative galaxy identification methods.
To create mock \HST{} images, we will be adapting the method used by \cite{nevin_accurate_2019}. 

\subsubsection{Radiative Transfer with {\sc SKIRT}}\label{sec:SKIRT}
The first step to create realistic mocks is to post-process TNG50 galaxies through full Monte Carlo continuum dust radiative transfer
calculations, using the publicly available code {\sc SKIRT}~\citep[version 9;][]{baes_efficient_2011,baes_skirt_2015,camps_skirt_2015,camps_skirt_2020}. We use the prescription described in \citet{vogelsberger_high-redshift_2020} and  \citet{shen_high-redshift_2020, shen_high-redshift_2022,Shen2024} , as done in \citet{nevin_accurate_2019}.

In this prescription, stellar particles in the simulations are assigned intrinsic emission using the stellar population synthesis method. 
Specifically, we adopt the Flexible Stellar Population Synthesis ({\sc Fsps}) code~\citep{conroy_propagation_2009, conroy_propagation_2010} to model the intrinsic spectral energy distributions (SEDs) of old stellar particles with $t_{\textrm{age}}>10$\,${\textrm {Myr}}$ (using the MILES spectral library and MIST isochrone library, this choice defines solar metallicity, $Z_{\odot}$) and the {\sc Mappings-III} SED library~\citep{groves_modeling_2008} to model those of young stellar particles with t$_{\textrm{age}}<10$\,${\textrm{Myr}}$. 
The {\sc Mappings-III} SED library self-consistently considers the dust attenuation in the birth clouds of young stars, which cannot be properly resolved in the simulations. 
We employ a K-D tree algorithm to calculate the smoothing length enclosing $64$~\footnote{It is an empirical choice here for adaptive softening and morphological characterizations are not sensitive to the number of neighbors used~\citep[e.g.][]{Torrey2015,RGomez2019}.} nearest stellar particles for all stellar particles within the galaxy. 
Given the spatial location and the smoothing length values of stellar particles, {\sc SKIRT} then creates a photon source distribution and emissivity profile through the entire space by interpolating over these kernels. 
At the beginning of radiative transfer calculations, photon packages are randomly released based on the source distribution characterized in this way. 
Wavelengths from the Lyman edge out to optical and IR wavelengths included in CANDELS filters are covered. 
The total number of released photon packages is set to be N$_{ p} = 10^{10}$. We also use an active galactic nuclei (AGN) template\footnote{\href{https://skirt.ugent.be/skirt8/class_quasar_s_e_d.html}{https://skirt.ugent.be/skirt8/class\_quasar\_s\_e\_d.html}} to assign emission to the black hole particles. 
The AGN emission is a broken power law (\citealt{schartmann_towards_2005}) characterized by the mass accretion rate and radiative efficiency of supermassive black holes in the simulation.

The emitted photon packages will further interact with the dust in the interstellar medium (ISM). To determine the distribution of dust in the ISM, we consider cold star-forming gas cells~(star-forming or with temperature $<8000\,{\textrm K}$) from the simulations and calculate the metal mass distribution based on their metallicities. 
We assume that dust is traced by metals in the ISM and adopt a constant dust-to-metal ratio among all galaxies at a fixed redshift. 
In \citet{vogelsberger_high-redshift_2020}, the dust-to-metal ratios at different redshifts has been calibrated based on the galaxy rest-frame UV luminosity functions at $z=2-10$~\citep[e.g.,][]{ouchi_large_2009,mclure_luminosity_2009,j_bouwens_alma_2016,finkelstein_observational_2016,oesch_dearth_2018}.  
For galaxies below this redshift range, we use a Milky Way dust-to-metal value of 0.4~\citep{Dwek1998}. 
We then turn the metal mass distribution into a dust mass distribution with the dust-to-metal ratio and map the dust distribution onto an adaptively-refined grid. 
The grid is refined with an Octree algorithm to maintain the fraction of dust mass within each grid cell to be smaller than $2\times10^{-6}$~\citep{Saftly2014}. 
The maximum refinement level is also adjusted to match the numerical resolution of the simulations. Besides, we assume a \citet{draine_dust_2007} dust mixture of amorphous silicate and graphitic grains, including varying amounts of Polycyclic Aromatic Hydrocarbons (PAHs) particles, which can reproduce the averaged extinction properties of the Milky Way.

Ultimately, after photons fully interacted with dust in the galaxy and escaped, they are collected by six simulated detectors $10\,{\textrm{Mpc}}$ away from the simulated galaxy along the positive (or negative) x,y,z-directions of the simulation coordinates. These six viewpoints will enlarge our training, validation, and test set sizes. We use a detector size that scales with redshift; the field of view is $100 (1+z)^{-1}$ kpc and 512 pixels per side, which is a physical size of 30 kpc at $z=1$ and 24 kpc at $z = 1.5$. We have selected the box size to scale with galaxy size, which also scales as $(1+z)^{-1}$ \citep{bouwens_galaxy_2004}. The flux in each pixel, as well as the integrated SED of the galaxy, are then recorded. Any galaxies that encountered errors or memory issues during the radiative transfer processes were discarded.

\subsubsection{Filter, Rebin, and PSF}\label{sec:filterrebinpsf}

From the radiative transferred images, we produce three-color (F814W, F160W, and F606W), filtered images using the Python code  {\sc SEDPY} \citep{johnson_bd-jsedpy_2021}. 
All five CANDELS fields have coverage in these bands.
\textbf{These filtered images are the same size as the {\sc SKIRT} output, $512\times512$ pixels.}
\textbf{Following {\sc SEDPY} filtering, we rebin each image from the $512\times512$ pixel size to the appropriate \HST/WFC3 and \HST/ACS pixel scales ($0\farcs03$ for F814W and F606W, and $0\farcs06$ for F160W).
This brings our images to sizes $202\times202$ or $101\times101$ pixels at $z = 1$, and $154\times154$ or $77\times77$ pixels at $z = 1.5$.
Due to the lower pixel scale, F160W band images have half the number of pixels as the other two bands, so we rebin them to match the rest of the filters and create images with matching pixel size in all three bands ($202\times202$ and $154\times154$).}

We avoid using the bluest bands available to encourage the CNN to learn overall morphology rather than focus on star forming clumps. 
Following filtering, the images are convolved with the point spread function (PSF) of each given \textit{HST} band, which we simulate using \texttt{TinyTim} \citep{krist_20_2011}. 
We implement cosmological surface brightness dimming by a factor of $(1 + z)^{-3}$ (in frequency space).
We repeat the process for each CANDELS filter, as each filter has different standard flux values and background noise \citep{koekemoer_candels_2011}. 

\begin{figure*}
    \centering
    \includegraphics[width=\textwidth]{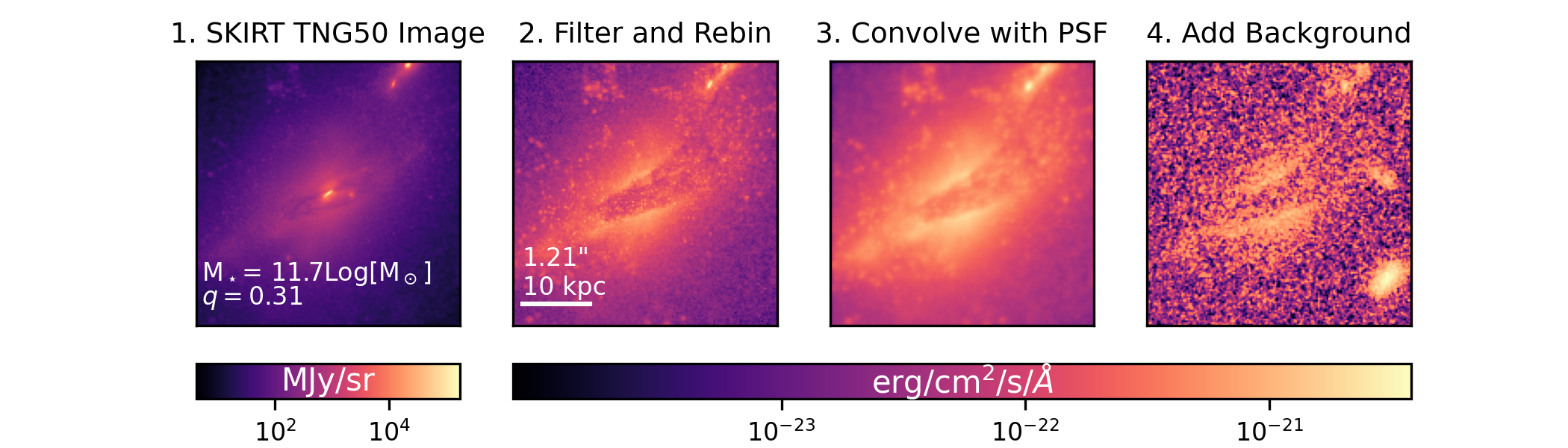}
    \caption{Main steps to create a mock F814W CANDELS image from the radiative transferred TNG50 data: 1) The left-most panel shows the image just after it has been processed by SKIRT; 2) Next, we apply an \HST ~F814W filter to the image so we are no longer seeing all wavelengths of light; we also rebin the image to the same pixel scale as the CANDELS mosaics; 3) Next we convolve with the PSF of the telescope to replicate what this galaxy would look like if observed by \HST. 4) Lastly we add real CANDELS backgrounds to include real CANDELS noise and background sources. This process will supply a more realistic training set, and is a crucial piece to building a robust network that can classify CANDELS galaxies. }
    \label{fig:hstize}
\end{figure*}

\subsubsection{Realistic Environments}\label{sec:realisticenvironments}
A key step to making the images fully realistic is ensuring the galaxies are placed in the simulated image as they would be in \textit{HST} observations, including in realistic field environments \citep{bottrell_deep_2019}. 
We incorporate background sources and noise from the CANDELS COSMOS field mosaics \citep{CANDELS} by \textbf{creating $100(1 + z)^{-1}$ kpc cutouts} from the mosaics that have no sources at the center. 
\textbf{A center pixel coordinate was randomly selected from the mosaic, and if any known source from \citet{barro_candelsshards_2019} was within 20 pixels in any direction \textbf{in the F160W mosaic}, it was discarded.
\revtwo{20 pixels corresponds to $\sim 10$kpc at $z\sim1$.
We want background sources to be allowed close enough to potentially look like false close pairs, e.g., pairs that are near on the sky but not in redshift separation, while avoiding placing galaxies directly on top of each other. 
10kpc is twice the minimum separation used for close pairs in CANDELS by \citep{duncan_observational_2019}, and roughly the size of a high mass galaxy at $z\sim1$ in HST imaging \citep{van_der_wel_3d-hstcandels_2014}.}
Additionally, the image had to be positioned so that all pixels contained the sky, and thus we were not choosing a background center right at the edge of the mosaic.}
We then overlay our mock images on top of that section of sky. 
Therefore, the noise is truly that of CANDELS imaging, and there are real background CANDELS galaxies and foreground stars in our images, an essential piece for our network to be able to distinguish from the central mergers.
All of the main steps to make an example mock image in the F814W filter can be seen in Figure~\ref{fig:hstize}. 

\subsection{Final Dataset Creation}
We split up our dataset into training (70$\%$), validation (15$\%$), and test sets (15$\%$). 
Each class is split by the ratios for the training/validation/test set above, then combined into the final training/validation/test set following the split.
This ensures each set is split equally between mergers and nonmergers, so that neither outcome dominates the network's final decisions.
In reality, there are far more nonmergers than mergers in the universe.
However, in order for the CNN to truly learn the difference between mergers and nonmergers, we do not want to bias its decision-making by providing an imbalanced dataset.
Recall from Section~\ref{sec:SKIRT} that each galaxy is viewed from six viewpoints (as if from each face of a cube).
When splitting the data, we make sure that all viewpoints of a given galaxy are always included in the same set, so that none of the viewpoints of galaxies included in the training set can appear when validating and testing the network.
Each image is normalized to have values between 0 and 1, typical for a CNN input.
\textbf{To do that, we employ the {\sc Astropy} visualization inverse hyperbolic sin stretch for normalization, with a \texttt{asinh\_a}$ =10^{-3}$, defining the transition from linear to logarithmic behavior.
} 
\textbf{This stretch} helps faint features, such as tidal tails, to become more visible.
\textbf{Following stretching, we clip each image to the 0.5 and 99.5 percentiles to lessen the impact of any extremely bright or dim pixels.
Finally, after clipping, we subtract the minimum pixel value and divide each image by its dynamic range to re-map brightnesses within the 0 to 1 range.
This final per-image normalization helps the network handle both bright, massive galaxies and faint, low-mass galaxies.}

\begin{figure*}
    \begin{centering}
    \includegraphics[width=0.7\textwidth]{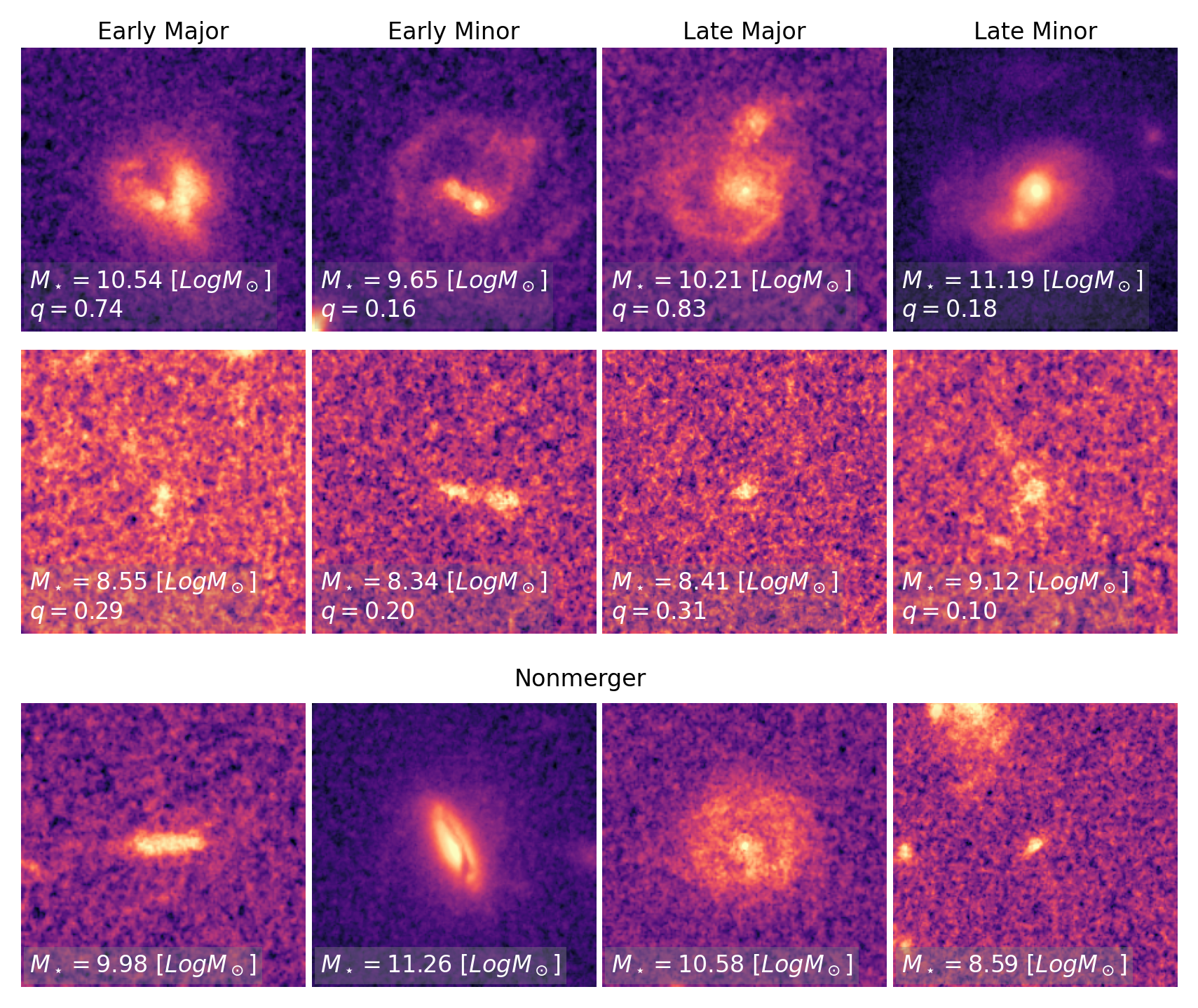}
    \caption{\textbf{Examples of our galaxies following the full mock image process as they appear to the CNN. They are all normalized to have pixel values between 0 and 1 and $224\times224$ pixels. The top section shows different types of mergers, with a higher mass galaxies in the top row and lower mass galaxies in the middle row. The bottom section shows nonmerger galaxies across a range of masses.}}
    \label{fig:visexample}
    \end{centering}
\end{figure*}

\textbf{After the mock image process, we resize our images one final time to fit standard pre-trained CNN specifications.
ConvNeXt-Nano (see Section \ref{sec:model} for information on the CNN model) commonly} takes images of a size $224\times224$ pixels, so we use the resize operation from \textsc{skimage} to reach this size, as done in \citet{bickley_effect_2024}.
This algorithm is stretching the images with interpolation, so we maintain the spatial resolution information of the pixels in the original mock image, \textbf{unlike the rebinning done in Section~\ref{sec:filterrebinpsf}.}
To additionally enhance our training set size, we use data augmentation on both the mergers and nonmergers. 
Each galaxy image can be randomly rotated up to 180$\degree$, randomly flipped horizontally or vertically, or randomly translated of up to 5\% and scaled between 85\% and 110\% of the original size.
The edges of the rotated images are filled with zeros ensuring that all images have the same shape.
We include both the original image and \textbf{three versions of the augmented images in training, for a final training size 4$\times$ the original dataset.}
\textbf{Following the mock image and data augmentation processes, the final dataset consists of: training set with 11808 mergers and 11820 nonmergers}, a validation dataset with 630 mergers and 624 nonmergers, and a test set of 630 mergers and 630 nonmergers\footnote{The final dataset is available on \href{https://doi.org/10.5281/zenodo.19374288}{Zenodo}.}.

\section{Methods}

\label{sec:methods}

\subsection{Convolutional Neural Network} \label{sec:model}

\begin{table*}[]
\centering
\begin{tabular}{*{2}l*{4}c}
\multicolumn{6}{c}{CNN Architecture} \\
\toprule
\textbf{Component} & \textbf{Stage} & \textbf{Blocks} & \textbf{Channels} & \textbf{Stride} & \textbf{Output Size} \\
\midrule
\multirow{6}{*}{Zoobot Encoder}
  & Stem                        & 1       & 80            & 4       & $56 \times 56$ \\
  & Stage 1                     & 2       & 80            & 1       & $56 \times 56$ \\
  & Stage 2                     & 2       & 160           & 2       & $28 \times 28$ \\
  & Stage 3                     & 8       & 320           & 2       & $14 \times 14$ \\
  & Stage 4                     & 2       & 640           & 2       & $7 \times 7$   \\
  & Global Avg Pool + LayerNorm & \nodata & 640           & \nodata & $1 \times 1$   \\
\midrule
\multirow{5}{*}{Custom Classification Head}
  & Dropout($0.12$)                & \nodata & 640           & \nodata & \nodata        \\
  & Linear                      & \nodata & $640 \to 256$ & \nodata & \nodata        \\
  & ReLU                        & \nodata & 256           & \nodata & \nodata        \\
  & Dropout($0.12$)                & \nodata & 256           & \nodata & \nodata        \\
  & Linear                      & \nodata & $256 \to 2$   & \nodata & \nodata        \\
\bottomrule
\end{tabular}
\caption{The architecture of our Zoobot ConvNeXt-Nano CNN~\citep{liu_convnet_2022} with an added classification head. The ConvNeXt Stem consists of a $4\times4$ convolution and \texttt{LayerNorm}. Each ConvNeXt Stage consists of downsampling (except Stage 1) followed by blocks containing a depthwise $7\times7$ convolution, \texttt{LayerNorm}, and a multi-layer perceptron with two $1\times1$ convolutions, $4\times$ channel expansion, and Gaussian Error Linear Unit~\citep[GELU;][]{GELU} activation function.} 
%$p = 0.12$.}
\label{tab:architecture}
\end{table*}

A CNN is a type of neural network specifically designed to work with images. 
By convolving different filters with an input image, it is able to extract features such as edges and shapes from the input image. 
While even simpler CNN architectures can achieve very high accuracies (above $99\%$) when trained to classify everyday objects and animals using very large benchmark datasets such as MNIST~\citep{deng_mnist_2012}, classification tasks in astronomy can be much harder.
Training data sets in astronomy are often much smaller, \textbf{lack reliable labels}, and include classes that look somewhat similar (e.g., mergers and nonmergers both look like galaxies), which makes training CNNs much harder.

Even with data augmentation, our dataset is on the smaller side, therefore, we use transfer learning to obtain smoother loss and accuracy curves. 
We use the \textbf{ConvNeXt-Nano model ~\citep{liu_convnet_2022} with pre-trained weights from Zoobot 2.9.0, \texttt{zoobot-encoder-ConvNeXt\_nano} \citep{walmsley_zoobot_2023}\footnote{All code for mock images and training and testing the CNN is available on \href{ https://github.com/alschechter/CosmicNoonMergerID}{Github}}.} Zoobot models are trained on millions of galaxy images from Galaxy Zoo using labels provided by citizen scientists.
Since Zoobot was trained on galaxies as opposed to images of everyday objects, its weights provide a better starting point than models trained on everyday objects for our goal of classifying mergers.
Using Zoobot's \texttt{FinetuneableZoobotClassifier} class we set \texttt{num\_classes = 2} to force a binary classification: galaxy merger or nonmerger, \textbf{following our added custom classification head. The architecture of our model is shown in Table~\ref{tab:architecture}.
We use Optuna \citep{akiba_optuna_2019} for hyperparameter optimization.
We train our CNN with the AdamW optimizer \citep{loshchilov_decoupled_2017}, initial learning rate of $3.15\times10^{-7}$, exponential learning rate decay of 0.88, weight decay of $3.88\times10^{-4}$, dropout of 0.12, label smoothing of 0.18, batch size of 64, and cross-entropy loss. 
We apply a cosine annealing learning rate scheduler with a restart period of 30 epochs and a minimum learning rate of $0.1\times$initial learning rate.}
We initiate early stopping during training if the validation set loss does not decrease by at least $1\times10^{-4}$ after 25 epochs, \textbf{or if the gap between training loss and validation loss exceeds 0.05 for 10 epochs}. 
The weights of the model from epochs 100, 120, 150, and any other epoch with the lowest validation loss are saved to obtain predictions on the test data; we use epoch 120.
Training takes approximately \textbf{six and a half hours} depending on the random seed with one Nvidia\_a100 GPU.
We train with three different random seeds to ensure model stability. \textbf{In Figure~\ref{fig:acc-loss}, we show the loss and accuracy curves for our model's best random seed (Seed 626).}

\subsection{Model Performance and Calibration}
To assess the performance of the CNN, we start with standard metrics, such as confusion matrices and Receiver Operating Characteristic (ROC) curves.
When discussing these metrics in the context of this study, we consider merger the positive class and nonmerger the negative class. Therefore a true positive (TP) is a merger correctly identified as a merger, and a true negative (TN) is a nonmerger correctly identified as a nonmerger.
A false positive (FP) is a nonmerger misidentified as a merger, and a false negative (FN) is a merger misidentified as a nonmerger.

A confusion matrix tracks how many \textbf{FP} and \textbf{FN} the network found, along with the accurate predictions. 
The rows of the confusion matrix correspond to the actual class of the galaxy (true merger or nonmerger) and each column corresponds to the network's predicted class. 
If the network performed perfectly, the confusion matrix would have values only on the diagonal, with zeros everywhere else, indicating that there were no \textbf{FPs} or \textbf{FNs}. 
In our case, a \textbf{FP} would be an isolated galaxy identified as a merger, and a \textbf{FN} would be a merger identified as a nonmerger. 

Some metrics to analyze how well the network is performing are accuracy, purity, and completeness.

\begin{equation}
    \textrm{Accuracy} = \frac{TP + TN}{TP + TN + FP + FN}
\end{equation}

\begin{equation}
    \textrm{Purity} = \frac{TP}{TP + FP}
\end{equation}

\begin{equation}
    \textrm{Completeness} = \frac{TP}{TP + FN}
\end{equation}
Accuracy is a measurement of the fraction of the time the network is making the correct prediction overall, whereas purity and completeness are specific to the positive (merger) class.
Purity and completeness are also sometimes known as precision and recall. 
Purity can be thought of as the percentage of predicted mergers that are correctly classified (e.g., how often a predicted merger is actually a merger). 
Completeness is the fraction of mergers that are correctly retrieved (e.g., out of the true mergers how many are correctly classified). 
Here, a low completeness would mean we are missing many mergers by classifying them as nonmergers. Often an increase in purity, can lead to a decrease in completeness, and vice versa. 
In the case of our CNN, having high purity and completeness would mean low contamination in our merger sample, since we would be classifying most galaxies correctly (with the purity of our merger sample being particularly important).

ROC curves are another tool to analyze the success of a CNN. 
This type of curve shows the \textbf{TP} rate (completeness) against the \textbf{FP} rate. 
The goal is to increase the area under the curve, meaning that there are more \textbf{TPs} and fewer false positives. 
A model in which the network is guessing randomly each time will have an ROC curve with a diagonal line with a slope of 1. 
A perfect network has area under the curve of 1, and a completely random one has area under the curve of 0.5.

A trustworthy neural network must be both accurate and appropriately confident in its predictions. 
To examine the confidence of a neural network, we use two calibration tools. 
First, we use the Brier Score \citep{brier_verification_1950}.
The Brier score is a measurement of ``how correct" a prediction is. 
For binary classification, it is defined as:
\begin{equation}
    \text{Brier Score} = \frac{1}{N} \sum_{t = 1}^{N}\left(p_t - o_t\right)^2,
\end{equation}
where $N$ is the number of galaxies in the test dataset, $p_t$ is the \textbf{predicted} probability assigned to each galaxy by the neural network (the value following the softmax activation function, which always lies between 0 and 1), and $o_t$ is the final class (1 or 0 corresponding to merger or nonmerger with a cutoff score of 0.5) assigned to each galaxy. 
A Brier score is always between 0 and 1 because it is a mean of squared differences between output probabilities and predicted classes. 
A lower Brier score indicates a network that is making accurate and well-calibrated predictions, as there is less of a difference between the probability and the final classification.

The second calibration metric we use is the Expected Calibration Error (ECE; \citealt{naeini_obtaining_2015}).
ECE measures a weighted average of the difference between the accuracy and predicted probabilities.
ECE is defined by splitting the \textbf{predicted probabilities} into $K$ equally spaced bins, and calculating a weighted average of the difference between the accuracy and output probabilities in each bin.
\begin{equation}
    \text{ECE} = \sum_{k =  1}^{K} \frac{\left|B_k\right|}{N} \left|\text{acc}\left(B_k\right) - p\left(B_k\right)\right|,
\end{equation}
where $N$ is the total number of galaxies in the test set, $B_k$ is the number of galaxies in the $k^{th}$ bin, $acc$ is the accuracy, or \textbf{fraction} of correctly classified galaxies, in each bin, and $p$ is the predicted probability of the CNN in each bin.
A lower ECE indicates a smaller difference between the accuracy and probability in each bin. This is the desired outcome for a well-calibrated network. 
We can visualize this difference through a reliability diagram \citep{degroot_comparison_1983, niculescu-mizil_predicting_2005}. 
These diagrams show the accuracy in each bin as a function of the probability in each bin. 
A perfectly calibrated model would have no difference, or ``gap", between the accuracy and probability, and thus would plot a 1:1 line. 
The difference from a 1:1 line shows the miscalibration of the network.

\subsection{CNN Interpretability}

Going beyond the more standard metrics, we use Gradient-weighted Class Activation Mapping \citep[Grad-CAM;][]{selvaraju_grad-cam_2020}, an XAI technique that highlights sections of an input image that are important to the final prediction.
Grad-CAM uses gradients from the final convolutional layers to highlight these influential regions so that they contain spatial information before it is lost in the fully connected layers.
Understanding where in the image our CNN is basing its decisions can help us build trust in the model, both showing when it perceives an image ``correctly", and to see why it makes an incorrect decision when it does.
Often, the decisions make sense when we can see what the CNN focused on.
Grad-CAM can be activated to either the merger or nonmerger class for any input image.
Activating a specific class shows which pixels are influential to that class, and not to other classes.

Finally we apply Uniform Manifold Approximation and Projection \citep[UMAP;][]{mcinnes_umap_2020}, a dimensionality reduction method. 
We use it to visualize the high dimensional latent space of \textbf{the final layer of the Zoobot encoder, before our custom classification head} in only two dimensions.
Galaxies that the network deems similar will lie close to each other in UMAP space, and galaxies that the network deems different from each other will lie far apart. 
We can use the UMAP distribution along with the physical quantities associated with each galaxy (e.g., stellar mass, merger stage, SFR) to investigate how the network related different galaxies.

\section{Results} \label{sec:results}
We present our CNN training metrics and results when the model is applied to our test set. We train our model with three different random seed initializations and present the mean and standard deviations of different performance metrics for our test set in Table \ref{tab:pcbe}. Accuracy, purity, and completeness are all $\sim65\%$, with the Brier score and ECE being low and indicating  good calibration overall of our models.
When presenting performance of an individual model (in the text and figures in the following sections), we use Seed 626, since that seed had the highest overall accuracy. 

\subsection{Model Performance and Calibration}\label{sec:results-performance}

\begin{table*}[]
    \centering
    \begin{tabular}{*{6}c}
    \multicolumn{6}{c}{Performance Metrics} \\
    \midrule
        Accuracy & Purity & Completeness & Brier Score & ECE & AUC\\
        \midrule
         $63.56\pm{1.17}\%$  &
         $62.73\pm{1.52}\%$ & 
         $66.08\pm{1.93}\%$ & 
         $0.22\pm{0.01}$ & 
         $0.03\pm{0.01}$ & 
         $0.69\pm{0.01}$\\
          
    \end{tabular}
    \caption{Accuracy, Purity, Completeness, Brier Score, ECE, and AUC for our models. The values shown are the mean and standard deviation of the three random seeds.}
    \label{tab:pcbe}
\end{table*}

\begin{figure}
    \centering
    \includegraphics[width=0.99\linewidth]{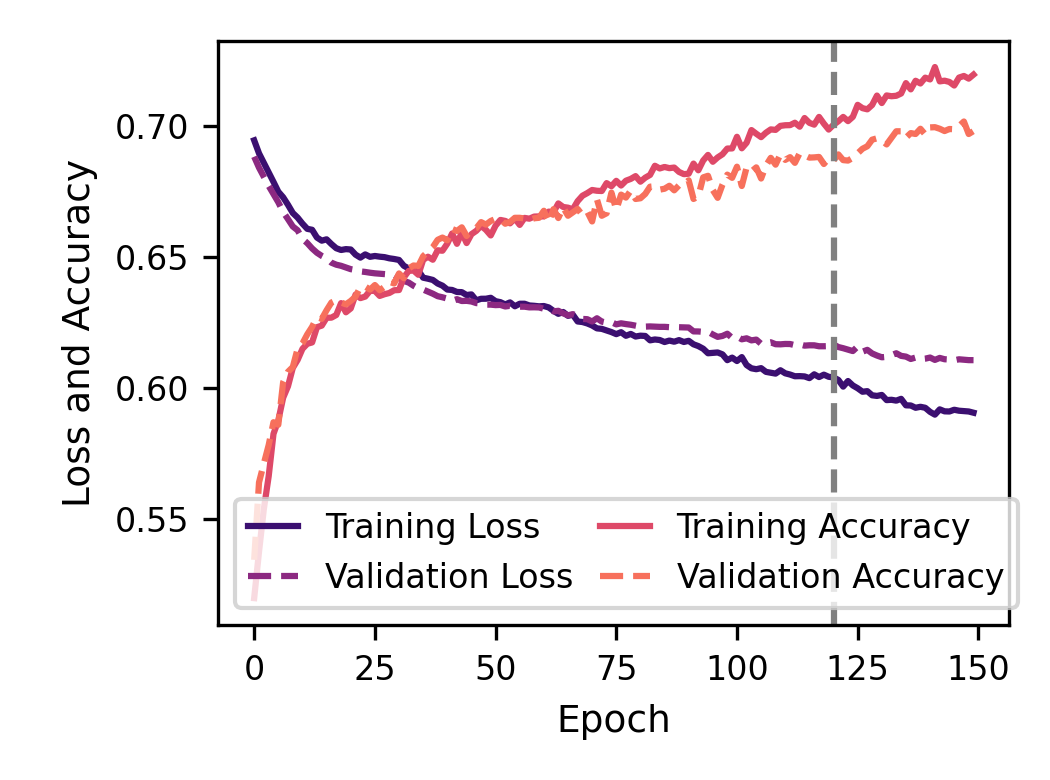}
    \caption{The loss and accuracy curves for our network.
    Though the training set curves in solid purple continued to improve, the validation set curves plateaued, so we implemented early stopping to avoid overfitting. We use the weights from \textbf{epoch 120} as our best model, noted by the grey dashed line.}
    \label{fig:acc-loss}
\end{figure}

\begin{figure}
    \centering
    \includegraphics[width=0.99\linewidth]{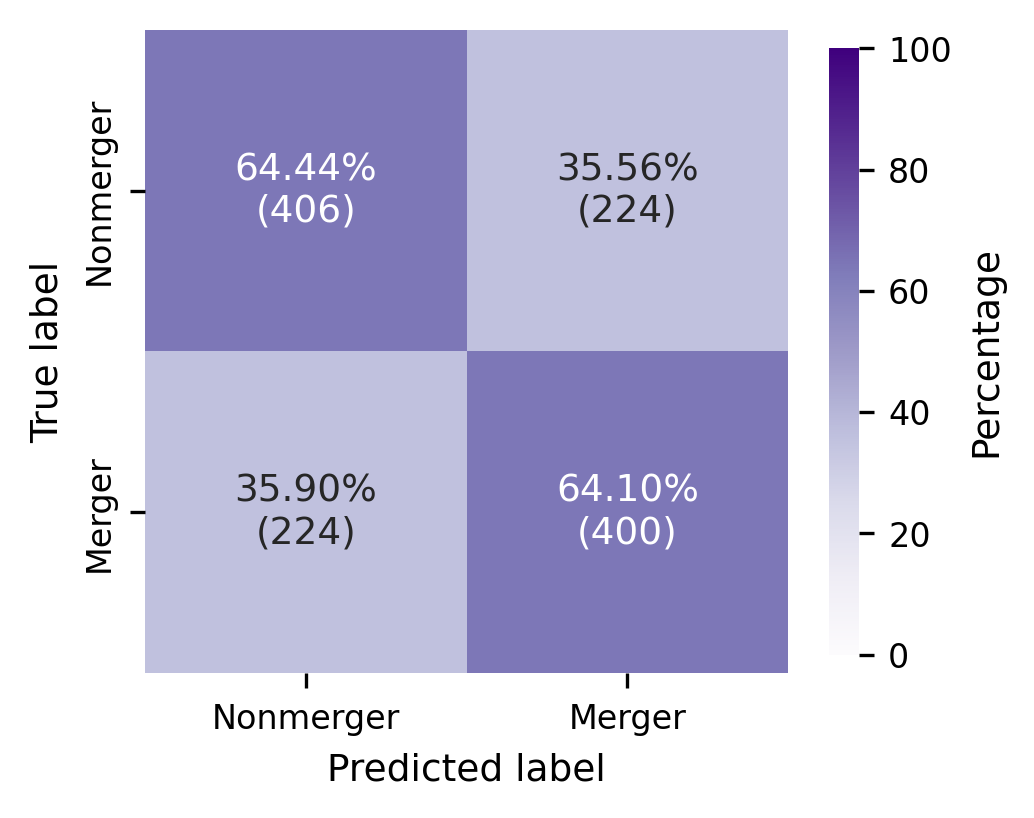}
    \caption{The confusion matrix of our Seed 626 network showing that we do classify the majority of galaxies correctly. The darker squares along the diagonal show the galaxies classified correctly. \textbf{Each cell shows both percentage of galaxies (by true class) along with the raw count.}}
    \label{fig:confusion}
\end{figure}

The loss and accuracy curves for one of our models (Seed 626) can be seen in Figure~\ref{fig:acc-loss}. 
\textbf{Overfitting is a problem for our model, and we tested different levels of layer-wise learning rate decay, dropout, dimensions of the classification head layers, learning rates, learning rate schedulers, label smoothing, and weight decay.
We find that after epoch $\sim100$ the performance on the test set is identical within a percentage point of accuracy, and it decreases following epoch $\sim150$.
Therefore, even though the training and validation losses do both continue to decrease, we use weights from epoch 120\footnote{The weights from epoch 120 are available on \href{ https://doi.org/10.5281/zenodo.19374288}{Zenodo}.}, before the overfitting gap between training and validation loss gets too large and the model begins to memorize rather that generalize.}
The confusion matrix for this model our test set is shown in Figure~\ref{fig:confusion}.
The overall accuracy of this model \textbf{(Seed 626) is 63.56$\pm{1.17}$\%, with a completeness of 66.08$\pm{1.93}$\%.}
The majority of our galaxies are classified correctly, but about a \textbf{third} of each class are not. 
\textbf{During training, both the training and validation sets exceeded $70\%$ accuracy around epoch 120, with losses continuing to decrease, although signs of overfitting began to emerge. The relatively small size of the test set likely contributes to a somewhat worse performance compared to training and validation sets. Additionally, increasing the size of the training set would likely improve the model’s ability to generalize to unseen data.} We discuss other possible sources of misclassifications in Section \ref{sec:miss}.

\begin{figure}
    \centering
    \includegraphics[width=0.99\linewidth]{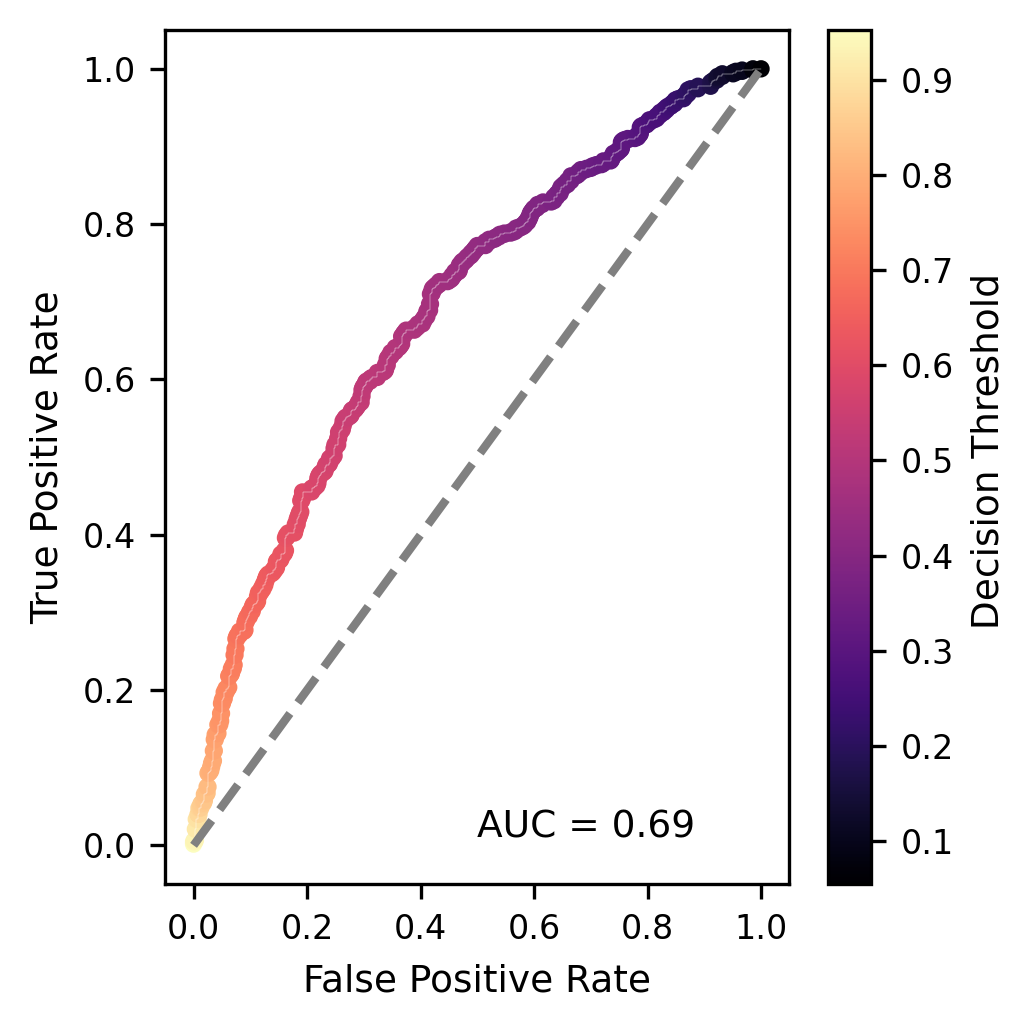}
    \caption{The receiver operating characteristic curve for our test set in Seed 626. The better the network performs, the closer the AUC is to 1. The grey dashed line indicates a network that randomly guesses each time, with an accuracy of 0.5.}
    \label{fig:roc}
\end{figure}

Our network's ROC curve for Seed 626 is shown in Figure~\ref{fig:roc}. 
It performs better than a random classifier, seen by our thick, colored line above the grey dashed (random) line.
\textbf{The area under the curve AUC = 0.69.} 
We use the default decision threshold of 0.5 (i.e. a predicted probability of more than or equal to 0.5 is a merger, and less than 0.5 is a nonmerger) throughout our analysis.
The colorbar in Figure~\ref{fig:roc} shows what changing that decision threshold does to the true and false positive rates.

\begin{figure}
    \centering
    \includegraphics[width=0.99\linewidth]{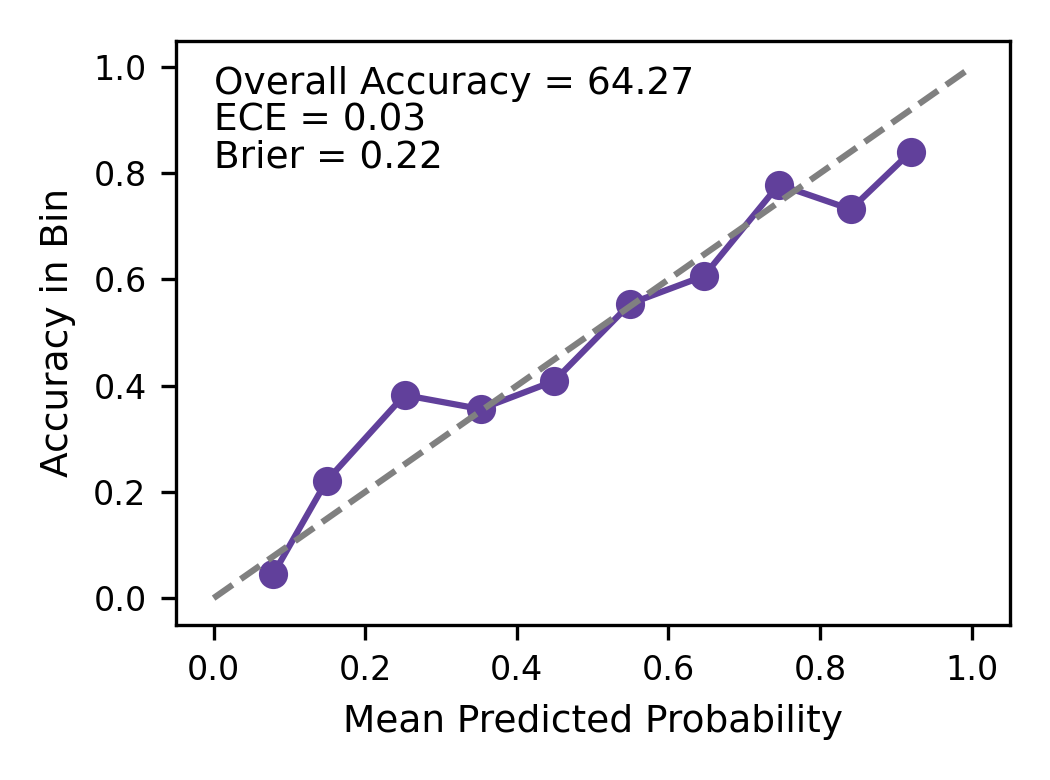}
    \caption{Calibration curve of our Seed 626 network with the test set data \textbf{for the merger class}. The Brier score and ECE, in addition to the overall accuracy, are denoted in the top center. A perfectly calibrated network would have all bins lying along the 1:1 dashed line.}
    \label{fig:reliability}
\end{figure}

In addition to simple metrics like accuracy, we also implement two calibration metrics and plot a calibration curve \textbf{for the merger class} in Figure~\ref{fig:reliability}.
\textbf{A calibration curve shows the relationship between predicted probabilities and model accuracy by grouping predictions into bins over the interval $[0,1]$. For each bin, the mean predicted probability (i.e., average confidence) is compared to the model accuracy. A well-calibrated network produces a curve that lies along the diagonal (grey dashed line). Deviations from this line indicate miscalibration: a curve below the diagonal corresponds to overconfidence (predicted probabilities are too high), while a curve above the diagonal indicates underconfidence (predicted probabilities are too low).}
The Brier score of our network is 0.22, and the ECE is 0.03.
Overall, the network is not severely miscalibrated.

\subsection{Effects of Orientation Angle, Merger Mass Ratio, and Stellar Mass}
\begin{figure}[ht]
    \centering
    \includegraphics[width=0.99\linewidth]{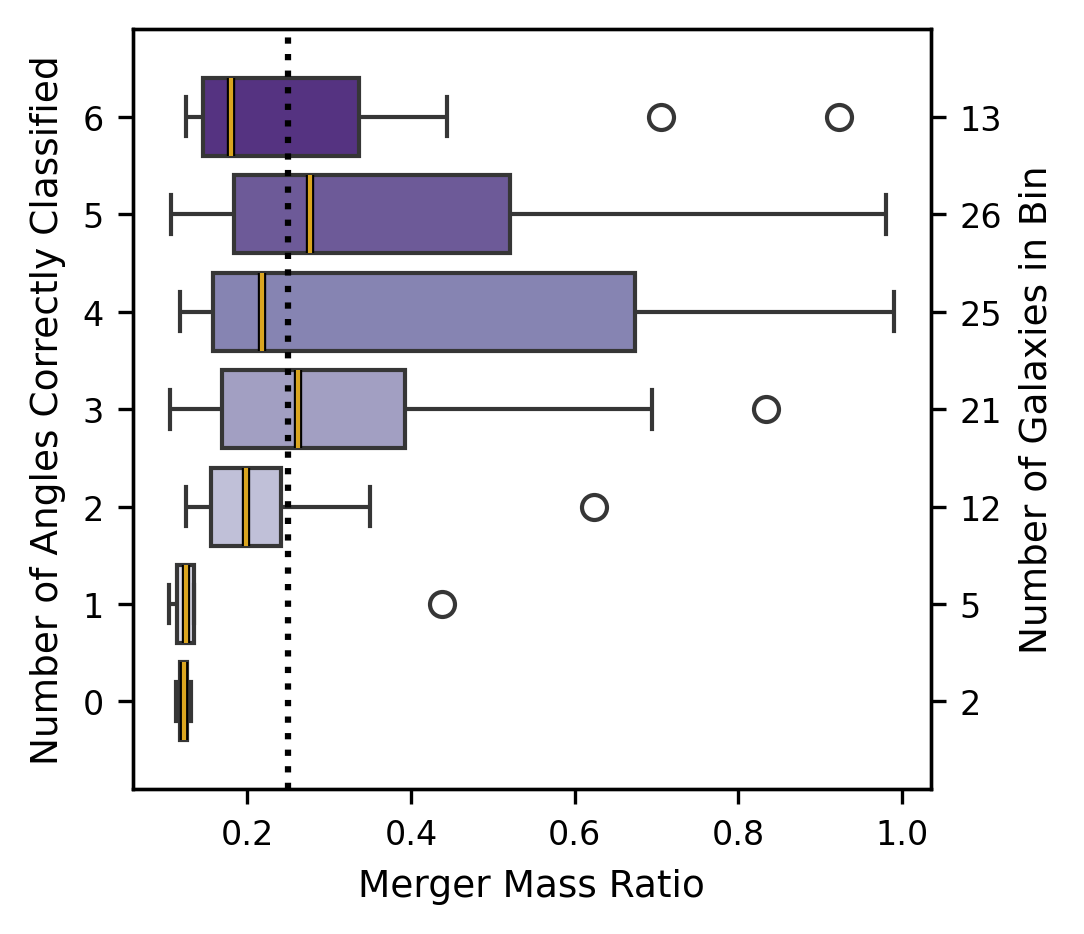}
    \caption{Box plot showing the number of angles (out of 6 possible angles) for which each merger was correctly classified as a function of the merger mass ratio in Seed 626. A box plot extends from the first quartile to the third quartile of the data, with the solid line marking the median. The whiskers extend to $1.5$ times the range of the first to third quartile. The empty circles are any points not included in the range of the whiskers. The black dotted line marks $q = 0.25$, the divider between a minor and major merger. The bottom, lightest box shows that \textbf{only two mergers} are misidentified from every angle.} 
    \label{fig:anglesmassratio}
\end{figure}

\begin{figure}[ht]
    \centering
    \includegraphics[width=0.99\linewidth]{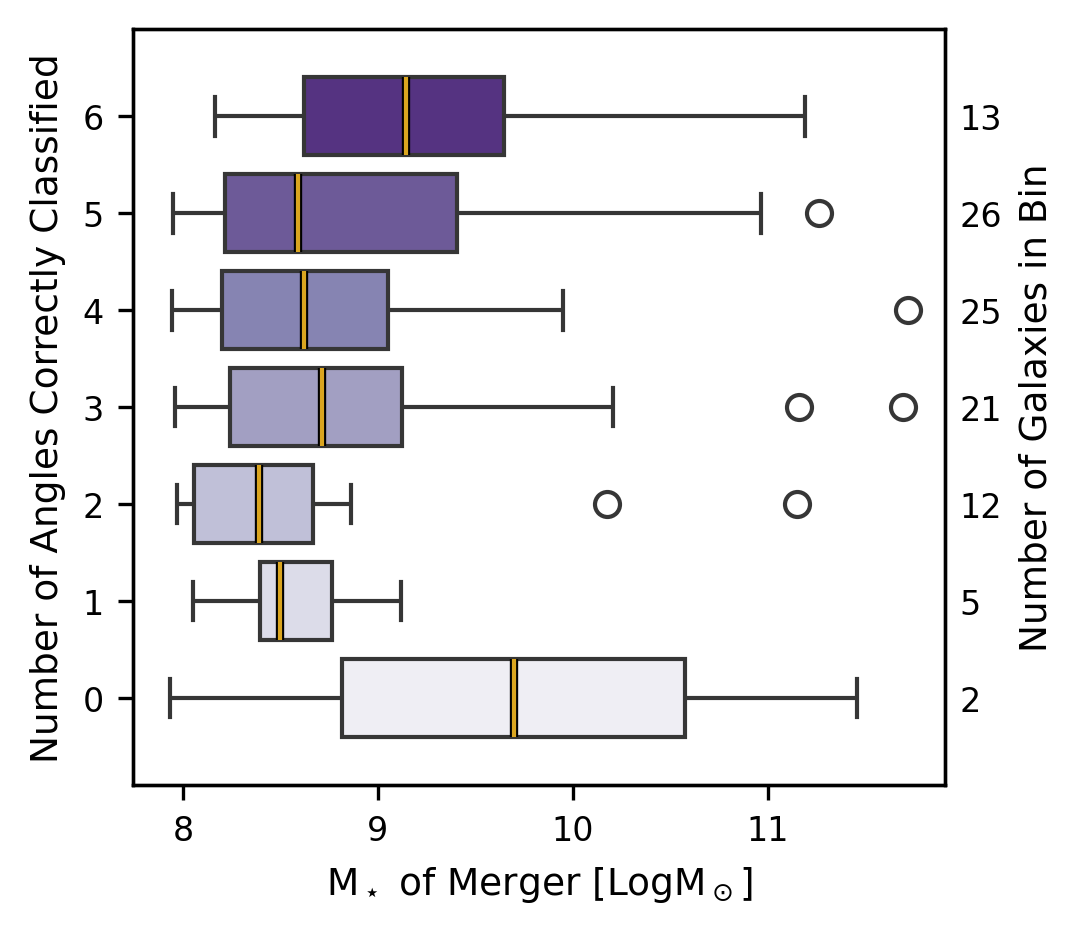}
    \includegraphics[width=0.99\linewidth]{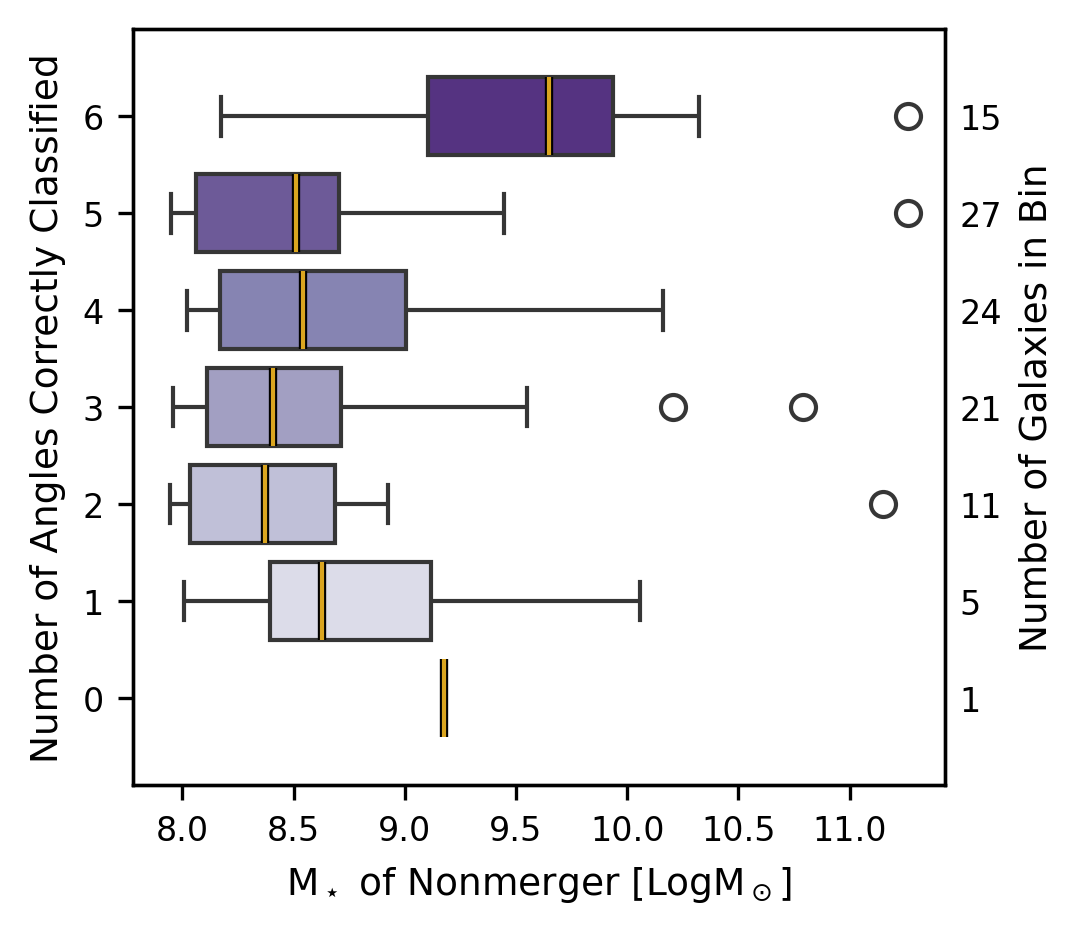}
    \caption{Box plot similar to Figure \ref{fig:anglesmassratio} showing the number of angles for which a galaxy was classified correctly for a range of stellar masses. The distributions for the mergers are shown on top, and the nonmergers on the bottom. A higher stellar mass did not automatically mean the galaxy was easy to classify, as seen by the outlier circles, and lowest box on the merger plot. \textbf{The lack of a major trend shows that mass is not the main factor leading to misclassifications.}}
    \label{fig:anglesstellarmass}
\end{figure}

To understand where the network performs well and where it does not, we seek to understand the effects of orientation angle, specifically in tandem with the merger mass ratio and stellar mass. 
Even a merger that is clear from one angle, perhaps a face-on disk with clear tidal tails, may look like a nonmerger from another. 
Each galaxy is viewed from 6 angles, so we now look at how many angles each galaxy in the test set was correctly classified from.
For the entire sample as a whole, the mean and standard deviation of the number of angles correct per galaxy for our highest performing random seed is $3.86\pm1.51$, with a median of 4. 
In Seed 626, \textbf{98.09\% (205 of 209) of galaxies were correctly identified from at least one angle, 61.24\% (128 of 209) of galaxies were correctly classified from the majority (4 or more) angles, with only 14.83\% (31 of 209) classified correctly from all 6 angles.
This implies that with our network, we are likely to misclassify a galaxy roughly a third of the time, just due to viewing angle alone.
We can see from the number counts on the right side of Figures~\ref{fig:anglesmassratio} and \ref{fig:anglesstellarmass} that most galaxies have some observation angles from which they are misclassified.
It is not the case overall that some galaxies are always correctly classified and some are never correctly classified purely based on observation angle, but rather that observation angle matters.
This holds for both mergers and nonmergers.}
However, considering how varied the galaxies in our sample can be, we break this down into more detail for the best model (Seed 626).

\begin{table}[]
    \centering
    \begin{tabular}{c|c}
    \multicolumn{2}{c}{Accuracies} \\
    \midrule
    \midrule
    All Mergers & $66.33\pm{3.86}\%$ \\
    Major & $68.91\pm{2.05}\%$ \\
    Minor & $63.74\pm{2.28}\%$ \\
    Early Stage &  $73.58\pm{2.04}\%$ \\
    Late Stage & $60.36\pm{1.84}\%$ \\
    Early \& Major & $76.67\pm{3.00}\%$ \\
    Early \& Minor & $71.11\pm{1.54}\%$ \\
    Late \& Major & $63.17\pm{1.55}\%$ \\
    Late \& Minor & $57.99\pm{2.87}\%$ \\
    Nonmergers & $61.06\pm{3.14}\%$ \\

    \end{tabular}
    \caption{The accuracy for our models, broken down by different types of galaxies. The values shown are the mean and standard deviation from our three random seeds. \textbf{We see that the network performed significantly better on early stage mergers compared to late stage mergers, and better on major mergers than minor mergers. Most misclassifications were from late stage mergers and nonmergers.}}
    \label{tab:acc}
\end{table}

We start by examining only the mergers.  
We show in Figure~\ref{fig:anglesmassratio} the number of angles for which a galaxy was correctly identified relative to its merger mass ratio.
The distributions in these bins (the number of correctly classified angles) are shown with box plots, in order to demonstrate the spread. 
Box plots show the middle 50\% of data inside the box, with the whiskers extending to the farthest datapoint within $1.5$ times that range. 
The outliers on these plots are galaxies with a mass ratio much higher than the bulk of the galaxies in that bin, simply indicating that mass ratio is more rare.
Encouragingly, only \textbf{two} merging galaxies were classified incorrectly every time (the lowest bin on Figure~\ref{fig:anglesmassratio}).
Both mergers that were classified as a nonmerger from all six viewpoints were minor mergers.
Despite that, we also see that plenty of minor mergers are identified often, if not from every angle. 
We conclude that observation angle does strongly affect our ability to identify mergers, especially minor mergers.
If observation angle was not a factor, we would only see galaxies being correctly identified from no angles or every angle.
Instead, we see many galaxies that are correctly classified at most angles, but not all.
This gives us a better understanding of how many mergers we may be missing when these networks are applied to real data, assuming we miss a similar percentage due to observation angle.

We next examine the relationship between the stellar mass of a galaxy and how many angles the network could correctly classify it from. 
In Figure~\ref{fig:anglesstellarmass}, we look at both mergers (top figure) and nonmergers (bottom figure).
It may be initially surprising that some high mass galaxies are often misclassified, since a larger mass could make a merger easier to see, but a large, spheroidal galaxy undergoing a merger, especially a minor one, could easily completely obscure its companion from some angles.
Additionally, we note that there are far more low-mass galaxies than high-mass galaxies in our sample. 
\textbf{Encouragingly, there is not a stark trend with stellar mass.}
Many merger studies are only able to consider high mass galaxies, so it is encouraging that with enough spatial resolution, we do have tools to identify at least a fraction of lower mass mergers, even at $z \sim 1$.

We next ask the question, does our network perform equally well on each type of merger in our merging sample?
We include multiple merger stages in our sample. 
The earliest mergers are still recognized as two separate subhalos by the TNG merger trees, and the latest stage mergers are after coalescence into one subhalo. 
We also include both major and minor mergers.
Our breakdown of merger type and accuracy can be seen in Table~\ref{tab:acc} \textbf{(the mean and standard deviation of accuracies from our models initialized with three different random seeds)}.

One would expect the major mergers to be easier for the network to identify than the minor mergers, as major mergers tend to lead to large-scale disruptions of morphology. 
Minor mergers, on the other hand, may not have much of an effect on the morphology of the larger galaxy.
\textbf{We do find that major mergers are easier to identify than minor mergers, with a difference of 5.17 percentage points (68.91\% vs 63.74\%).}
However, identifying minor mergers is a crucial step in understanding how galaxies evolve and grow (e.g., \citealt{ newman_can_2012, kaviraj_importance_2014, martin_role_2018}.

The network identified early stage mergers more accurately than late stage mergers.
We expect early stage mergers that still have two obvious bulges and may maintain some organized structure before they truly encounter the other galaxy in the merger to be easier to classify.
However, multiple papers have found that CNNs can be successful at finding post-merger galaxies at $z < 1$ (e.g., \citealt{bickley_convolutional_2021,ferreira_galaxy_2024, bickley_effect_2024}).
CNNs will be key in analyzing the volume of data coming from large imaging survey telescopes \textbf{and therefore we need to be specific in our merger identification tasks.
CNNs specifically trained to find late stage mergers vastly outperformed our $60.36\%$.}

\textbf{Finally, we examine the interplay between merger stage and mass ratio.
We see that early stage mergers were much easier to detect than late stage mergers for both major (by 13.50 percentage points) and minor mergers (by 13.12 percentage points). 
\revtwo{Examples of late stage mergers that are not immediately obvious can be seen in Figure~\ref{fig:visexample} (top and middle rows, third and fourth columns.)}
Though the late stage merger results are poor for both mass ratio classes, the early stage minor merger results provide a positive outlook: even down to $M_\star>10^8M_\odot$, we are able to detect early stage, minor mergers at $z\sim1$ with 71.11\% accuracy, provided there are enough examples in the training set. 
This is promising for surveys with higher resolution, such as those from JWST, which study galaxy evolution and growth.
}

\begin{figure*}
    \centering
    \includegraphics[width=\linewidth]{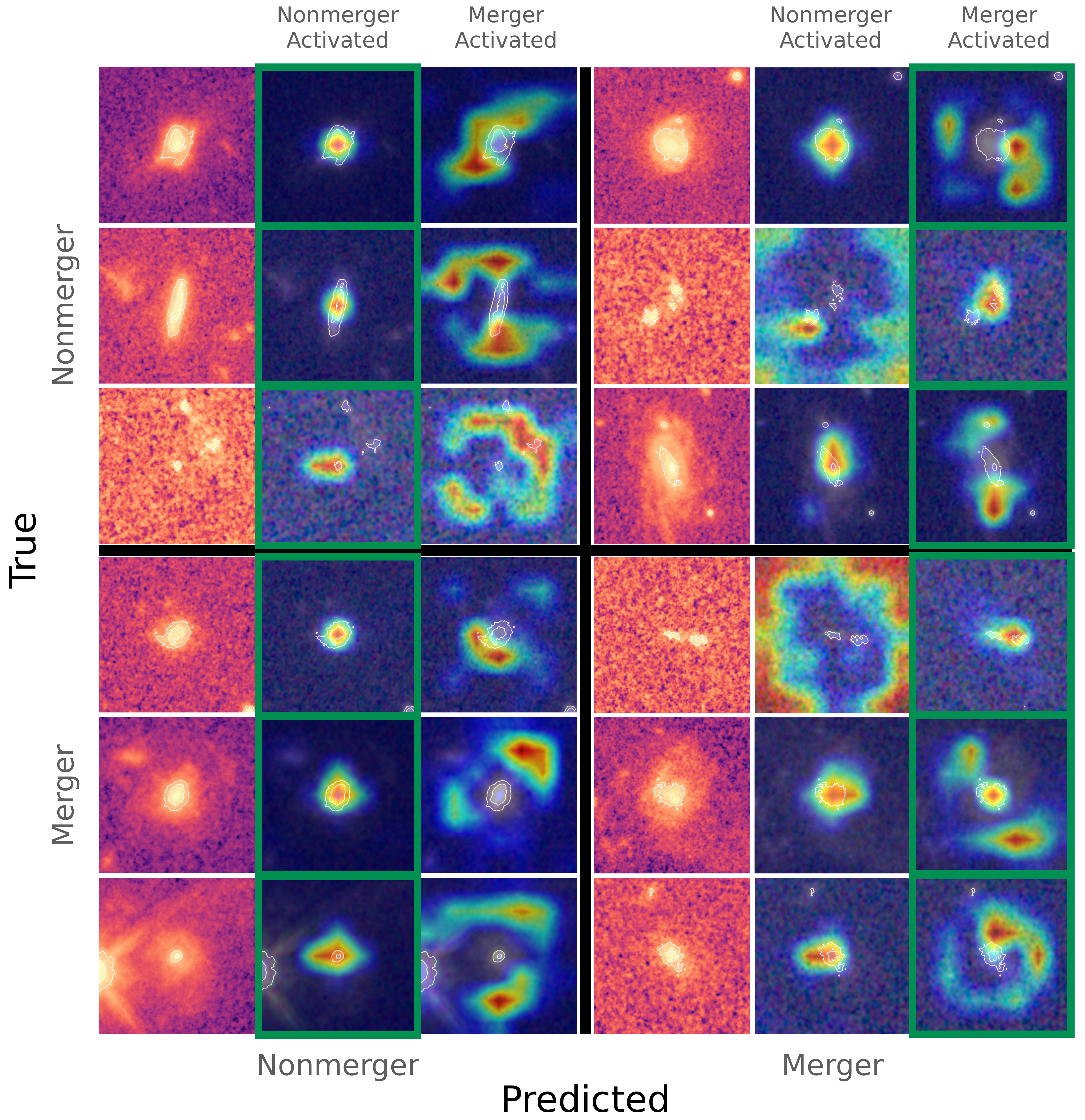}
    \caption{Visual confusion matrix of galaxies in the test set. \textbf{TNs} are in the upper left, and \textbf{TPs} are in the lower right. The contours are 3$\sigma$ and 5$\sigma$ overlayed simply to guide the eye to where the structure is in the Grad-CAM images. The left image is the input image, the middle is the Grad-CAM with class ``nonmerger" activated, and the right is the Grad-CAM with class ``merger" activated. The class that the network predicted is boxed in green. \textbf{The red pixels are the most important to that class and the blue the least.} The Grad-CAMs show that \textbf{the network often, but not always, focuses more on the center of the galaxy when predicting a nonmergers and the more extended features when predicting a merger. We also see that the network has learned to ignore other sources, such as stars, seen in the bottom image in the FN quadrant.}}
    \label{fig:Grad-CAM_CM}
\end{figure*}

\subsection{Understanding Misclassifications} \label{sec:miss}
In Figure~\ref{fig:Grad-CAM_CM} we show example galaxies and their Grad-CAMs set up in the same layout as a confusion matrix. 
For each example, the input galaxy image is on the left, the Grad-CAM with class ``nonmerger" activated is in the center, and the Grad-CAM with class ``merger" activated is on the right.
The \textbf{TNs and TPs} are on the diagonal, with the misclassifications on the off-diagonals. 
When discussing Figure~\ref{fig:Grad-CAM_CM}, activating a class refers to asking the network to highlight which pixels are important for that class.
\textbf{The Grad-CAMs offer confidence that the network has learned not to focus on background noise or sources, as seen in the bottom figure of the \textbf{FN} quadrant, an important quality of successful merger identification using CNNs \citep{bottrell_deep_2019}.
We can see that overall, when predicting a nonmerger the network tends to focus on the central pixels of the image and thus the more central regions of the galaxies.
On the other hand, when predicting a merger, the Grad-CAM highlights borader swaths of pixels, and showing the network tends to focus on more of the extended features of the galaxy.
The Grad-CAMs confirm that the network learned fainter physical features, such as tidal tails, that are not as obvious to the naked eye (e.g., bottom image in the TP quadrant).
}

\textbf{There are some cases where this overall trend of central pixels for nonmergers and extended pixels for mergers does not hold. 
The images where the Grad-CAMs show the opposite pattern (i.e., the nonmerger activation is spread in the outskirts and the merger activation is focused on the center) tend to be low mass, faint galaxies.
The two central bulges are sometimes the only feature that can be seen above the noise limit.
These cases, with faint galaxies with two bulges, are often predicted as a merger, and is an important source of contamination to consider in future merger samples.
Examples of this can be seen in the middle image in the FP quadrant and the top image of the TP quadrant.}

We can also see in this image that some misclassifications make sense to the human eye, while some are more confusing. 
The \textbf{FNs} are seen in the lower left.
The top image has a central structure but is overall quite smooth, with no obvious secondary bulge.
Though the galaxy appears to have some structure from the contours, the single bright nucleus and relatively clean background make sense for why this merger was misclassified as a nonmerger.
\textbf{However, the middle galaxy in the FN quadrant does contain more extended features visible even to the human eye. 
Here, we speculate the network may have struggled to separate a merger from an isolated galalxy with an added background source, and ended up ignoring the extended features.
The bottom FN image was likely confusing due to the star on the left.} 

The top right corner represents FPs. 
\textbf{The top FP image shows the merger activation focusing on extended features that are not obvious to the human eye. 
This is a case of a misclassification that does not make immediate sense.
However, the two images below show morphological features that could belong to mergers, so we understand why the network classified them as mergers.
The middle galaxy is faint, and three clumps are the apparent structure. 
If that is all the information the network has, it makes sense that it may have thought these were multiple bulges, though they may be star formation clumps.
The bottom image in the FP quadrant is a brighter galaxy, but it does have a few clumps and extended features in the image.
Indeed, the merger activation highlights the top clump and extended feature at the bottom. 
Again, though it is a nonmerger, it is easy to see why the network predicted a merger.}

\begin{figure}
    \includegraphics[width=0.99\linewidth]{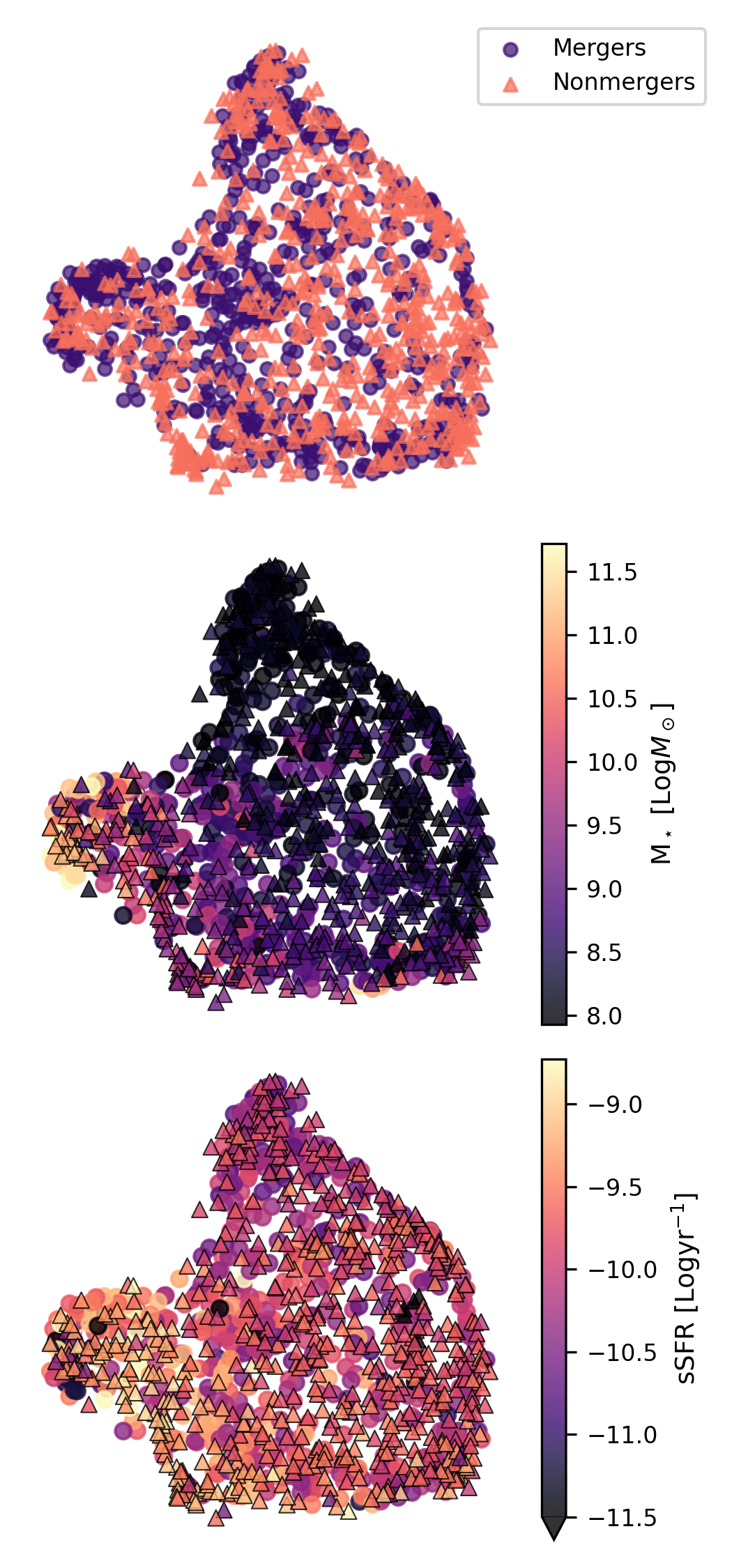}
    \caption{\textbf{UMAPs of the test set color-coded by true class (\emph{top}, mergers as circles and nonmergers as triangles), stellar mass (\emph{middle}), and specific star formation rate (\emph{bottom}).} The true nonmergers are triangles and the true mergers are circles. We exclude axes because the important information in a UMAP is in the relative distance between points and clustering patterns, not in absolute distances. There are clear trends with both stellar mass and star formation rate, showing that the network picks up on these quantities even with no input information about them. }
    \label{fig:UMAP}
\end{figure}

UMAP \citep{mcinnes_umap_2020} is a non-linear, dimensionality reduction technique that we use to visualize the latent space of a CNN.
In Figure~\ref{fig:UMAP}, we show UMAPs of the test set images with colors representing values of different physical quantities of the galaxies. 
\textbf{As the latent space, we use the 640-dimensional final layer of the Zoobot ConvNeXt encoder, before the added classification head.}

\textbf{On the top plot of Figure~\ref{fig:UMAP} we show the two classes (mergers - purple circles, nonmergers - orange triangles). We do not see a clear distinction between the classes overall, but the model shows clear trends based on different physical properties of galaxies. Namely,} we see a clear gradient in the UMAP when colored by the stellar mass of each galaxy (middle plot of Figure~\ref{fig:UMAP}). 
The low-mass galaxies are on the top and right of the UMAP, with stellar mass increasing towards the left.
No stellar mass information was provided during training, but the network was able to recognize this physically meaningful quantity.
There is again a clear gradient in the UMAP when colored by specific star formation rate (bottom plot of Figure~\ref{fig:UMAP}).
Similarly to stellar mass, the lower sSFR galaxies are on the top and right, and sSFR increases towards the left, even though no sSFR information was input to the model.
The exception is extremely low sSFR, which is more scattered throughout. 
Because of this gradient, we speculate some of the nonmergers misclassified as mergers may be due to high, clumpy star formation, likely in the nonmergers seen on the left.
There were no obvious trends for UMAPs relative to the merger stage or merger mass ratio.

\begin{figure*}
    \includegraphics[width=0.95\textwidth]{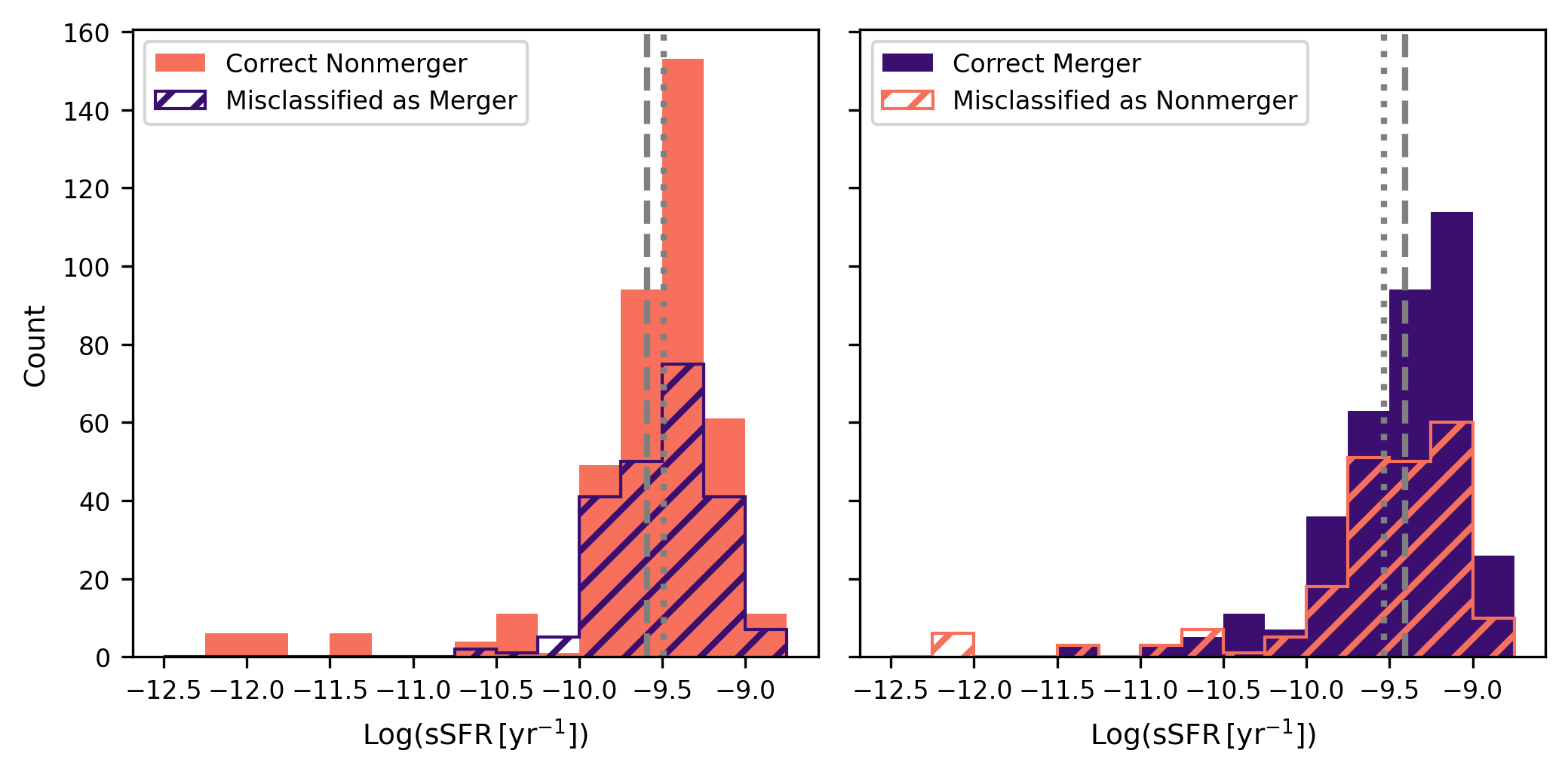}
    \caption{ \emph{Left:} sSFRs of nonmergers that were classified correctly as nonmergers in orange and incorrectly classified as mergers in purple. \emph{Right:} sSFRs of mergers that were correctly classified as mergers in purple and incorrectly as nonmergers in orange. The dashed and dotted lines represent the mean of the correct and missclasified distributions, respectively. The incorrectly classified nonmergers have higher sSFRs than the correctly classified images and vice versa for the incorrectly classified mergers.}
    \label{fig:rightwrong}
\end{figure*}

To further investigate our speculation from UMAP, we plot the specific star formation rates in Figure \ref{fig:rightwrong}.
The nonmergers incorrectly classified as mergers had a higher mean sSFR than the correctly classified nonmergers (dotted and dashed lines on the left plot, respectively).
The mergers incorrectly classified as nonmergers had a lower mean sSFR than the correctly classified mergers (dotted and dashed lines on the right plot, respectively).
This agrees with what the UMAPs showed: the trend with sSFR could account for some of the misclassifications.
The network may have learned that mergers tend to have higher sSFRs than nonmergers, or picked up on a feature in the image correlated to sSFR, leading to this result.

\section{Discussion} \label{sec:discussion}

\subsection{Comparison to Other Studies} 
Galaxy merger identification is an inherently difficult task. 
Many methods have been developed for this task, from visual identification to non-parametric methods such as CAS and Gini-M$_{20}$.
These non-parametric methods are designed to quantify the distribution of light in an image, for example, where the light is concentrated and the range of galaxy brightness throughout the image, into a single value.
The threshold value separates mergers and nonmergers. 
These statistics alone can only capture $\sim50\%$ of mergers, but when combined can be incredibly powerful \citep{nevin_accurate_2019, snyder_automated_2019, wilkinson_limitations_2024}.

For example, \citet{nevin_accurate_2019} successfully classify mock SDSS mergers and nonmergers with linear discriminant analysis (LDA). 
They achieve accuracies of 85\% for major mergers and 81\% for minor mergers.
\citet{wilkinson_limitations_2024} builds on this result, creating mock galaxy images at $z < 0.2$ with spatial resolutions varying from lower than SDSS to higher than Rubin ten year co-adds, applying non-parametric methods, and additionally combining them with LDA and random forest methods. 
Even for their pristine imaging (before downgrading resolution with atmospheric blurring and sky noise), single non-parametric statistics provide maximum completeness (number of true mergers identified correctly/number of total mergers in the sample) of only $\sim$55\%.
This increases to 73\% with an LDA, and 86\% with a random forest.
Random forests have also been used for higher-redshift samples.
At a redshift of $z = 4$, \citet{snyder_automated_2019} achieves a $\sim$70\% completeness for mock \HST{} and \jwst{} images, roughly twice that when only using two non-parametric statistics (Gini - M$_{20}$ or C-A). 
All three of these works use galaxies with $M_\star\sim10^{10}M_\odot$.
From $0.5 < z < 4$, the random forest in \citet{rose_identifying_2023} applied to galaxies with $10^5M_\odot < M_\star < 10^{12}M_\odot$ attains an accuracy of $\sim$60\% on mock \jwst{} CEERS imaging. 
They extend this result to $4 < z < 5$ in \citet{rose_ceers_2024} and correctly classify 59\% of nonmergers and 67\% of mergers.

Many works have shown that CNNs are successful at classifying galaxy mergers and often outperform non-parametric methods and LDA and random forest at higher redshifts.
With $M_\star>10^{10}M_\odot$ at $z \sim 0.01$, \citet{pearson_identifying_2019} classifies mergers and nonmergers in simulated SDSS images from the EAGLE simulation, achieving 65.2\% accuracy. 
They discuss that the simulation includes a more complete sample of mergers, not just those easy to identify by eye. 
\textbf{We achieve comparable accuracy here with much lower masses, and higher redshift.}
Therefore, it is by default a harder task for the CNN than identifying mergers in SDSS observations that were also visually identified, where they achieved 95.1\% accuracy.
\textbf{\citet{margalef-bentabol_galaxy_2024} emphasizes this point, stating that $\sim80\%$ is the best that modern merger classifiers can achieve when trained from simulations.}

\textbf{Late stage mergers are often the trickiest to identify by-eye.
CNNs provide an avenue for expanding accurate merger identification into this late stage regime.
\citet{bickley_convolutional_2021} ($M_\star>10^{10} M_\odot, z<0.5$) achieves 88\% accuracy and \citet{ferreira_galaxy_2024, ferreira_galaxy_2025} ($M_\star>10^{10} M_\odot, z<0.3$) achieves 84\% accuracy when classifying nonmergers from late stage mergers in the CFIS and UNIONS Surveys, respectively.
With low redshift and high mass used in these works, late stage mergers can be identified.
However, our network especially struggled with late stage minor mergers (60.36\% accuracy), likely because of the higher redshift and lower mass.}

Mergers and nonmergers around cosmic noon were first classified with a CNN in \citet{ciprijanovic_deepmerge_2020}, who used mock images at $z = 2$ from the original Illustris simulation.
They showed that adding noise to images does decrease accuracy (79\% accuracy on pristine images and 76\% on noisy images), but noise is a key aspect for realistic mock images.
Their Grad-CAMs highlighted larger areas of the image for the merger class and focused on more compact regions for nonmergers.
Relevant for this work, \citet{ciprijanovic_deepmerge_2020} uses the same 500Myr merging window as our work here. 
However, their stellar mass lower limit is $M_\star = 10^{9.5}M_\odot$.
They note that most of their misclassifications are coming from the low mass end of their sample.
Our sample stretches to even lower stellar masses, so stellar mass may be responsible for some of our misclassifications (see Section \ref{sec:miss}).

CANDELS galaxies with $M_\star>10^{10}M_\odot$ have been classified out to \z{} = 3 \citep{ferreira_galaxy_2020}. 
That mass limit was specifically chosen to use with the largest box size of TNG, TNG300, in order to maximize training set size. 
They achieve 87\% accuracy on pre-mergers, 78\% on post-mergers, 94\% on nonmergers.
It is mentioned that the spatial resolution in CANDELS is higher than that of some of their mock images, due to the simulation's spatial resolution.
Our work uses the smaller box, but higher resolution run of TNG50, and thus we stretch to galaxies with stellar masses down to 100 times smaller. 
That wider mass range comes with a drop in overall accuracy.

\textbf{Classifying mergers from all nonmergers is a more difficult task than separating e.g., mergers from star forming nonmergers.
\citet{ferreira_simulation-driven_2022} classifies mergers from star-forming galaxies in CANDELS, trained on TNG100, with accuracy of at least 80\% out to $z\sim2$ in galaxies of $M_\star>10^{9.5}M_\odot$. 
This result, combined with ours, provides insight into how to conduct future merger classification projects-- in order to maximize accuracy and limit contamination in merger samples, we should train on large datasets and limit the nonmerger sample beyond just mass (e.g., by sSFR or Asymmetry).}

Pushing past cosmic noon, \citet{rose_ceers_2024} classifies mergers and nonmergers in CEERS with mock images from TNG100.
Different from other works, their test set of mock images is unbalanced, with far more nonmergers than mergers, to better represent the real universe. 
They split their data into three redshift bins between $3 < z < 5$ and achieve $\sim60-70\%$ accuracies in all of them, \textbf{similar to our results}.
Their mass range stretches to even lower masses than this work, with galaxies included at $M_\star>10^{7}M_\odot$.
The Grad-CAMs of these mock \jwst{} galaxies do not show clear patterns when all six filters are included. 
They note they see evidence of the network focusing on the galaxy in some single filter Grad-CAM images, but the activation in other images are seemingly random.

\subsection{Why Do CNNs Struggle with Merger Identification?}
We showed that CNNs can find non local ($z \sim 1$) mergers at low mass ratios ($q > 1:10$) and low stellar masses ($M_\star > 10^8 M_\odot$) when enough training examples are provided. TNG50 provides a high resolution training set enabling this classification.
While the network outperforms previous CNNs at this low mass range for specific classes, it still misclassifies many galaxies, especially in the late stage, minor regime.
We now explore what makes merger identification an inherently difficult task, especially at $z > 1$. 
First we discuss the morphology of galaxies in combination with image resolution and depth. 
We then focus on the effect of viewing orientation and the inherent ceiling on merger identification.

\subsubsection{Impact of Morphology and Image Quality}
While a task like identifying two stellar bulges at low redshift and high masses may be trivial for a CNN, distinguishing between clumpy star-forming regions and two stellar bulges at higher redshifts and low masses is not so trivial.
Additionally, tidal features can be key to distinguishing a merger from a clumpy, isolated galaxy, but these features can be dim, especially compared to the brightnesses of star-forming regions and stellar bulges.
Normalization of the images with a log stretch does help this issue, ensuring that all pixel values are between 0 and 1 and the values are well dispersed within that range. 
With high spatial resolution data from \nancy{}, Rubin, \jwst{}, and \Euclid{}, merger identification at high-$z$ should become easier, provided we create accurate training sets, and trust our ML algorithms.
Ensuring that the network can understand the difference in these brightness scales and the physical features that they tie to requires many training images.
In our case, TNG50 did not include enough mergers at $M_\star > 10^{10}M_\odot$ for the network to understand how to classify high-mass mergers. 
Conversely, \citet{margalef-bentabol_galaxy_2024} did not include as many $M_\star\sim10^9M_\odot$ galaxies as $M_\star>10^{10}M_\odot$ galaxies, and their completeness scores were highly mass dependent (see their Figure 8).
Training a network with realistic mock images improves performance on real data \citep{bottrell_deep_2019}.
In addition to supplying a deep network with high-quality mock images, it is also necessary to have an adequate number of training examples spanning different stellar masses, merger stages, and mass ratios.

\subsubsection{Impact of Viewing Angle on Accuracy} \label{sec:angle}
Orientation angle has a known effect on classifying the morphology of galaxies. 
Extending this to mergers, the orientation angle can make it easier or more difficult to visually identify a companion galaxy or merger.
\citet{bickley_effect_2024} suggests that there may be an upper limit on the accuracy of a CNN for merger identification around 90 percent, partially due to orientation. 
They note that at some level, there is no way to distinguish a clumpy nonmerger from a merger, or to view every galaxy from a favorable angle. 
\cite{wilkinson_limitations_2024} conduct a thorough investigation of the limitations of their merger identification scheme by viewing angle. 
They classify a $3\times10^{10}M_\odot$ major merger ($q = 0.26$) viewed at 648 angles.
They find for this easy-to-identify major merger at \z{} = 0.05 that $\sim$8\% of the viewing angles are unable to identify the merger, even with their well trained LDA and random forest algorithms.

\textbf{With their entire sample, they study four angles of each galaxy, similar to our six in Section~\ref{sec:angle}. 
They find their methods detect mergers in each of the four angles 49\% (LDA) and 57\% (random forest) of the time.
This is much better performance than our 14.83\% of the time.
However, if we consider the fraction of our galaxies that could be identified from at least four angles, we exceed \citet{wilkinson_limitations_2024} with 61.24\%.
\citet{wilkinson_limitations_2024} also state that if a merger is identified in at least one angle by an LDA or random forest, it is likely to be missed by an individual non-parametric morphology statistic about 50\% of the time, purely due to observation angle. 
With our model, if a galaxy is identified as a merger from at least one angle, meaning it has a physical merger signature, it is likely to be missed about 35\% of the time due to observation angle, based on the counts in Figure~\ref{fig:anglesmassratio}.}

\textbf{Therefore, while each model and method has its own limitations, there may be an inherent accuracy limit due to observation angle, before taking redshift, stellar mass, mass ratio, or telescope parameters into account.}
When we account for the fact that our sample is composed of lower mass (by up to 100 times) and higher redshift (which has an effect on image resolution) galaxies and mergers with larger mass ratios (i.e. minor mergers), we should expect an even lower accuracy upper limit.
If the \citet{wilkinson_limitations_2024} $\sim$90\% accuracy upper limit applies to  $z < 1$ galaxies, then the upper limit will undoubtedly be even lower than 90\% at $z =1 - 1.5$.

\subsubsection{Impact of Star Formation} \label{sec:impactSFR}
As seen in Figure \ref{fig:UMAP}, the CNN recognizes star formation to be important to its final classifications. 
With the multiband images that the CNN is fed, it can learn that a bluer galaxy, \textbf{within our class-balanced training and test set,} and thus a galaxy undergoing recent star formation, is more likely to be a merger than a nonmerger.
This could be seen as a drawback, and \citet{bottrell_deep_2019} use one band to specifically avoid the CNN classifying bluer galaxies as mergers and instead force it to classify based on morphology. 
However, we argue this trend with star formation rate shows that CNN is learning something physically meaningful: a bluer galaxy is more likely to be a merger than a nonmerger \textbf{when the numbers in each class are balanced.}
Indeed, the mass-matched mergers in our TNG50 dataset, on average, have higher star formation rates than the nonmergers \citep{schechter_enhanced_2025}.
Therefore, it makes sense that the CNN would identify a bluer galaxy as a merger more often.
\textbf{This has a worrying implication for applying to real data, when nonmergers greatly outnumber mergers, and thus many blue, nonmerging galaxies would contaminate a merging sample.}

We confirm in Figure \ref{fig:rightwrong} (left plot) that the nonmergers that are incorrectly classified have a higher mean sSFR than the nonmergers that are correctly classified. 
The right plot on Figure \ref{fig:rightwrong} shows that mergers misclassified as nonmergers have a lower mean sSFR than the mergers correctly classified, especially with a few low sSFR galaxies causing a longer tail of the missclassified distribution.

In order to confirm that color information is not hindering our model, we additionally ran two versions of a greyscale CNN. 
One model was fed a single image that summed the flux from all three filters in preprocessing, and thus did not have the breakdown of astronomical filters that the three-channel CNN did. 
The other was given a single filter \textbf{F160W} image.
\textbf{The model trained on summed images performed similarly to the three-channel CNN (accuracy of 61.96\%, compared to the three-channel 63.56\%), but the F160W-only model barely learned (accuracy 54.70\%).
See details and performance metrics of these models in Appendix~\ref{sec:app}.
Neither greyscale model added any physical insight into classifications, or removed the sSFR bias successfully}.
Taking input channels away did not decrease a reliance on star formation, and appears to have removed some key features the model used when separating mergers from nonmergers.

\subsection{Looking Ahead With Applications to Real Data}

In this work we used fully radiative transferred images, as it is important to train on mock images as similar to real observations as possible \citep{bottrell_deep_2019}.
However, no matter how careful the mock image procedure is, there will always be differences between mock images and real data such as noise, image artefacts, and morphology of galaxies depending on the simulation used for training.
\citet{ciprijanovic_deepmerge_2021} applied domain adaptation techniques, which enable the network to find and utilize similar features from the training domain (mock images of simulated galaxies) and the target domain (real observations) i.e., domain-invariant features.
These techniques help overcome the limitation any differences in the mock images and real observations. 
Since radiative transfer is expensive, some papers opt to not use radiative transfer with only a small reduction in accuracy (e.g., \citealt{bottrell_deep_2019}).
At $z > 1$ we argue it is important to use radiative transfer as dust is impacting our observations and AGN, which can be very dusty, are more common.
\textbf{In a follow up paper (Schechter et al. in prep), we will analyze the importance of using radiative transfer in the mock image process for merger identification} and explore using domain adaptation in place of full radiative transfer for merger classification with CNNs in order to save computational time.

As we broach the high-$z$ merger universe, we want to build trust in our use of AI and understand where AI cannot provide all of the answers.
Using XAI techniques is crucial as we are looking at imaging where there is no obvious correct answer.
Additionally, understanding which galaxies cause miscalibration can provide insight into which tasks ML is well-suited for and for which tasks more classical methods may still be preferred. 
For example, combining spectroscopic pairs for early stage mergers with CNNs specifically trained for late stage mergers can give a more complete catalog of all mergers in a sample \citep{ferreira_galaxy_2024-1}.
This also makes the CNN's task easier, because it only has to learn what a late stage merger looks like, instead of generalizing to any stage of merger. 
Knowing which tools to apply to which problems is crucial as we enter an era of big data in astronomy.

UMAPs show us that the network is sensitive to the stellar mass and star formation rates of the galaxies.
In the future, we could potentially improve performance by combining galaxies with $M_\star < 10^{9.5}M_\odot$ from TNG50 with a sample of galaxies with $M_\star > 10^{9.5}M_\odot$ from the larger simulation box size, TNG100, to create a more mass-balanced and larger training set. This would be similar to the treatment in \citet{margalef-bentabol_galaxy_2024}, where authors combine TNG100 and TGN300 galaxies for their training set.
Our nonmerger sample is currently only mass-matched due to the small box size of TNG50.
To match in any other quantity required widening the mass match threshold too much, due to the small number of galaxies with $M_\star > 10^{10}M_\odot$. 
However, if we draw high-mass galaxies from TNG100, we could find SFR and mass matched nonmergers to try to reduce misclassifications. 
This would also provide a larger training set overall, which would be helpful since our current training set is small by ML standards.
With a training set of similar numbers of major and high-mass mergers as minor and low-mass mergers, we could potentially improve the distinction between mergers and nonmergers for all subcategories of galaxies. 

\textbf{Finally, we must recognize that in observations, nonmergers far outnumber mergers.
Therefore, it is likely that merger samples created by any given model will be contaminated with nonmergers.
In addition to domain adaptation and XAI techniques, \citet{ferreira_galaxy_2024} propose that to minimize contamination, we utilize jury-based classifications for mergers to overcome the class imbalance. 
They train multiple models on separate subsets of their training set, and use each model's decision on a test set galaxy as a vote.
When only counting unanimous mergers, they can achieve a remarkable purity of 95\%.
However, with a sample of unanimous mergers, they are likely excluding less-obvious mergers.
Combining a jury system with Grad-CAM or UMAP, to uncover the physical properties of mergers and nonmergers that are classified unanimously versus those classified correctly by only a subset of models, will lead to both more accurate galaxy merger catalogs from large survey telescopes, and better knowledge of how interlopers may be affecting the final sample.
}

\FloatBarrier
\section{Conclusion} \label{sec:conclusion}
We trained a CNN on mock \HST{} CANDELS images to identify galaxy mergers at $1 \leq z \leq 1.5$. 
We used the highest spatial resolution simulation in the TNG suite, TNG50, enabling more detailed structures of galaxies and galaxies of lower masses to be included in our training set. 
With mergers in stages from pre- to post-coalescence, $M_\star > 10^8M_\odot$, and $q \geq 1:10$, we successfully classified less obvious galaxy mergers close to cosmic noon.
Our main results are as follows:
\begin{enumerate}
    \item Our network was \textbf{63.56\% accurate, and was able to identify major mergers 68.91\% and minor mergers 63.74\%} of the time. 
    \item Early stage mergers (two clear galaxies) were identified \textbf{with higher accuracy} than late stage mergers (around coalescence) \textbf{by 13.22 percentage points (73.58\% vs 60.36\%). The network struggled the most with late stage, minor mergers, achieving only 57.99\% accuracy.} 
    \item Orientation angle matters when searching for mergers, as there are some angles where the merger is unlikely to be correctly identified. 
    \textbf{However, 98.09\% of galaxies were correctly identified from at least one out of six possible angles. The observation angle distribution is not bimodal-- most galaxies have some angles they can be correctly classified from, and others where they cannot, rather than being identified from all or none.} 
    \item In order for a single model to accurately classify galaxies in a wide range of stellar masses (e.g., $\sim10^8M_\odot - 10^{12}M_\odot$), the training set must include sufficient examples from the entire mass range (even if that is not representative of the observed galaxy mass distribution).
    \item \textbf{If the training set contains different star formation distributions between mergers and nonmergers, the CNN can exploit this property for classification. Therefore, training sets that are matched in both stellar mass and SFR are necessary to confidently separate these mergers from nonmergers.}
\end{enumerate}

Accurately identifying galaxy mergers is a key step in understanding how galaxies build stellar mass and evolve in morphologies over time. 
To do this with large imaging surveys we need trustworthy ML algorithms.
Understanding how to build better training sets as well as where \textbf{and how often} high-$z$ merger identification is failing is crucial as we step into the era of \jwst{}, \nancy{} Rubin, and \textit{Euclid}.

\begin{acknowledgments}
The authors would like to thank the anonymous reviewer for a thorough report that greatly improved this paper. The computations in this paper were run on the FASRC Cannon cluster supported by the FAS Division of Science Research Computing Group at Harvard University. This work utilized the Alpine high performance computing resource at the University of Colorado Boulder. Alpine is jointly funded by the University of Colorado Boulder, the University of Colorado Anschutz, and Colorado State University and with support from NSF grants OAC-2201538 and OAC-2322260. This work utilized the Blanca condo computing resource at the University of Colorado Boulder. Blanca is jointly funded by computing users and the University of Colorado Boulder.

A.L.S. would like to thank Michelle Ntampaka for helpful discussions in the early stages of this work.

A.L.S. and J.M.C. acknowledge support from NASA’s Astrophysics Data Analysis program, grant number 80NSSC21K0646, and NSF AST-1847938.

X.S. acknowledges the support from the NASA theory grant JWST-AR-04814. 

The work of A.S. was supported by the National Science Foundation MPS-Ascend Postdoctoral Research Fellowship under grant No. 2213288.

L.B. acknowledges support from NASA Astrophysics Theory program, grant 80NSSC22K0808, and NSF AAG 2307171.

A.\'C: This work was produced by Fermi Forward Discovery Group, LLC under Contract No. 89243024CSC000002 with the U.S. Department of Energy, Office of Science, Office of High Energy Physics. The United States Government retains and the publisher, by accepting the work for publication, acknowledges that the United States Government retains a non-exclusive, paid-up, irrevocable, world-wide license to publish or reproduce the published form of this work, or allow others to do so, for United States Government purposes. The Department of Energy will provide public access to these results of federally sponsored research in accordance with the DOE Public Access Plan (\url{http://energy.gov/downloads/doe-public-access-plan}).

We acknowledge the use of LLMs in troubleshooting code and occasional rewording of parts of a sentence. 
The main code and complete sentences are original work.

We acknowledge the Deep Skies Lab as a community of multi-domain experts and collaborators who’ve facilitated an environment of open discussion, idea generation, and collaboration. This community was important for the development of this project.

\textbf{Author Contributions:} The following authors contributed in different ways to the manuscript.

\noindent Schechter: Writing manuscript, mock image creation post radiative transfer, all ML code and analysis

\noindent \'{C}iprijanovi\'{c}: editing manuscript, mentoring and overseeing ML code and analysis

\noindent Shen: Radiative transfer, writing section 2.2.1

\noindent Nevin: Editing manuscript, creating TNG merger catalog, mentoring and overseeing ML code and analysis

\noindent Comerford: Editing manuscript, mentoring and overall direction of paper

\noindent Stemo: Assistance with mock image creation

\noindent Blecha: Initial ideas and direction of paper

\revtwo{\noindent Fraley: Troubleshooting mock image code}

\end{acknowledgments}

\bibliography{references_zotero,references}{}

@ARTICLE{Torrey2015,
       author = {{Torrey}, Paul and {Snyder}, Gregory F. and {Vogelsberger}, Mark and {Hayward}, Christopher C. and {Genel}, Shy and {Sijacki}, Debora and {Springel}, Volker and {Hernquist}, Lars and {Nelson}, Dylan and {Kriek}, Mariska and {Pillepich}, Annalisa and {Sales}, Laura V. and {McBride}, Cameron K.},
        title = "{Synthetic galaxy images and spectra from the Illustris simulation}",
      journal = {\mnras},
         year = 2015,
        month = mar,
       volume = {447},
       number = {3},
        pages = {2753-2771},
          doi = {10.1093/mnras/stu2592},
archivePrefix = {arXiv},
       eprint = {1411.3717},
 primaryClass = {astro-ph.GA},
       adsurl = {https://ui.adsabs.harvard.edu/abs/2015MNRAS.447.2753T}
}

@ARTICLE{RGomez2019,
       author = {{Rodriguez-Gomez}, Vicente and {Snyder}, Gregory F. and {Lotz}, Jennifer M. and {Nelson}, Dylan and {Pillepich}, Annalisa and {Springel}, Volker and {Genel}, Shy and {Weinberger}, Rainer and {Tacchella}, Sandro and {Pakmor}, R{\"u}diger and {Torrey}, Paul and {Marinacci}, Federico and {Vogelsberger}, Mark and {Hernquist}, Lars and {Thilker}, David A.},
        title = "{The optical morphologies of galaxies in the IllustrisTNG simulation: a comparison to Pan-STARRS observations}",
      journal = {\mnras},
         year = 2019,
        month = mar,
       volume = {483},
       number = {3},
        pages = {4140-4159},
          doi = {10.1093/mnras/sty3345},
archivePrefix = {arXiv},
       eprint = {1809.08239},
 primaryClass = {astro-ph.GA},
       adsurl = {https://ui.adsabs.harvard.edu/abs/2019MNRAS.483.4140R}
}

@ARTICLE{Shen2024,
       author = {{Shen}, Xuejian and {Vogelsberger}, Mark and {Borrow}, Josh and {Hu}, Yongao and {Erickson}, Evan and {Kannan}, Rahul and {Smith}, Aaron and {Garaldi}, Enrico and {Hernquist}, Lars and {Morishita}, Takahiro and {Tacchella}, Sandro and {Zier}, Oliver and {Sun}, Guochao and {Eilers}, Anna-Christina and {Wang}, Hui},
        title = "{The THESAN project: galaxy sizes during the epoch of reionization}",
      journal = {\mnras},
         year = 2024,
        month = oct,
       volume = {534},
       number = {2},
        pages = {1433-1458},
          doi = {10.1093/mnras/stae2156},
archivePrefix = {arXiv},
       eprint = {2402.08717},
 primaryClass = {astro-ph.GA},
       adsurl = {https://ui.adsabs.harvard.edu/abs/2024MNRAS.534.1433S}
}

@ARTICLE{Dwek1998,
       author = {{Dwek}, Eli},
        title = "{The Evolution of the Elemental Abundances in the Gas and Dust Phases of the Galaxy}",
      journal = {\apj},
         year = 1998,
        month = jul,
       volume = {501},
        pages = {643},
          doi = {10.1086/305829},
archivePrefix = {arXiv},
       eprint = {astro-ph/9707024},
 primaryClass = {astro-ph},
       adsurl = {https://ui.adsabs.harvard.edu/abs/1998ApJ...501..643D}
}

@ARTICLE{Saftly2014,
       author = {{Saftly}, W. and {Baes}, M. and {Camps}, P.},
        title = "{Hierarchical octree and k-d tree grids for 3D radiative transfer simulations}",
      journal = {\aap},
         year = 2014,
        month = jan,
       volume = {561},
          eid = {A77},
        pages = {A77},
          doi = {10.1051/0004-6361/201322593},
archivePrefix = {arXiv},
       eprint = {1311.0705},
 primaryClass = {astro-ph.IM},
       adsurl = {https://ui.adsabs.harvard.edu/abs/2014A&A...561A..77S}
}

@misc{roman,
      title={The Wide Field Infrared Survey Telescope: 100 Hubbles for the 2020s}, 
      author={Rachel Akeson and Lee Armus and Etienne Bachelet and Vanessa Bailey and Lisa Bartusek and Andrea Bellini and Dominic Benford and David Bennett and Aparna Bhattacharya and Ralph Bohlin and Martha Boyer and Valerio Bozza and Geoffrey Bryden and Sebastiano Calchi Novati and Kenneth Carpenter and Stefano Casertano and Ami Choi and David Content and Pratika Dayal and Alan Dressler and Olivier Doré and S. Michael Fall and Xiaohui Fan and Xiao Fang and Alexei Filippenko and Steven Finkelstein and Ryan Foley and Steven Furlanetto and Jason Kalirai and B. Scott Gaudi and Karoline Gilbert and Julien Girard and Kevin Grady and Jenny Greene and Puragra Guhathakurta and Chen Heinrich and Shoubaneh Hemmati and David Hendel and Calen Henderson and Thomas Henning and Christopher Hirata and Shirley Ho and Eric Huff and Anne Hutter and Rolf Jansen and Saurabh Jha and Samson Johnson and David Jones and Jeremy Kasdin and Patrick Kelly and Robert Kirshner and Anton Koekemoer and Jeffrey Kruk and Nikole Lewis and Bruce Macintosh and Piero Madau and Sangeeta Malhotra and Kaisey Mandel and Elena Massara and Daniel Masters and Julie McEnery and Kristen McQuinn and Peter Melchior and Mark Melton and Bertrand Mennesson and Molly Peeples and Matthew Penny and Saul Perlmutter and Alice Pisani and Andrés Plazas and Radek Poleski and Marc Postman and Clément Ranc and Bernard Rauscher and Armin Rest and Aki Roberge and Brant Robertson and Steven Rodney and James Rhoads and Jason Rhodes and Russell Ryan Jr. au2 and Kailash Sahu and David Sand and Dan Scolnic and Anil Seth and Yossi Shvartzvald and Karelle Siellez and Arfon Smith and David Spergel and Keivan Stassun and Rachel Street and Louis-Gregory Strolger and Alexander Szalay and John Trauger and M. A. Troxel and Margaret Turnbull and Roeland van der Marel and Anja von der Linden and Yun Wang and David Weinberg and Benjamin Williams and Rogier Windhorst and Edward Wollack and Hao-Yi Wu and Jennifer Yee and Neil Zimmerman},
      year={2019},
      eprint={1902.05569},
      archivePrefix={arXiv},
      primaryClass={astro-ph.IM},
      url={https://arxiv.org/abs/1902.05569}, 
}

@article{euclid,
   title={Euclid preparation: I. The Euclid Wide Survey},
   volume={662},
   ISSN={1432-0746},
   url={http://dx.doi.org/10.1051/0004-6361/202141938},
   DOI={10.1051/0004-6361/202141938},
   journal={Astronomy \&; Astrophysics},
   publisher={EDP Sciences},
   author={Scaramella, R. and Amiaux, J. and Mellier, Y. and Burigana, C. and Carvalho, C. S. and Cuillandre, J.-C. and Da Silva, A. and Derosa, A. and Dinis, J. and Maiorano, E. and Maris, M. and Tereno, I. and Laureijs, R. and Boenke, T. and Buenadicha, G. and Dupac, X. and Gaspar Venancio, L. M. and Gómez-Álvarez, P. and Hoar, J. and Lorenzo Alvarez, J. and Racca, G. D. and Saavedra-Criado, G. and Schwartz, J. and Vavrek, R. and Schirmer, M. and Aussel, H. and Azzollini, R. and Cardone, V. F. and Cropper, M. and Ealet, A. and Garilli, B. and Gillard, W. and Granett, B. R. and Guzzo, L. and Hoekstra, H. and Jahnke, K. and Kitching, T. and Maciaszek, T. and Meneghetti, M. and Miller, L. and Nakajima, R. and Niemi, S. M. and Pasian, F. and Percival, W. J. and Pottinger, S. and Sauvage, M. and Scodeggio, M. and Wachter, S. and Zacchei, A. and Aghanim, N. and Amara, A. and Auphan, T. and Auricchio, N. and Awan, S. and Balestra, A. and Bender, R. and Bodendorf, C. and Bonino, D. and Branchini, E. and Brau-Nogue, S. and Brescia, M. and Candini, G. P. and Capobianco, V. and Carbone, C. and Carlberg, R. G. and Carretero, J. and Casas, R. and Castander, F. J. and Castellano, M. and Cavuoti, S. and Cimatti, A. and Cledassou, R. and Congedo, G. and Conselice, C. J. and Conversi, L. and Copin, Y. and Corcione, L. and Costille, A. and Courbin, F. and Degaudenzi, H. and Douspis, M. and Dubath, F. and Duncan, C. A. J. and Dusini, S. and Farrens, S. and Ferriol, S. and Fosalba, P. and Fourmanoit, N. and Frailis, M. and Franceschi, E. and Franzetti, P. and Fumana, M. and Gillis, B. and Giocoli, C. and Grazian, A. and Grupp, F. and Haugan, S. V. H. and Holmes, W. and Hormuth, F. and Hudelot, P. and Kermiche, S. and Kiessling, A. and Kilbinger, M. and Kohley, R. and Kubik, B. and Kümmel, M. and Kunz, M. and Kurki-Suonio, H. and Lahav, O. and Ligori, S. and Lilje, P. B. and Lloro, I. and Mansutti, O. and Marggraf, O. and Markovic, K. and Marulli, F. and Massey, R. and Maurogordato, S. and Melchior, M. and Merlin, E. and Meylan, G. and Mohr, J. J. and Moresco, M. and Morin, B. and Moscardini, L. and Munari, E. and Nichol, R. C. and Padilla, C. and Paltani, S. and Peacock, J. and Pedersen, K. and Pettorino, V. and Pires, S. and Poncet, M. and Popa, L. and Pozzetti, L. and Raison, F. and Rebolo, R. and Rhodes, J. and Rix, H.-W. and Roncarelli, M. and Rossetti, E. and Saglia, R. and Schneider, P. and Schrabback, T. and Secroun, A. and Seidel, G. and Serrano, S. and Sirignano, C. and Sirri, G. and Skottfelt, J. and Stanco, L. and Starck, J. L. and Tallada-Crespí, P. and Tavagnacco, D. and Taylor, A. N. and Teplitz, H. I. and Toledo-Moreo, R. and Torradeflot, F. and Trifoglio, M. and Valentijn, E. A. and Valenziano, L. and Verdoes Kleijn, G. A. and Wang, Y. and Welikala, N. and Weller, J. and Wetzstein, M. and Zamorani, G. and Zoubian, J. and Andreon, S. and Baldi, M. and Bardelli, S. and Boucaud, A. and Camera, S. and Di Ferdinando, D. and Fabbian, G. and Farinelli, R. and Galeotta, S. and Graciá-Carpio, J. and Maino, D. and Medinaceli, E. and Mei, S. and Neissner, C. and Polenta, G. and Renzi, A. and Romelli, E. and Rosset, C. and Sureau, F. and Tenti, M. and Vassallo, T. and Zucca, E. and Baccigalupi, C. and Balaguera-Antolínez, A. and Battaglia, P. and Biviano, A. and Borgani, S. and Bozzo, E. and Cabanac, R. and Cappi, A. and Casas, S. and Castignani, G. and Colodro-Conde, C. and Coupon, J. and Courtois, H. M. and Cuby, J. and de la Torre, S. and Desai, S. and Dole, H. and Fabricius, M. and Farina, M. and Ferreira, P. G. and Finelli, F. and Flose-Reimberg, P. and Fotopoulou, S. and Ganga, K. and Gozaliasl, G. and Hook, I. M. and Keihanen, E. and Kirkpatrick, C. C. and Liebing, P. and Lindholm, V. and Mainetti, G. and Martinelli, M. and Martinet, N. and Maturi, M. and McCracken, H. J. and Metcalf, R. B. and Morgante, G. and Nightingale, J. and Nucita, A. and Patrizii, L. and Potter, D. and Riccio, G. and Sánchez, A. G. and Sapone, D. and Schewtschenko, J. A. and Schultheis, M. and Scottez, V. and Teyssier, R. and Tutusaus, I. and Valiviita, J. and Viel, M. and Vriend, W. and Whittaker, L.},
   year={2022},
   month=jun, pages={A112} }

@ARTICLE{jwst,
       author = {{Gardner}, Jonathan P. and {Mather}, John C. and {Clampin}, Mark and {Doyon}, Rene and {Greenhouse}, Matthew A. and {Hammel}, Heidi B. and {Hutchings}, John B. and {Jakobsen}, Peter and {Lilly}, Simon J. and {Long}, Knox S. and {Lunine}, Jonathan I. and {McCaughrean}, Mark J. and {Mountain}, Matt and {Nella}, John and {Rieke}, George H. and {Rieke}, Marcia J. and {Rix}, Hans-Walter and {Smith}, Eric P. and {Sonneborn}, George and {Stiavelli}, Massimo and {Stockman}, H.~S. and {Windhorst}, Rogier A. and {Wright}, Gillian S.},
        title = "{The James Webb Space Telescope}",
      journal = {\ssr},
         year = 2006,
        month = apr,
       volume = {123},
       number = {4},
        pages = {485-606},
          doi = {10.1007/s11214-006-8315-7},
archivePrefix = {arXiv},
       eprint = {astro-ph/0606175},
 primaryClass = {astro-ph},
       adsurl = {https://ui.adsabs.harvard.edu/abs/2006SSRv..123..485G}
}

@article{GELU,
  title={Gaussian Error Linear Units (GELUs)},
  author={Dan Hendrycks and Kevin Gimpel},
  journal={arXiv: Learning},
  year={2016},
  url={https://api.semanticscholar.org/CorpusID:125617073}
}

@misc{CANDELS,
doi = {10.17909/T94S3X},
url = {http://archive.stsci.edu/doi/resolve/resolve.html?doi=10.17909/T94S3X},
author = {{Faber}, Sandra},
title = {The Cosmic Assembly Near-IR Deep Extragalactic Legacy Survey ("CANDELS")},
publisher = {STScI/MAST},
year = {2011}
}

@article{van_der_wel_3d-hstcandels_2014,
	title = {{3D}-{HST}+{CANDELS}: {The} {Evolution} of the {Galaxy} {Size}-{Mass} {Distribution} since z = 3},
	volume = {788},
	issn = {0004-637X},
	shorttitle = {{3D}-{HST}+{CANDELS}},
	url = {https://ui.adsabs.harvard.edu/abs/2014ApJ...788...28V},
	doi = {10.1088/0004-637X/788/1/28},
	urldate = {2026-05-25},
	journal = {The Astrophysical Journal},
	publisher = {IOP},
	author = {van der Wel, A. and Franx, M. and van Dokkum, P. G. and Skelton, R. E. and Momcheva, I. G. and Whitaker, K. E. and Brammer, G. B. and Bell, E. F. and Rix, H.-W. and Wuyts, S. and Ferguson, H. C. and Holden, B. P. and Barro, G. and Koekemoer, A. M. and Chang, Yu-Yen and McGrath, E. J. and Häussler, B. and Dekel, A. and Behroozi, P. and Fumagalli, M. and Leja, J. and Lundgren, B. F. and Maseda, M. V. and Nelson, E. J. and Wake, D. A. and Patel, S. G. and Labbé, I. and Faber, S. M. and Grogin, N. A. and Kocevski, D. D.},
	month = jun,
	year = {2014},
	note = {ADS Bibcode: 2014ApJ...788...28V},
	pages = {28},
}

@article{ferreira_galaxy_2026,
	title = {Galaxy mergers in {UNIONS} ─ {II}. {Predicting} time-scales in the post-merger regime},
	volume = {546},
	issn = {0035-8711},
	url = {https://ui.adsabs.harvard.edu/abs/2026MNRAS.546ag178F},
	doi = {10.1093/mnras/stag178},
	urldate = {2026-04-12},
	journal = {Monthly Notices of the Royal Astronomical Society},
	publisher = {OUP},
	author = {Ferreira, Leonardo and Ellison, Sara L. and Patton, David R. and Byrne-Mamahit, Shoshannah and Wilkinson, Scott and Bickley, Robert W.},
	month = mar,
	year = {2026},
	note = {ADS Bibcode: 2026MNRAS.546ag178F},
	pages = {stag178},
}

@article{lopez-sanjuan_alhambra_2015,
	title = {The {ALHAMBRA} survey: accurate merger fractions derived by {PDF} analysis of photometrically close pairs},
	volume = {576},
	issn = {0004-6361},
	shorttitle = {The {ALHAMBRA} survey},
	url = {https://ui.adsabs.harvard.edu/abs/2015A&A...576A..53L},
	doi = {10.1051/0004-6361/201424913},
	urldate = {2026-04-10},
	journal = {Astronomy and Astrophysics},
	publisher = {EDP},
	author = {López-Sanjuan, C. and Cenarro, A. J. and Varela, J. and Viironen, K. and Molino, A. and Benítez, N. and Arnalte-Mur, P. and Ascaso, B. and Díaz-García, L. A. and Fernández-Soto, A. and Jiménez-Teja, Y. and Márquez, I. and Masegosa, J. and Moles, M. and Pović, M. and Aguerri, J. A. L. and Alfaro, E. and Aparicio-Villegas, T. and Broadhurst, T. and Cabrera-Caño, J. and Castander, F. J. and Cepa, J. and Cerviño, M. and Cristóbal-Hornillos, D. and Del Olmo, A. and González Delgado, R. M. and Husillos, C. and Infante, L. and Martínez, V. J. and Perea, J. and Prada, F. and Quintana, J. M.},
	month = apr,
	year = {2015},
	note = {ADS Bibcode: 2015A\&A...576A..53L},
	pages = {A53},
}

@article{lopez-sanjuan_spectro-photometric_2010,
	title = {Spectro-photometric close pairs in {GOODS}-{S}: major and minor companions of intermediate-mass galaxies},
	volume = {518},
	issn = {0004-6361},
	shorttitle = {Spectro-photometric close pairs in {GOODS}-{S}},
	url = {https://ui.adsabs.harvard.edu/abs/2010A&A...518A..20L},
	doi = {10.1051/0004-6361/201014236},
	urldate = {2026-04-10},
	journal = {Astronomy and Astrophysics},
	publisher = {EDP},
	author = {López-Sanjuan, C. and Balcells, M. and Pérez-González, P. G. and Barro, G. and Gallego, J. and Zamorano, J.},
	month = jul,
	year = {2010},
	note = {ADS Bibcode: 2010A\&A...518A..20L},
	pages = {A20},
}

@inproceedings{akiba_optuna_2019,
	address = {New York, NY, USA},
	series = {{KDD} '19},
	title = {Optuna: {A} {Next}-generation {Hyperparameter} {Optimization} {Framework}},
	isbn = {978-1-4503-6201-6},
	shorttitle = {Optuna},
	url = {https://dl.acm.org/doi/10.1145/3292500.3330701},
	doi = {10.1145/3292500.3330701},
	urldate = {2026-04-05},
	booktitle = {Proceedings of the 25th {ACM} {SIGKDD} {International} {Conference} on {Knowledge} {Discovery} \& {Data} {Mining}},
	publisher = {Association for Computing Machinery},
	author = {Akiba, Takuya and Sano, Shotaro and Yanase, Toshihiko and Ohta, Takeru and Koyama, Masanori},
	month = jul,
	year = {2019},
	pages = {2623--2631},
}

@misc{loshchilov_decoupled_2017,
	title = {Decoupled {Weight} {Decay} {Regularization}},
	url = {https://ui.adsabs.harvard.edu/abs/2017arXiv171105101L},
	doi = {10.48550/arXiv.1711.05101},
	urldate = {2026-04-05},
	publisher = {arXiv},
	author = {Loshchilov, Ilya and Hutter, Frank},
	month = nov,
	year = {2017},
	note = {ADS Bibcode: 2017arXiv171105101L},
}

@misc{liu_convnet_2022,
	title = {A {ConvNet} for the 2020s},
	url = {https://ui.adsabs.harvard.edu/abs/2022arXiv220103545L},
	doi = {10.48550/arXiv.2201.03545},
	urldate = {2026-04-05},
	publisher = {arXiv},
	author = {Liu, Zhuang and Mao, Hanzi and Wu, Chao-Yuan and Feichtenhofer, Christoph and Darrell, Trevor and Xie, Saining},
	month = jan,
	year = {2022},
	note = {ADS Bibcode: 2022arXiv220103545L},
}

@article{ferreira_galaxy_2025,
	title = {Galaxy evolution in the {Post}-{Merger} {Regime} - {I}. {Most} merger-induced in situ stellar mass growth happens post-coalescence},
	volume = {538},
	issn = {0035-8711},
	url = {https://ui.adsabs.harvard.edu/abs/2025MNRAS.538L..31F},
	doi = {10.1093/mnrasl/slaf004},
	urldate = {2026-04-03},
	journal = {Monthly Notices of the Royal Astronomical Society},
	publisher = {OUP},
	author = {Ferreira, Leonardo and Ellison, Sara L. and Patton, David R. and Byrne-Mamahit, Shoshannah and Wilkinson, Scott and Bickley, Robert and Conselice, Christopher J. and Bottrell, Connor},
	month = mar,
	year = {2025},
	note = {ADS Bibcode: 2025MNRAS.538L..31F},
	pages = {L31--L36},
}

@article{la_marca_dust_2024,
	title = {Dust and power: {Unravelling} the merger-active galactic nucleus connection in the second half of cosmic history},
	volume = {690},
	issn = {0004-6361},
	shorttitle = {Dust and power},
	url = {https://ui.adsabs.harvard.edu/abs/2024A&A...690A.326L},
	doi = {10.1051/0004-6361/202348188},
	urldate = {2026-03-30},
	journal = {Astronomy and Astrophysics},
	publisher = {EDP},
	author = {La Marca, A. and Margalef-Bentabol, B. and Wang, L. and Gao, F. and Goulding, A. D. and Martin, G. and Rodriguez-Gomez, V. and Trager, S. C. and Yang, G. and Davé, R. and Dubois, Y.},
	month = oct,
	year = {2024},
	note = {ADS Bibcode: 2024A\&A...690A.326L},
	pages = {A326},
}

@article{lee_morphology_2024,
	title = {Morphology of {Galaxies} in {JWST} {Fields}: {Initial} {Distribution} and {Evolution} of {Galaxy} {Morphology}},
	volume = {966},
	issn = {0004-637X},
	shorttitle = {Morphology of {Galaxies} in {JWST} {Fields}},
	url = {https://ui.adsabs.harvard.edu/abs/2024ApJ...966..113L},
	doi = {10.3847/1538-4357/ad3448},
	urldate = {2026-03-30},
	journal = {The Astrophysical Journal},
	publisher = {IOP},
	author = {Lee, Jeong Hwan and Park, Changbom and Hwang, Ho Seong and Kwon, Minseong},
	month = may,
	year = {2024},
	note = {ADS Bibcode: 2024ApJ...966..113L},
	pages = {113},
}

@article{huertas-company_cosmos-web_2025,
	title = {{COSMOS}-{Web}: {The} emergence of the {Hubble} sequence},
	volume = {704},
	issn = {0004-6361},
	shorttitle = {{COSMOS}-{Web}},
	url = {https://ui.adsabs.harvard.edu/abs/2025A&A...704A..94H},
	doi = {10.1051/0004-6361/202553782},
	urldate = {2026-03-30},
	journal = {Astronomy and Astrophysics},
	publisher = {EDP},
	author = {Huertas-Company, M. and Shuntov, M. and Dong, Y. and Walmsley, M. and Ilbert, O. and McCracken, H. J. and Akins, H. B. and Allen, N. and Casey, C. M. and Costantin, L. and Daddi, E. and Dekel, A. and Franco, M. and Garland, I. L. and Géron, T. and Gozaliasl, G. and Hirschmann, M. and Kartaltepe, J. S. and Koekemoer, A. M. and Lintott, C. and Liu, D. and Lucas, R. and Masters, K. and Pacucci, F. and Paquereau, L. and Pérez-González, P. G. and Rhodes, J. D. and Robertson, B. E. and Simmons, B. and Smethurst, R. and Toft, S. and Yang, L.},
	month = dec,
	year = {2025},
	note = {ADS Bibcode: 2025A\&A...704A..94H},
	pages = {A94},
}

@article{bickley_convolutional_2021,
	title = {Convolutional neural network identification of galaxy post-mergers in {UNIONS} using {IllustrisTNG}},
	volume = {504},
	issn = {0035-8711},
	url = {https://ui.adsabs.harvard.edu/abs/2021MNRAS.504..372B/abstract},
	doi = {10.1093/mnras/stab806},
	language = {en},
	number = {1},
	urldate = {2022-04-19},
	journal = {Monthly Notices of the Royal Astronomical Society, Volume 504, Issue 1, pp.372-392},
	author = {Bickley, Robert W. and Bottrell, Connor and Hani, Maan H. and Ellison, Sara L. and Teimoorinia, Hossen and Yi, Kwang Moo and Wilkinson, Scott and Gwyn, Stephen and Hudson, Michael J.},
	month = jun,
	year = {2021},
	pages = {372},
}

@article{vogelsberger_introducing_2014,
	title = {Introducing the {Illustris} {Project}: simulating the coevolution of dark and visible matter in the {Universe}},
	volume = {444},
	issn = {0035-8711},
	shorttitle = {Introducing the {Illustris} {Project}},
	url = {https://ui.adsabs.harvard.edu/abs/2014MNRAS.444.1518V},
	doi = {10.1093/mnras/stu1536},
	urldate = {2025-07-21},
	journal = {Monthly Notices of the Royal Astronomical Society},
	publisher = {OUP},
	author = {Vogelsberger, Mark and Genel, Shy and Springel, Volker and Torrey, Paul and Sijacki, Debora and Xu, Dandan and Snyder, Greg and Nelson, Dylan and Hernquist, Lars},
	month = oct,
	year = {2014},
	note = {ADS Bibcode: 2014MNRAS.444.1518V},
	pages = {1518--1547},
}

@article{schechter_enhanced_2025,
	title = {Enhanced {Star} {Formation} and {Black} {Hole} {Accretion} {Rates} in {Galaxy} {Mergers} in {IllustrisTNG50}},
	volume = {989},
	issn = {0004-637X},
	url = {https://ui.adsabs.harvard.edu/abs/2025ApJ...989..149S},
	doi = {10.3847/1538-4357/ade791},
	urldate = {2025-09-10},
	journal = {The Astrophysical Journal},
	author = {Schechter, Aimee and Genel, Shy and Terrazas, Bryan and Comerford, Julia M. and Hartley, Abigail and Somerville, Rachel S. and Nevin, Rebecca and Simon, Joseph and Nelson, Erica},
	month = aug,
	year = {2025},
	note = {ADS Bibcode: 2025ApJ...989..149S},
	pages = {149},
}

@article{huertas-company_galaxy_2024,
	title = {Galaxy morphology from z ∼ 6 through the lens of {JWST}},
	volume = {685},
	issn = {0004-6361},
	url = {https://ui.adsabs.harvard.edu/abs/2024A&A...685A..48H},
	doi = {10.1051/0004-6361/202346800},
	urldate = {2025-10-20},
	journal = {Astronomy and Astrophysics},
	publisher = {EDP},
	author = {Huertas-Company, M. and Iyer, K. G. and Angeloudi, E. and Bagley, M. B. and Finkelstein, S. L. and Kartaltepe, J. and McGrath, E. J. and Sarmiento, R. and Vega-Ferrero, J. and Arrabal Haro, P. and Behroozi, P. and Buitrago, F. and Cheng, Y. and Costantin, L. and Dekel, A. and Dickinson, M. and Elbaz, D. and Grogin, N. A. and Hathi, N. P. and Holwerda, B. W. and Koekemoer, A. M. and Lucas, R. A. and Papovich, C. and Pérez-González, P. G. and Pirzkal, N. and Seillé, L. -M. and de la Vega, A. and Wuyts, S. and Yang, G. and Yung, L. Y. A.},
	month = may,
	year = {2024},
	note = {ADS Bibcode: 2024A\&A...685A..48H},
	pages = {A48},
}

@article{ferreira_jwst_2023,
	title = {The {JWST} {Hubble} {Sequence}: {The} {Rest}-frame {Optical} {Evolution} of {Galaxy} {Structure} at 1.5 {\textless} z {\textless} 6.5},
	volume = {955},
	issn = {0004-637X},
	shorttitle = {The {JWST} {Hubble} {Sequence}},
	url = {https://ui.adsabs.harvard.edu/abs/2023ApJ...955...94F},
	doi = {10.3847/1538-4357/acec76},
	urldate = {2025-10-20},
	journal = {The Astrophysical Journal},
	publisher = {IOP},
	author = {Ferreira, Leonardo and Conselice, Christopher J. and Sazonova, Elizaveta and Ferrari, Fabricio and Caruana, Joseph and Tohill, Clár-Bríd and Lucatelli, Geferson and Adams, Nathan and Irodotou, Dimitrios and Marshall, Madeline A. and Roper, Will J. and Lovell, Christopher C. and Verma, Aprajita and Austin, Duncan and Trussler, James and Wilkins, Stephen M.},
	month = oct,
	year = {2023},
	note = {ADS Bibcode: 2023ApJ...955...94F},
	pages = {94},
}

@article{ivezic_lsst_2019,
	title = {{LSST}: {From} {Science} {Drivers} to {Reference} {Design} and {Anticipated} {Data} {Products}},
	volume = {873},
	issn = {0004-637X},
	shorttitle = {{LSST}},
	url = {https://ui.adsabs.harvard.edu/abs/2019ApJ...873..111I},
	doi = {10.3847/1538-4357/ab042c},
	urldate = {2025-09-10},
	journal = {The Astrophysical Journal},
	author = {Ivezić, \{{\textbackslash}v\{Z\}\}eljko and Kahn, Steven M. and Tyson, J. Anthony and Abel, Bob and Acosta, Emily and Allsman, Robyn and Alonso, David and AlSayyad, Yusra and Anderson, Scott F. and Andrew, John and Angel, James Roger P. and Angeli, George Z. and Ansari, Reza and Antilogus, Pierre and Araujo, Constanza and Armstrong, Robert and Arndt, Kirk T. and Astier, Pierre and Aubourg, Éric and Auza, Nicole and Axelrod, Tim S. and Bard, Deborah J. and Barr, Jeff D. and Barrau, Aurelian and Bartlett, James G. and Bauer, Amanda E. and Bauman, Brian J. and Baumont, Sylvain and Bechtol, Ellen and Bechtol, Keith and Becker, Andrew C. and Becla, Jacek and Beldica, Cristina and Bellavia, Steve and Bianco, Federica B. and Biswas, Rahul and Blanc, Guillaume and Blazek, Jonathan and Blandford, Roger D. and Bloom, Josh S. and Bogart, Joanne and Bond, Tim W. and Booth, Michael T. and Borgland, Anders W. and Borne, Kirk and Bosch, James F. and Boutigny, Dominique and Brackett, Craig A. and Bradshaw, Andrew and Brandt, William Nielsen and Brown, Michael E. and Bullock, James S. and Burchat, Patricia and Burke, David L. and Cagnoli, Gianpietro and Calabrese, Daniel and Callahan, Shawn and Callen, Alice L. and Carlin, Jeffrey L. and Carlson, Erin L. and Chandrasekharan, Srinivasan and Charles-Emerson, Glenaver and Chesley, Steve and Cheu, Elliott C. and Chiang, Hsin-Fang and Chiang, James and Chirino, Carol and Chow, Derek and Ciardi, David R. and Claver, Charles F. and Cohen-Tanugi, Johann and Cockrum, Joseph J. and Coles, Rebecca and Connolly, Andrew J. and Cook, Kem H. and Cooray, Asantha and Covey, Kevin R. and Cribbs, Chris and Cui, Wei and Cutri, Roc and Daly, Philip N. and Daniel, Scott F. and Daruich, Felipe and Daubard, Guillaume and Daues, Greg and Dawson, William and Delgado, Francisco and Dellapenna, Alfred and de Peyster, Robert and de Val-Borro, Miguel and Digel, Seth W. and Doherty, Peter and Dubois, Richard and Dubois-Felsmann, Gregory P. and Durech, Josef and Economou, Frossie and Eifler, Tim and Eracleous, Michael and Emmons, Benjamin L. and Fausti Neto, Angelo and Ferguson, Henry and Figueroa, Enrique and Fisher-Levine, Merlin and Focke, Warren and Foss, Michael D. and Frank, James and Freemon, Michael D. and Gangler, Emmanuel and Gawiser, Eric and Geary, John C. and Gee, Perry and Geha, Marla and Gessner, Charles J. B. and Gibson, Robert R. and Gilmore, D. Kirk and Glanzman, Thomas and Glick, William and Goldina, Tatiana and Goldstein, Daniel A. and Goodenow, Iain and Graham, Melissa L. and Gressler, William J. and Gris, Philippe and Guy, Leanne P. and Guyonnet, Augustin and Haller, Gunther and Harris, Ron and Hascall, Patrick A. and Haupt, Justine and Hernandez, Fabio and Herrmann, Sven and Hileman, Edward and Hoblitt, Joshua and Hodgson, John A. and Hogan, Craig and Howard, James D. and Huang, Dajun and Huffer, Michael E. and Ingraham, Patrick and Innes, Walter R. and Jacoby, Suzanne H. and Jain, Bhuvnesh and Jammes, Fabrice and Jee, M. James and Jenness, Tim and Jernigan, Garrett and Jevremović, Darko and Johns, Kenneth and Johnson, Anthony S. and Johnson, Margaret W. G. and Jones, R. Lynne and Juramy-Gilles, Claire and Jurić, Mario and Kalirai, Jason S. and Kallivayalil, Nitya J. and Kalmbach, Bryce and Kantor, Jeffrey P. and Karst, Pierre and Kasliwal, Mansi M. and Kelly, Heather and Kessler, Richard and Kinnison, Veronica and Kirkby, David and Knox, Lloyd and Kotov, Ivan V. and Krabbendam, Victor L. and Krughoff, K. Simon and Kubánek, Petr and Kuczewski, John and Kulkarni, Shri and Ku, John and Kurita, Nadine R. and Lage, Craig S. and Lambert, Ron and Lange, Travis and Langton, J. Brian and Le Guillou, Laurent and Levine, Deborah and Liang, Ming and Lim, Kian-Tat and Lintott, Chris J. and Long, Kevin E. and Lopez, Margaux and Lotz, Paul J. and Lupton, Robert H. and Lust, Nate B. and MacArthur, Lauren A. and Mahabal, Ashish and Mandelbaum, Rachel and Markiewicz, Thomas W. and Marsh, Darren S. and Marshall, Philip J. and Marshall, Stuart and May, Morgan and McKercher, Robert and McQueen, Michelle and Meyers, Joshua and Migliore, Myriam and Miller, Michelle and Mills, David J.},
	month = mar,
	year = {2019},
	note = {ADS Bibcode: 2019ApJ...873..111I},
	pages = {111},
}

@article{ferreira_simulation-driven_2022,
	title = {A {Simulation}-driven {Deep} {Learning} {Approach} for {Separating} {Mergers} and {Star}-forming {Galaxies}: {The} {Formation} {Histories} of {Clumpy} {Galaxies} in {All} of the {CANDELS} {Fields}},
	volume = {931},
	issn = {0004-637X},
	shorttitle = {A {Simulation}-driven {Deep} {Learning} {Approach} for {Separating} {Mergers} and {Star}-forming {Galaxies}},
	url = {https://ui.adsabs.harvard.edu/abs/2022ApJ...931...34F},
	doi = {10.3847/1538-4357/ac66ea},
	urldate = {2025-08-26},
	journal = {The Astrophysical Journal},
	author = {Ferreira, Leonardo and Conselice, Christopher J. and Kuchner, Ulrike and Tohill, Clár-Bríd},
	month = may,
	year = {2022},
	note = {ADS Bibcode: 2022ApJ...931...34F},
	pages = {34},
}

@article{toomre_galactic_1972,
	title = {Galactic {Bridges} and {Tails}},
	volume = {178},
	issn = {0004-637X},
	url = {https://ui.adsabs.harvard.edu/abs/1972ApJ...178..623T},
	doi = {10.1086/151823},
	urldate = {2025-08-15},
	journal = {The Astrophysical Journal},
	publisher = {IOP},
	author = {Toomre, Alar and Toomre, Juri},
	month = dec,
	year = {1972},
	note = {ADS Bibcode: 1972ApJ...178..623T},
	pages = {623--666},
}

@article{martin_role_2018,
	title = {The role of mergers in driving morphological transformation over cosmic time},
	volume = {480},
	issn = {0035-8711},
	url = {https://ui.adsabs.harvard.edu/abs/2018MNRAS.480.2266M},
	doi = {10.1093/mnras/sty1936},
	urldate = {2025-08-15},
	journal = {Monthly Notices of the Royal Astronomical Society},
	publisher = {OUP},
	author = {Martin, G. and Kaviraj, S. and Devriendt, J. E. G. and Dubois, Y. and Pichon, C.},
	month = oct,
	year = {2018},
	note = {ADS Bibcode: 2018MNRAS.480.2266M},
	pages = {2266--2283},
}

@article{deng_mnist_2012,
	title = {The {MNIST} {Database} of {Handwritten} {Digit} {Images} for {Machine} {Learning} {Research} [{Best} of the {Web}]},
	volume = {29},
	issn = {1558-0792},
	url = {https://ieeexplore.ieee.org/document/6296535},
	doi = {10.1109/MSP.2012.2211477},
	number = {6},
	urldate = {2025-07-24},
	journal = {IEEE Signal Processing Magazine},
	author = {Deng, Li},
	month = nov,
	year = {2012},
	pages = {141--142},
}

@misc{simonyan_deep_2013,
	title = {Deep {Inside} {Convolutional} {Networks}: {Visualising} {Image} {Classification} {Models} and {Saliency} {Maps}},
	shorttitle = {Deep {Inside} {Convolutional} {Networks}},
	url = {https://ui.adsabs.harvard.edu/abs/2013arXiv1312.6034S},
	doi = {10.48550/arXiv.1312.6034},
	urldate = {2025-07-21},
	publisher = {arXiv},
	author = {Simonyan, Karen and Vedaldi, Andrea and Zisserman, Andrew},
	month = dec,
	year = {2013},
	note = {ADS Bibcode: 2013arXiv1312.6034S},
}

@article{buitrago_early-type_2013,
	title = {Early-type galaxies have been the predominant morphological class for massive galaxies since only z ∼ 1},
	volume = {428},
	issn = {0035-8711},
	url = {https://ui.adsabs.harvard.edu/abs/2013MNRAS.428.1460B},
	doi = {10.1093/mnras/sts124},
	urldate = {2025-07-21},
	journal = {Monthly Notices of the Royal Astronomical Society},
	publisher = {OUP},
	author = {Buitrago, Fernando and Trujillo, Ignacio and Conselice, Christopher J. and Häußler, Boris},
	month = jan,
	year = {2013},
	note = {ADS Bibcode: 2013MNRAS.428.1460B},
	pages = {1460--1478},
}

@article{newman_can_2012,
	title = {Can {Minor} {Merging} {Account} for the {Size} {Growth} of {Quiescent} {Galaxies}? {New} {Results} from the {CANDELS} {Survey}},
	volume = {746},
	issn = {0004-637X},
	shorttitle = {Can {Minor} {Merging} {Account} for the {Size} {Growth} of {Quiescent} {Galaxies}?},
	url = {https://ui.adsabs.harvard.edu/abs/2012ApJ...746..162N},
	doi = {10.1088/0004-637X/746/2/162},
	urldate = {2025-07-15},
	journal = {The Astrophysical Journal},
	publisher = {IOP},
	author = {Newman, Andrew B. and Ellis, Richard S. and Bundy, Kevin and Treu, Tommaso},
	month = feb,
	year = {2012},
	note = {ADS Bibcode: 2012ApJ...746..162N},
	pages = {162},
}

@misc{mcinnes_umap_2020,
	title = {{UMAP}: {Uniform} {Manifold} {Approximation} and {Projection} for {Dimension} {Reduction}},
	shorttitle = {{UMAP}},
	url = {http://arxiv.org/abs/1802.03426},
	doi = {10.48550/arXiv.1802.03426},
	urldate = {2025-07-15},
	publisher = {arXiv},
	author = {McInnes, Leland and Healy, John and Melville, James},
	month = sep,
	year = {2020},
	note = {arXiv:1802.03426 [stat]},
}

@article{walmsley_zoobot_2023,
	title = {Zoobot: {Adaptable} {Deep} {Learning} {Models} for {Galaxy} {Morphology}},
	volume = {8},
	issn = {2475-9066},
	shorttitle = {Zoobot},
	url = {https://joss.theoj.org/papers/10.21105/joss.05312},
	doi = {10.21105/joss.05312},
	language = {en},
	number = {85},
	urldate = {2025-07-09},
	journal = {Journal of Open Source Software},
	author = {Walmsley, Mike and Allen, Campbell and Aussel, Ben and Bowles, Micah and Gregorowicz, Kasia and Slijepcevic, Inigo Val and Lintott, Chris J. and Scaife, Anna M. m and Jabłońska, Maja and Karchev, Kosio and Lanzieri, Denise and Mohan, Devina and O’Ryan, David and Saiguhan, Bharath and Suárez, Crisel and Guerra-Varas, Nicolás and Velu, Renuka},
	month = may,
	year = {2023},
	pages = {5312},
}

@article{rose_identifying_2023,
	title = {Identifying {Galaxy} {Mergers} in {Simulated} {CEERS} {NIRCam} {Images} {Using} {Random} {Forests}},
	volume = {942},
	issn = {0004-637X},
	url = {https://ui.adsabs.harvard.edu/abs/2023ApJ...942...54R},
	doi = {10.3847/1538-4357/ac9f10},
	urldate = {2025-04-24},
	journal = {The Astrophysical Journal},
	publisher = {IOP},
	author = {Rose, Caitlin and Kartaltepe, Jeyhan S. and Snyder, Gregory F. and Rodriguez-Gomez, Vicente and Yung, L. Y. Aaron and Arrabal Haro, Pablo and Bagley, Micaela B. and Calabró, Antonello and Cleri, Nikko J. and Cooper, M. C. and Costantin, Luca and Croton, Darren and Dickinson, Mark and Finkelstein, Steven L. and Häußler, Boris and Holwerda, Benne W. and Koekemoer, Anton M. and Kurczynski, Peter and Lucas, Ray A. and Mantha, Kameswara Bharadwaj and Papovich, Casey and Pérez-González, Pablo G. and Pirzkal, Nor and Somerville, Rachel S. and Straughn, Amber N. and Tacchella, Sandro},
	month = jan,
	year = {2023},
	note = {ADS Bibcode: 2023ApJ...942...54R},
	pages = {54},
}

@misc{johnson_bd-jsedpy_2021,
	title = {bd-j/sedpy: sedpy v0.2.0},
	shorttitle = {bd-j/sedpy},
	url = {https://zenodo.org/records/4582723},
	doi = {10.5281/zenodo.4582723},
	urldate = {2025-04-24},
	publisher = {Zenodo},
	author = {Johnson, Benjamin D.},
	month = mar,
	year = {2021},
}

@article{duncan_observational_2019,
	title = {Observational {Constraints} on the {Merger} {History} of {Galaxies} since z ≈ 6: {Probabilistic} {Galaxy} {Pair} {Counts} in the {CANDELS} {Fields}},
	volume = {876},
	issn = {0004-637X},
	shorttitle = {Observational {Constraints} on the {Merger} {History} of {Galaxies} since z ≈ 6},
	url = {https://ui.adsabs.harvard.edu/abs/2019ApJ...876..110D},
	doi = {10.3847/1538-4357/ab148a},
	urldate = {2025-04-07},
	journal = {The Astrophysical Journal},
	publisher = {IOP},
	author = {Duncan, Kenneth and Conselice, Christopher J. and Mundy, Carl and Bell, Eric and Donley, Jennifer and Galametz, Audrey and Guo, Yicheng and Grogin, Norman A. and Hathi, Nimish and Kartaltepe, Jeyhan and Kocevski, Dale and Koekemoer, Anton M. and Pérez-González, Pablo G. and Mantha, Kameswara B. and Snyder, Gregory F. and Stefanon, Mauro},
	month = may,
	year = {2019},
	note = {ADS Bibcode: 2019ApJ...876..110D},
	pages = {110},
}

@article{dominguez_sanchez_identification_2023,
	title = {Identification of tidal features in deep optical galaxy images with convolutional neural networks},
	volume = {521},
	issn = {0035-8711},
	url = {https://ui.adsabs.harvard.edu/abs/2023MNRAS.521.3861D},
	doi = {10.1093/mnras/stad750},
	urldate = {2025-04-07},
	journal = {Monthly Notices of the Royal Astronomical Society},
	publisher = {OUP},
	author = {Domínguez Sánchez, H. and Martin, G. and Damjanov, I. and Buitrago, F. and Huertas-Company, M. and Bottrell, C. and Bernardi, M. and Knapen, J. H. and Vega-Ferrero, J. and Hausen, R. and Kado-Fong, E. and Población-Criado, D. and Souchereau, H. and Leste, O. K. and Robertson, B. and Sahelices, B. and Johnston, K. V.},
	month = may,
	year = {2023},
	note = {ADS Bibcode: 2023MNRAS.521.3861D},
	pages = {3861--3872},
}

@article{avirett-mackenzie_post-merger_2024,
	title = {A post-merger enhancement only in star-forming {Type} 2 {Seyfert} galaxies: the deep learning view},
	volume = {528},
	issn = {0035-8711},
	shorttitle = {A post-merger enhancement only in star-forming {Type} 2 {Seyfert} galaxies},
	url = {https://ui.adsabs.harvard.edu/abs/2024MNRAS.528.6915A},
	doi = {10.1093/mnras/stae183},
	urldate = {2025-04-07},
	journal = {Monthly Notices of the Royal Astronomical Society},
	publisher = {OUP},
	author = {Avirett-Mackenzie, M. S. and Villforth, C. and Huertas-Company, M. and Wuyts, S. and Alexander, D. M. and Bonoli, S. and Lapi, A. and Lopez, I. E. and Ramos Almeida, C. and Shankar, F.},
	month = mar,
	year = {2024},
	note = {ADS Bibcode: 2024MNRAS.528.6915A},
	pages = {6915--6933},
}

@article{omori_galaxy_2023,
	title = {Galaxy mergers in {Subaru} {HSC}-{SSP}: {A} deep representation learning approach for identification, and the role of environment on merger incidence},
	volume = {679},
	issn = {0004-6361},
	shorttitle = {Galaxy mergers in {Subaru} {HSC}-{SSP}},
	url = {https://ui.adsabs.harvard.edu/abs/2023A&A...679A.142O},
	doi = {10.1051/0004-6361/202346743},
	urldate = {2025-04-07},
	journal = {Astronomy and Astrophysics},
	author = {Omori, Kiyoaki Christopher and Bottrell, Connor and Walmsley, Mike and Yesuf, Hassen M. and Goulding, Andy D. and Ding, Xuheng and Popping, Gergö and Silverman, John D. and Takeuchi, Tsutomu T. and Toba, Yoshiki},
	month = nov,
	year = {2023},
	note = {ADS Bibcode: 2023A\&A...679A.142O},
	pages = {A142},
}

@article{smethurst_galaxy_2025,
	title = {Galaxy {Zoo} {JWST}: {Up} to 75\% of discs are featureless at 3 {\textless} z {\textless} 7},
	issn = {0035-8711},
	shorttitle = {Galaxy {Zoo} {JWST}},
	url = {https://ui.adsabs.harvard.edu/abs/2025MNRAS.tmp..500S},
	doi = {10.1093/mnras/staf506},
	urldate = {2025-04-07},
	journal = {Monthly Notices of the Royal Astronomical Society},
	publisher = {OUP},
	author = {Smethurst, R. J. and Simmons, B. D. and Géron, T. and Dickinson, H. and Fortson, L. and Garland, I. L. and Kruk, S. and Jewell, S. M. and Lintott, C. J. and Makechemu, J. S. and Mantha, K. B. and Masters, K. L. and O'Ryan, D. and Roberts, H. and Thorne, M. R. and Walmsley, M. and Calabrò, M. and Holwerda, B. and Kartaltepe, J. S. and Koekemoer, A. M. and Lyu, Y. and Lucas, R. and Pacucci, F. and Tarrasse, M.},
	month = mar,
	year = {2025},
	note = {ADS Bibcode: 2025MNRAS.tmp..500S},
}

@article{willett_galaxy_2017,
	title = {Galaxy {Zoo}: morphological classifications for 120 000 galaxies in {HST} legacy imaging},
	volume = {464},
	issn = {0035-8711},
	shorttitle = {Galaxy {Zoo}},
	url = {https://ui.adsabs.harvard.edu/abs/2017MNRAS.464.4176W},
	doi = {10.1093/mnras/stw2568},
	urldate = {2025-04-07},
	journal = {Monthly Notices of the Royal Astronomical Society},
	publisher = {OUP},
	author = {Willett, Kyle W. and Galloway, Melanie A. and Bamford, Steven P. and Lintott, Chris J. and Masters, Karen L. and Scarlata, Claudia and Simmons, B. D. and Beck, Melanie and Cardamone, Carolin N. and Cheung, Edmond and Edmondson, Edward M. and Fortson, Lucy F. and Griffith, Roger L. and Häußler, Boris and Han, Anna and Hart, Ross and Melvin, Thomas and Parrish, Michael and Schawinski, Kevin and Smethurst, R. J. and Smith, Arfon M.},
	month = feb,
	year = {2017},
	note = {ADS Bibcode: 2017MNRAS.464.4176W},
	pages = {4176--4203},
}

@article{simmons_galaxy_2017,
	title = {Galaxy {Zoo}: quantitative visual morphological classifications for 48 000 galaxies from {CANDELS}★},
	volume = {464},
	issn = {0035-8711},
	shorttitle = {Galaxy {Zoo}},
	url = {https://doi.org/10.1093/mnras/stw2587},
	doi = {10.1093/mnras/stw2587},
	number = {4},
	urldate = {2025-04-07},
	journal = {Monthly Notices of the Royal Astronomical Society},
	author = {Simmons, B. D. and Lintott, Chris and Willett, Kyle W. and Masters, Karen L. and Kartaltepe, Jeyhan S. and Häußler, Boris and Kaviraj, Sugata and Krawczyk, Coleman and Kruk, S. J. and McIntosh, Daniel H. and Smethurst, R. J. and Nichol, Robert C. and Scarlata, Claudia and Schawinski, Kevin and Conselice, Christopher J. and Almaini, Omar and Ferguson, Henry C. and Fortson, Lucy and Hartley, William and Kocevski, Dale and Koekemoer, Anton M. and Mortlock, Alice and Newman, Jeffrey A. and Bamford, Steven P. and Grogin, N. A. and Lucas, Ray A. and Hathi, Nimish P. and McGrath, Elizabeth and Peth, Michael and Pforr, Janine and Rizer, Zachary and Wuyts, Stijn and Barro, Guillermo and Bell, Eric F. and Castellano, Marco and Dahlen, Tomas and Dekel, Avishai and Ownsworth, Jamie and Faber, Sandra M. and Finkelstein, Steven L. and Fontana, Adriano and Galametz, Audrey and Grützbauch, Ruth and Koo, David and Lotz, Jennifer and Mobasher, Bahram and Mozena, Mark and Salvato, Mara and Wiklind, Tommy},
	month = feb,
	year = {2017},
	pages = {4420--4447},
}

@article{lintott_galaxy_2008,
	title = {Galaxy {Zoo}: morphologies derived from visual inspection of galaxies from the {Sloan} {Digital} {Sky} {Survey}},
	volume = {389},
	issn = {0035-8711},
	shorttitle = {Galaxy {Zoo}},
	url = {https://ui.adsabs.harvard.edu/abs/2008MNRAS.389.1179L},
	doi = {10.1111/j.1365-2966.2008.13689.x},
	urldate = {2025-04-07},
	journal = {Monthly Notices of the Royal Astronomical Society},
	publisher = {OUP},
	author = {Lintott, Chris J. and Schawinski, Kevin and Slosar, Anže and Land, Kate and Bamford, Steven and Thomas, Daniel and Raddick, M. Jordan and Nichol, Robert C. and Szalay, Alex and Andreescu, Dan and Murray, Phil and Vandenberg, Jan},
	month = sep,
	year = {2008},
	note = {ADS Bibcode: 2008MNRAS.389.1179L},
	pages = {1179--1189},
}

@article{darg_galaxy_2010,
	title = {Galaxy {Zoo}: the fraction of merging galaxies in the {SDSS} and their morphologies},
	volume = {401},
	issn = {0035-8711},
	shorttitle = {Galaxy {Zoo}},
	url = {https://ui.adsabs.harvard.edu/abs/2010MNRAS.401.1043D},
	doi = {10.1111/j.1365-2966.2009.15686.x},
	urldate = {2025-04-07},
	journal = {Monthly Notices of the Royal Astronomical Society},
	publisher = {OUP},
	author = {Darg, D. W. and Kaviraj, S. and Lintott, C. J. and Schawinski, K. and Sarzi, M. and Bamford, S. and Silk, J. and Proctor, R. and Andreescu, D. and Murray, P. and Nichol, R. C. and Raddick, M. J. and Slosar, A. and Szalay, A. S. and Thomas, D. and Vandenberg, J.},
	month = jan,
	year = {2010},
	note = {ADS Bibcode: 2010MNRAS.401.1043D},
	pages = {1043--1056},
}

@inproceedings{niculescu-mizil_predicting_2005,
	address = {New York, NY, USA},
	series = {{ICML} '05},
	title = {Predicting good probabilities with supervised learning},
	isbn = {978-1-59593-180-1},
	url = {https://dl.acm.org/doi/10.1145/1102351.1102430},
	doi = {10.1145/1102351.1102430},
	urldate = {2025-03-11},
	booktitle = {Proceedings of the 22nd international conference on {Machine} learning},
	publisher = {Association for Computing Machinery},
	author = {Niculescu-Mizil, Alexandru and Caruana, Rich},
	month = aug,
	year = {2005},
	pages = {625--632},
}

@article{degroot_comparison_1983,
	title = {The {Comparison} and {Evaluation} of {Forecasters}},
	volume = {32},
	issn = {0039-0526},
	url = {https://www.jstor.org/stable/2987588},
	doi = {10.2307/2987588},
	number = {1},
	urldate = {2025-03-11},
	journal = {Journal of the Royal Statistical Society. Series D (The Statistician)},
	publisher = {[Royal Statistical Society, Wiley]},
	author = {DeGroot, Morris H. and Fienberg, Stephen E.},
	year = {1983},
	pages = {12--22},
}

@article{naeini_obtaining_2015,
	title = {Obtaining {Well} {Calibrated} {Probabilities} {Using} {Bayesian} {Binning}},
	volume = {29},
	copyright = {Copyright (c)},
	issn = {2374-3468},
	url = {https://ojs.aaai.org/index.php/AAAI/article/view/9602},
	doi = {10.1609/aaai.v29i1.9602},
	language = {en},
	number = {1},
	urldate = {2025-03-11},
	journal = {Proceedings of the AAAI Conference on Artificial Intelligence},
	author = {Naeini, Mahdi Pakdaman and Cooper, Gregory and Hauskrecht, Milos},
	month = feb,
	year = {2015},
	note = {Number: 1},
}

@article{brier_verification_1950,
	title = {Verification of {Forecasts} {Expressed} in {Terms} of {Probability}},
	volume = {78},
	issn = {0027-0644},
	url = {https://ui.adsabs.harvard.edu/abs/1950MWRv...78....1B},
	doi = {10.1175/1520-0493(1950)078<0001:VOFEIT>2.0.CO;2},
	urldate = {2025-03-11},
	journal = {Monthly Weather Review},
	publisher = {AMS},
	author = {Brier, Glenn W.},
	month = jan,
	year = {1950},
	note = {ADS Bibcode: 1950MWRv...78....1B},
	pages = {1},
}

@article{shen_high-redshift_2022,
	title = {High-redshift predictions from {IllustrisTNG} - {III}. {Infrared} luminosity functions, obscured star formation, and dust temperature of high-redshift galaxies},
	volume = {510},
	issn = {0035-8711},
	url = {https://ui.adsabs.harvard.edu/abs/2022MNRAS.510.5560S},
	doi = {10.1093/mnras/stab3794},
	urldate = {2025-03-03},
	journal = {Monthly Notices of the Royal Astronomical Society},
	publisher = {OUP},
	author = {Shen, Xuejian and Vogelsberger, Mark and Nelson, Dylan and Tacchella, Sandro and Hernquist, Lars and Springel, Volker and Marinacci, Federico and Torrey, Paul},
	month = mar,
	year = {2022},
	note = {ADS Bibcode: 2022MNRAS.510.5560S},
	pages = {5560--5578},
}

@misc{camps_skirt_2020,
	title = {{SKIRT} 9: redesigning an advanced dust radiative transfer code to allow kinematics, line transfer and polarization by aligned dust grains},
	shorttitle = {{SKIRT} 9},
	url = {http://arxiv.org/abs/2003.00721},
	doi = {10.48550/arXiv.2003.00721},
	urldate = {2025-02-20},
	publisher = {arXiv},
	author = {Camps, Peter and Baes, Maarten},
	month = mar,
	year = {2020},
	note = {arXiv:2003.00721 [astro-ph]},
}

@article{bouwens_galaxy_2004,
	title = {Galaxy {Size} {Evolution} at {High} {Redshift} and {Surface} {Brightness} {Selection} {Effects}: {Constraints} from the {Hubble} {Ultra} {Deep} {Field}},
	volume = {611},
	issn = {0004-637X},
	shorttitle = {Galaxy {Size} {Evolution} at {High} {Redshift} and {Surface} {Brightness} {Selection} {Effects}},
	url = {https://ui.adsabs.harvard.edu/abs/2004ApJ...611L...1B},
	doi = {10.1086/423786},
	urldate = {2025-02-20},
	journal = {The Astrophysical Journal},
	publisher = {IOP},
	author = {Bouwens, R. J. and Illingworth, G. D. and Blakeslee, J. P. and Broadhurst, T. J. and Franx, M.},
	month = aug,
	year = {2004},
	note = {ADS Bibcode: 2004ApJ...611L...1B},
	pages = {L1--L4},
}

@article{baes_skirt_2015,
	title = {{SKIRT}: {The} design of a suite of input models for {Monte} {Carlo} radiative transfer simulations},
	volume = {12},
	issn = {2213-1337},
	shorttitle = {{SKIRT}},
	url = {https://ui.adsabs.harvard.edu/abs/2015A&C....12...33B},
	doi = {10.1016/j.ascom.2015.05.006},
	urldate = {2025-02-20},
	journal = {Astronomy and Computing},
	author = {Baes, M. and Camps, P.},
	month = sep,
	year = {2015},
	note = {ADS Bibcode: 2015A\&C....12...33B},
	pages = {33--44},
}

@article{dominguez_sanchez_improving_2018,
	title = {Improving galaxy morphologies for {SDSS} with {Deep} {Learning}},
	volume = {476},
	issn = {0035-8711},
	url = {https://ui.adsabs.harvard.edu/abs/2018MNRAS.476.3661D},
	doi = {10.1093/mnras/sty338},
	urldate = {2025-02-20},
	journal = {Monthly Notices of the Royal Astronomical Society},
	publisher = {OUP},
	author = {Domínguez Sánchez, H. and Huertas-Company, M. and Bernardi, M. and Tuccillo, D. and Fischer, J. L.},
	month = feb,
	year = {2018},
	note = {ADS Bibcode: 2018MNRAS.476.3661D},
	pages = {3661--3676},
}

@article{dieleman_rotation-invariant_2015,
	title = {Rotation-invariant convolutional neural networks for galaxy morphology prediction},
	volume = {450},
	issn = {0035-8711},
	url = {https://ui.adsabs.harvard.edu/abs/2015MNRAS.450.1441D},
	doi = {10.1093/mnras/stv632},
	urldate = {2025-02-20},
	journal = {Monthly Notices of the Royal Astronomical Society},
	publisher = {OUP},
	author = {Dieleman, Sander and Willett, Kyle W. and Dambre, Joni},
	month = jun,
	year = {2015},
	note = {ADS Bibcode: 2015MNRAS.450.1441D},
	pages = {1441--1459},
}

@article{pawlik_shape_2016,
	title = {Shape asymmetry: a morphological indicator for automatic detection of galaxies in the post-coalescence merger stages},
	volume = {456},
	issn = {0035-8711},
	shorttitle = {Shape asymmetry},
	url = {https://ui.adsabs.harvard.edu/abs/2016MNRAS.456.3032P},
	doi = {10.1093/mnras/stv2878},
	urldate = {2025-02-20},
	journal = {Monthly Notices of the Royal Astronomical Society},
	publisher = {OUP},
	author = {Pawlik, M. M. and Wild, V. and Walcher, C. J. and Johansson, P. H. and Villforth, C. and Rowlands, K. and Mendez-Abreu, J. and Hewlett, T.},
	month = mar,
	year = {2016},
	note = {ADS Bibcode: 2016MNRAS.456.3032P},
	pages = {3032--3052},
}

@article{ciprijanovic_deepmerge_2020,
	title = {{DeepMerge}: {Classifying} high-redshift merging galaxies with deep neural networks},
	volume = {32},
	issn = {2213-1337},
	shorttitle = {{DeepMerge}},
	url = {https://ui.adsabs.harvard.edu/abs/2020A&C....3200390C},
	doi = {10.1016/j.ascom.2020.100390},
	urldate = {2025-02-12},
	journal = {Astronomy and Computing},
	author = {Ćiprijanović, A. and Snyder, G. F. and Nord, B. and Peek, J. E. G.},
	month = jul,
	year = {2020},
	note = {ADS Bibcode: 2020A\&C....3200390C},
	pages = {100390},
}

@article{bottrell_deep_2019,
	title = {Deep learning predictions of galaxy merger stage and the importance of observational realism},
	volume = {490},
	issn = {0035-8711},
	url = {https://ui.adsabs.harvard.edu/abs/2019MNRAS.490.5390B},
	doi = {10.1093/mnras/stz2934},
	urldate = {2025-02-12},
	journal = {Monthly Notices of the Royal Astronomical Society},
	publisher = {OUP},
	author = {Bottrell, Connor and Hani, Maan H. and Teimoorinia, Hossen and Ellison, Sara L. and Moreno, Jorge and Torrey, Paul and Hayward, Christopher C. and Thorp, Mallory and Simard, Luc and Hernquist, Lars},
	month = dec,
	year = {2019},
	note = {ADS Bibcode: 2019MNRAS.490.5390B},
	pages = {5390--5413},
}

@article{lotz_new_2004,
	title = {A {New} {Nonparametric} {Approach} to {Galaxy} {Morphological} {Classification}},
	volume = {128},
	issn = {0004-6256},
	url = {https://ui.adsabs.harvard.edu/abs/2004AJ....128..163L},
	doi = {10.1086/421849},
	urldate = {2025-02-12},
	journal = {The Astronomical Journal},
	publisher = {IOP},
	author = {Lotz, Jennifer M. and Primack, Joel and Madau, Piero},
	month = jul,
	year = {2004},
	note = {ADS Bibcode: 2004AJ....128..163L},
	pages = {163--182},
}

@article{conselice_relationship_2003,
	title = {The {Relationship} between {Stellar} {Light} {Distributions} of {Galaxies} and {Their} {Formation} {Histories}},
	volume = {147},
	issn = {0067-0049},
	url = {https://ui.adsabs.harvard.edu/abs/2003ApJS..147....1C},
	doi = {10.1086/375001},
	urldate = {2025-02-12},
	journal = {The Astrophysical Journal Supplement Series},
	publisher = {IOP},
	author = {Conselice, Christopher J.},
	month = jul,
	year = {2003},
	note = {ADS Bibcode: 2003ApJS..147....1C},
	pages = {1--28},
}

@article{kartaltepe_candels_2015,
	title = {{CANDELS} {Visual} {Classifications}: {Scheme}, {Data} {Release}, and {First} {Results}},
	volume = {221},
	issn = {0067-0049},
	shorttitle = {{CANDELS} {Visual} {Classifications}},
	url = {https://ui.adsabs.harvard.edu/abs/2015ApJS..221...11K},
	doi = {10.1088/0067-0049/221/1/11},
	urldate = {2025-01-28},
	journal = {The Astrophysical Journal Supplement Series},
	publisher = {IOP},
	author = {Kartaltepe, Jeyhan S. and Mozena, Mark and Kocevski, Dale and McIntosh, Daniel H. and Lotz, Jennifer and Bell, Eric F. and Faber, Sandy and Ferguson, Harry and Koo, David and Bassett, Robert and Bernyk, Maksym and Blancato, Kirsten and Bournaud, Frederic and Cassata, Paolo and Castellano, Marco and Cheung, Edmond and Conselice, Christopher J. and Croton, Darren and Dahlen, Tomas and de Mello, Duilia F. and DeGroot, Laura and Donley, Jennifer and Guedes, Javiera and Grogin, Norman and Hathi, Nimish and Hilton, Matt and Hollon, Brett and Koekemoer, Anton and Liu, Nick and Lucas, Ray A. and Martig, Marie and McGrath, Elizabeth and McPartland, Conor and Mobasher, Bahram and Morlock, Alice and O'Leary, Erin and Peth, Mike and Pforr, Janine and Pillepich, Annalisa and Rosario, David and Soto, Emmaris and Straughn, Amber and Telford, Olivia and Sunnquist, Ben and Trump, Jonathan and Weiner, Benjamin and Wuyts, Stijn and Inami, Hanae and Kassin, Susan and Lani, Caterina and Poole, Gregory B. and Rizer, Zachary},
	month = nov,
	year = {2015},
	note = {ADS Bibcode: 2015ApJS..221...11K},
	pages = {11},
}

@article{grogin_candels_2011,
	title = {{CANDELS}: {The} {Cosmic} {Assembly} {Near}-infrared {Deep} {Extragalactic} {Legacy} {Survey}},
	volume = {197},
	issn = {0067-0049},
	shorttitle = {{CANDELS}},
	url = {https://ui.adsabs.harvard.edu/abs/2011ApJS..197...35G},
	doi = {10.1088/0067-0049/197/2/35},
	urldate = {2025-01-28},
	journal = {The Astrophysical Journal Supplement Series},
	publisher = {IOP},
	author = {Grogin, Norman A. and Kocevski, Dale D. and Faber, S. M. and Ferguson, Henry C. and Koekemoer, Anton M. and Riess, Adam G. and Acquaviva, Viviana and Alexander, David M. and Almaini, Omar and Ashby, Matthew L. N. and Barden, Marco and Bell, Eric F. and Bournaud, Frédéric and Brown, Thomas M. and Caputi, Karina I. and Casertano, Stefano and Cassata, Paolo and Castellano, Marco and Challis, Peter and Chary, Ranga-Ram and Cheung, Edmond and Cirasuolo, Michele and Conselice, Christopher J. and Roshan Cooray, Asantha and Croton, Darren J. and Daddi, Emanuele and Dahlen, Tomas and Davé, Romeel and de Mello, Duília F. and Dekel, Avishai and Dickinson, Mark and Dolch, Timothy and Donley, Jennifer L. and Dunlop, James S. and Dutton, Aaron A. and Elbaz, David and Fazio, Giovanni G. and Filippenko, Alexei V. and Finkelstein, Steven L. and Fontana, Adriano and Gardner, Jonathan P. and Garnavich, Peter M. and Gawiser, Eric and Giavalisco, Mauro and Grazian, Andrea and Guo, Yicheng and Hathi, Nimish P. and Häussler, Boris and Hopkins, Philip F. and Huang, Jia-Sheng and Huang, Kuang-Han and Jha, Saurabh W. and Kartaltepe, Jeyhan S. and Kirshner, Robert P. and Koo, David C. and Lai, Kamson and Lee, Kyoung-Soo and Li, Weidong and Lotz, Jennifer M. and Lucas, Ray A. and Madau, Piero and McCarthy, Patrick J. and McGrath, Elizabeth J. and McIntosh, Daniel H. and McLure, Ross J. and Mobasher, Bahram and Moustakas, Leonidas A. and Mozena, Mark and Nandra, Kirpal and Newman, Jeffrey A. and Niemi, Sami-Matias and Noeske, Kai G. and Papovich, Casey J. and Pentericci, Laura and Pope, Alexandra and Primack, Joel R. and Rajan, Abhijith and Ravindranath, Swara and Reddy, Naveen A. and Renzini, Alvio and Rix, Hans-Walter and Robaina, Aday R. and Rodney, Steven A. and Rosario, David J. and Rosati, Piero and Salimbeni, Sara and Scarlata, Claudia and Siana, Brian and Simard, Luc and Smidt, Joseph and Somerville, Rachel S. and Spinrad, Hyron and Straughn, Amber N. and Strolger, Louis-Gregory and Telford, Olivia and Teplitz, Harry I. and Trump, Jonathan R. and van der Wel, Arjen and Villforth, Carolin and Wechsler, Risa H. and Weiner, Benjamin J. and Wiklind, Tommy and Wild, Vivienne and Wilson, Grant and Wuyts, Stijn and Yan, Hao-Jing and Yun, Min S.},
	month = dec,
	year = {2011},
	note = {ADS Bibcode: 2011ApJS..197...35G},
	pages = {35},
}

@article{koekemoer_candels_2011,
	title = {{CANDELS}: {The} {Cosmic} {Assembly} {Near}-infrared {Deep} {Extragalactic} {Legacy} {Survey}—{The} {Hubble} {Space} {Telescope} {Observations}, {Imaging} {Data} {Products}, and {Mosaics}},
	volume = {197},
	issn = {0067-0049},
	shorttitle = {{CANDELS}},
	url = {https://ui.adsabs.harvard.edu/abs/2011ApJS..197...36K},
	doi = {10.1088/0067-0049/197/2/36},
	urldate = {2025-01-28},
	journal = {The Astrophysical Journal Supplement Series},
	publisher = {IOP},
	author = {Koekemoer, Anton M. and Faber, S. M. and Ferguson, Henry C. and Grogin, Norman A. and Kocevski, Dale D. and Koo, David C. and Lai, Kamson and Lotz, Jennifer M. and Lucas, Ray A. and McGrath, Elizabeth J. and Ogaz, Sara and Rajan, Abhijith and Riess, Adam G. and Rodney, Steve A. and Strolger, Louis and Casertano, Stefano and Castellano, Marco and Dahlen, Tomas and Dickinson, Mark and Dolch, Timothy and Fontana, Adriano and Giavalisco, Mauro and Grazian, Andrea and Guo, Yicheng and Hathi, Nimish P. and Huang, Kuang-Han and van der Wel, Arjen and Yan, Hao-Jing and Acquaviva, Viviana and Alexander, David M. and Almaini, Omar and Ashby, Matthew L. N. and Barden, Marco and Bell, Eric F. and Bournaud, Frédéric and Brown, Thomas M. and Caputi, Karina I. and Cassata, Paolo and Challis, Peter J. and Chary, Ranga-Ram and Cheung, Edmond and Cirasuolo, Michele and Conselice, Christopher J. and Roshan Cooray, Asantha and Croton, Darren J. and Daddi, Emanuele and Davé, Romeel and de Mello, Duilia F. and de Ravel, Loic and Dekel, Avishai and Donley, Jennifer L. and Dunlop, James S. and Dutton, Aaron A. and Elbaz, David and Fazio, Giovanni G. and Filippenko, Alexei V. and Finkelstein, Steven L. and Frazer, Chris and Gardner, Jonathan P. and Garnavich, Peter M. and Gawiser, Eric and Gruetzbauch, Ruth and Hartley, Will G. and Häussler, Boris and Herrington, Jessica and Hopkins, Philip F. and Huang, Jia-Sheng and Jha, Saurabh W. and Johnson, Andrew and Kartaltepe, Jeyhan S. and Khostovan, Ali A. and Kirshner, Robert P. and Lani, Caterina and Lee, Kyoung-Soo and Li, Weidong and Madau, Piero and McCarthy, Patrick J. and McIntosh, Daniel H. and McLure, Ross J. and McPartland, Conor and Mobasher, Bahram and Moreira, Heidi and Mortlock, Alice and Moustakas, Leonidas A. and Mozena, Mark and Nandra, Kirpal and Newman, Jeffrey A. and Nielsen, Jennifer L. and Niemi, Sami and Noeske, Kai G. and Papovich, Casey J. and Pentericci, Laura and Pope, Alexandra and Primack, Joel R. and Ravindranath, Swara and Reddy, Naveen A. and Renzini, Alvio and Rix, Hans-Walter and Robaina, Aday R. and Rosario, David J. and Rosati, Piero and Salimbeni, Sara and Scarlata, Claudia and Siana, Brian and Simard, Luc and Smidt, Joseph and Snyder, Diana and Somerville, Rachel S. and Spinrad, Hyron and Straughn, Amber N. and Telford, Olivia and Teplitz, Harry I. and Trump, Jonathan R. and Vargas, Carlos and Villforth, Carolin and Wagner, Cory R. and Wandro, Pat and Wechsler, Risa H. and Weiner, Benjamin J. and Wiklind, Tommy and Wild, Vivienne and Wilson, Grant and Wuyts, Stijn and Yun, Min S.},
	month = dec,
	year = {2011},
	note = {ADS Bibcode: 2011ApJS..197...36K},
	pages = {36},
}

@article{margalef-bentabol_galaxy_2024,
	title = {Galaxy merger challenge: {A} comparison study between machine learning-based detection methods},
	volume = {687},
	issn = {0004-6361, 1432-0746},
	shorttitle = {Galaxy merger challenge},
	url = {http://arxiv.org/abs/2403.15118},
	doi = {10.1051/0004-6361/202348239},
	urldate = {2025-01-16},
	journal = {Astronomy \& Astrophysics},
	author = {Margalef-Bentabol, B. and Wang, L. and Marca, A. La and Blanco-Prieto, C. and Chudy, D. and Domínguez-Sánchez, H. and Goulding, A. D. and Guzmán-Ortega, A. and Huertas-Company, M. and Martin, G. and Pearson, W. J. and Rodriguez-Gomez, V. and Walmsley, M. and Bickley, R. W. and Bottrell, C. and Conselice, C. and O'Ryan, D.},
	month = jul,
	year = {2024},
	note = {arXiv:2403.15118 [astro-ph]},
	pages = {A24},
}

@article{wilkinson_limitations_2024,
	title = {The limitations (and potential) of non-parametric morphology statistics for post-merger identification},
	volume = {528},
	issn = {0035-8711},
	url = {https://ui.adsabs.harvard.edu/abs/2024MNRAS.528.5558W},
	doi = {10.1093/mnras/stae287},
	urldate = {2025-01-13},
	journal = {Monthly Notices of the Royal Astronomical Society},
	publisher = {OUP},
	author = {Wilkinson, Scott and Ellison, Sara L. and Bottrell, Connor and Bickley, Robert W. and Byrne-Mamahit, Shoshannah and Ferreira, Leonardo and Patton, David R.},
	month = mar,
	year = {2024},
	note = {ADS Bibcode: 2024MNRAS.528.5558W},
	pages = {5558--5585},
}

@article{rose_ceers_2024,
	title = {{CEERS} {Key} {Paper}. {IX}. {Identifying} {Galaxy} {Mergers} in {CEERS} {NIRCam} {Images} {Using} {Random} {Forests} and {Convolutional} {Neural} {Networks}},
	volume = {976},
	issn = {0004-637X},
	url = {https://ui.adsabs.harvard.edu/abs/2024ApJ...976L...8R},
	doi = {10.3847/2041-8213/ad8dd4},
	urldate = {2024-12-12},
	journal = {The Astrophysical Journal},
	publisher = {IOP},
	author = {Rose, Caitlin and Kartaltepe, Jeyhan S. and Snyder, Gregory F. and Huertas-Company, Marc and Yung, L. Y. Aaron and Arrabal Haro, Pablo and Bagley, Micaela B. and Bisigello, Laura and Calabrò, Antonello and Cleri, Nikko J. and Dickinson, Mark and Ferguson, Henry C. and Finkelstein, Steven L. and Fontana, Adriano and Grazian, Andrea and Grogin, Norman A. and Holwerda, Benne W. and Iyer, Kartheik G. and Kewley, Lisa J. and Kirkpatrick, Allison and Kocevski, Dale D. and Koekemoer, Anton M. and Lotz, Jennifer M. and Lucas, Ray A. and Napolitano, Lorenzo and Papovich, Casey and Pentericci, Laura and Pérez-González, Pablo G. and Pirzkal, Nor and Ravindranath, Swara and Somerville, Rachel S. and Straughn, Amber N. and Trump, Jonathan R. and Wilkins, Stephen M. and Yang, Guang},
	month = nov,
	year = {2024},
	note = {ADS Bibcode: 2024ApJ...976L...8R},
	pages = {L8},
}

@article{bickley_effect_2024,
	title = {The effect of image quality on galaxy merger identification with deep learning},
	volume = {534},
	issn = {0035-8711},
	url = {https://ui.adsabs.harvard.edu/abs/2024MNRAS.534.2533B},
	doi = {10.1093/mnras/stae2246},
	urldate = {2024-12-11},
	journal = {Monthly Notices of the Royal Astronomical Society},
	publisher = {OUP},
	author = {Bickley, Robert W. and Wilkinson, Scott and Ferreira, Leonardo and Ellison, Sara L. and Bottrell, Connor and Jyoti, Debarpita},
	month = nov,
	year = {2024},
	note = {ADS Bibcode: 2024MNRAS.534.2533B},
	pages = {2533--2550},
}

@article{ferreira_galaxy_2024,
	title = {Galaxy mergers in {UNIONS} - {I}. {A} simulation-driven hybrid deep learning ensemble for pure galaxy merger classification},
	volume = {533},
	issn = {0035-8711},
	url = {https://ui.adsabs.harvard.edu/abs/2024MNRAS.533.2547F},
	doi = {10.1093/mnras/stae1885},
	urldate = {2024-10-31},
	journal = {Monthly Notices of the Royal Astronomical Society},
	publisher = {OUP},
	author = {Ferreira, Leonardo and Bickley, Robert W. and Ellison, Sara L. and Patton, David R. and Byrne-Mamahit, Shoshannah and Wilkinson, Scott and Bottrell, Connor and Fabbro, Sébastien and Gwyn, Stephen D. J. and McConnachie, Alan},
	month = sep,
	year = {2024},
	note = {ADS Bibcode: 2024MNRAS.533.2547F},
	pages = {2547--2569},
}

@misc{ferreira_galaxy_2024-1,
	title = {Galaxy evolution in the post-merger regime {I} -- {Most} merger-induced in-situ stellar mass growth happens post-coalescence},
	url = {http://arxiv.org/abs/2410.06356},
	urldate = {2024-10-31},
	publisher = {arXiv},
	author = {Ferreira, Leonardo and Ellison, Sara L. and Patton, David R. and Byrne-Mamahit, Shoshannah and Wilkinson, Scott and Bickley, Robert and Conselice, Christopher J. and Bottrell, Connor},
	month = oct,
	year = {2024},
	note = {arXiv:2410.06356},
}

@article{mihos_gasdynamics_1996,
	title = {Gasdynamics and {Starbursts} in {Major} {Mergers}},
	volume = {464},
	issn = {0004-637X},
	url = {https://ui.adsabs.harvard.edu/abs/1996ApJ...464..641M},
	doi = {10.1086/177353},
	urldate = {2024-09-12},
	journal = {The Astrophysical Journal},
	publisher = {IOP},
	author = {Mihos, J. Christopher and Hernquist, Lars},
	month = jun,
	year = {1996},
	note = {ADS Bibcode: 1996ApJ...464..641M},
	pages = {641},
}

@article{planck_collaboration_planck_2016,
	title = {Planck 2015 results. {XIII}. {Cosmological} parameters},
	volume = {594},
	issn = {0004-6361},
	url = {https://ui.adsabs.harvard.edu/abs/2016A&A...594A..13P},
	doi = {10.1051/0004-6361/201525830},
	urldate = {2024-03-04},
	journal = {Astronomy and Astrophysics},
	author = {{Planck Collaboration} and Ade, P. A. R. and Aghanim, N. and Arnaud, M. and Ashdown, M. and Aumont, J. and Baccigalupi, C. and Banday, A. J. and Barreiro, R. B. and Bartlett, J. G. and Bartolo, N. and Battaner, E. and Battye, R. and Benabed, K. and Benoît, A. and Benoit-Lévy, A. and Bernard, J. -P. and Bersanelli, M. and Bielewicz, P. and Bock, J. J. and Bonaldi, A. and Bonavera, L. and Bond, J. R. and Borrill, J. and Bouchet, F. R. and Boulanger, F. and Bucher, M. and Burigana, C. and Butler, R. C. and Calabrese, E. and Cardoso, J. -F. and Catalano, A. and Challinor, A. and Chamballu, A. and Chary, R. -R. and Chiang, H. C. and Chluba, J. and Christensen, P. R. and Church, S. and Clements, D. L. and Colombi, S. and Colombo, L. P. L. and Combet, C. and Coulais, A. and Crill, B. P. and Curto, A. and Cuttaia, F. and Danese, L. and Davies, R. D. and Davis, R. J. and de Bernardis, P. and de Rosa, A. and de Zotti, G. and Delabrouille, J. and Désert, F. -X. and Di Valentino, E. and Dickinson, C. and Diego, J. M. and Dolag, K. and Dole, H. and Donzelli, S. and Doré, O. and Douspis, M. and Ducout, A. and Dunkley, J. and Dupac, X. and Efstathiou, G. and Elsner, F. and Enßlin, T. A. and Eriksen, H. K. and Farhang, M. and Fergusson, J. and Finelli, F. and Forni, O. and Frailis, M. and Fraisse, A. A. and Franceschi, E. and Frejsel, A. and Galeotta, S. and Galli, S. and Ganga, K. and Gauthier, C. and Gerbino, M. and Ghosh, T. and Giard, M. and Giraud-Héraud, Y. and Giusarma, E. and Gjerløw, E. and González-Nuevo, J. and Górski, K. M. and Gratton, S. and Gregorio, A. and Gruppuso, A. and Gudmundsson, J. E. and Hamann, J. and Hansen, F. K. and Hanson, D. and Harrison, D. L. and Helou, G. and Henrot-Versillé, S. and Hernández-Monteagudo, C. and Herranz, D. and Hildebrandt, S. R. and Hivon, E. and Hobson, M. and Holmes, W. A. and Hornstrup, A. and Hovest, W. and Huang, Z. and Huffenberger, K. M. and Hurier, G. and Jaffe, A. H. and Jaffe, T. R. and Jones, W. C. and Juvela, M. and Keihänen, E. and Keskitalo, R. and Kisner, T. S. and Kneissl, R. and Knoche, J. and Knox, L. and Kunz, M. and Kurki-Suonio, H. and Lagache, G. and Lähteenmäki, A. and Lamarre, J. -M. and Lasenby, A. and Lattanzi, M. and Lawrence, C. R. and Leahy, J. P. and Leonardi, R. and Lesgourgues, J. and Levrier, F. and Lewis, A. and Liguori, M. and Lilje, P. B. and Linden-Vørnle, M. and López-Caniego, M. and Lubin, P. M. and Macías-Pérez, J. F. and Maggio, G. and Maino, D. and Mandolesi, N. and Mangilli, A. and Marchini, A. and Maris, M. and Martin, P. G. and Martinelli, M. and Martínez-González, E. and Masi, S. and Matarrese, S. and McGehee, P. and Meinhold, P. R. and Melchiorri, A. and Melin, J. -B. and Mendes, L. and Mennella, A. and Migliaccio, M. and Millea, M. and Mitra, S. and Miville-Deschênes, M. -A. and Moneti, A. and Montier, L. and Morgante, G. and Mortlock, D. and Moss, A. and Munshi, D. and Murphy, J. A. and Naselsky, P. and Nati, F. and Natoli, P. and Netterfield, C. B. and Nørgaard-Nielsen, H. U. and Noviello, F. and Novikov, D. and Novikov, I. and Oxborrow, C. A. and Paci, F. and Pagano, L. and Pajot, F. and Paladini, R. and Paoletti, D. and Partridge, B. and Pasian, F. and Patanchon, G. and Pearson, T. J. and Perdereau, O. and Perotto, L. and Perrotta, F. and Pettorino, V. and Piacentini, F. and Piat, M. and Pierpaoli, E. and Pietrobon, D. and Plaszczynski, S. and Pointecouteau, E. and Polenta, G. and Popa, L. and Pratt, G. W. and Prézeau, G. and Prunet, S. and Puget, J. -L. and Rachen, J. P. and Reach, W. T. and Rebolo, R. and Reinecke, M. and Remazeilles, M. and Renault, C. and Renzi, A. and Ristorcelli, I. and Rocha, G. and Rosset, C. and Rossetti, M. and Roudier, G. and Rouillé d'Orfeuil, B. and Rowan-Robinson, M. and Rubiño-Martín, J. A. and Rusholme, B. and Said, N. and Salvatelli, V. and Salvati, L. and Sandri, M. and Santos, D. and Savelainen, M. and Savini, G. and Scott, D. and Seiffert, M. D. and Serra, P. and Shellard, E. P. S. and Spencer, L. D. and Spinelli, M. and Stolyarov, V. and Stompor, R. and Sudiwala, R. and Sunyaev, R. and Sutton, D. and Suur-Uski, A. -S. and Sygnet, J. -F. and Tauber, J. A. and Terenzi, L. and Toffolatti, L. and Tomasi, M. and Tristram, M. and Trombetti, T. and Tucci, M. and Tuovinen, J. and Türler, M. and Umana, G. and Valenziano, L. and Valiviita, J. and Van Tent, F. and Vielva, P. and Villa, F. and Wade, L. A. and Wandelt, B. D. and Wehus, I. K. and White, M. and White, S. D. M. and Wilkinson, A. and Yvon, D. and Zacchei, A. and Zonca, A.},
	month = sep,
	year = {2016},
	note = {ADS Bibcode: 2016A\&A...594A..13P},
	pages = {A13},
}

@article{naiman_first_2018,
	title = {First results from the {IllustrisTNG} simulations: a tale of two elements - chemical evolution of magnesium and europium},
	volume = {477},
	issn = {0035-8711},
	shorttitle = {First results from the {IllustrisTNG} simulations},
	url = {https://ui.adsabs.harvard.edu/abs/2018MNRAS.477.1206N},
	doi = {10.1093/mnras/sty618},
	urldate = {2024-03-04},
	journal = {Monthly Notices of the Royal Astronomical Society},
	author = {Naiman, Jill P. and Pillepich, Annalisa and Springel, Volker and Ramirez-Ruiz, Enrico and Torrey, Paul and Vogelsberger, Mark and Pakmor, Rüdiger and Nelson, Dylan and Marinacci, Federico and Hernquist, Lars and Weinberger, Rainer and Genel, Shy},
	month = jun,
	year = {2018},
	note = {ADS Bibcode: 2018MNRAS.477.1206N},
	pages = {1206--1224},
}

@article{marinacci_first_2018,
	title = {First results from the {IllustrisTNG} simulations: radio haloes and magnetic fields},
	volume = {480},
	issn = {0035-8711},
	shorttitle = {First results from the {IllustrisTNG} simulations},
	url = {https://ui.adsabs.harvard.edu/abs/2018MNRAS.480.5113M},
	doi = {10.1093/mnras/sty2206},
	urldate = {2024-03-04},
	journal = {Monthly Notices of the Royal Astronomical Society},
	author = {Marinacci, Federico and Vogelsberger, Mark and Pakmor, Rüdiger and Torrey, Paul and Springel, Volker and Hernquist, Lars and Nelson, Dylan and Weinberger, Rainer and Pillepich, Annalisa and Naiman, Jill and Genel, Shy},
	month = nov,
	year = {2018},
	note = {ADS Bibcode: 2018MNRAS.480.5113M},
	pages = {5113--5139},
}

@article{nelson_first_2018,
	title = {First results from the {IllustrisTNG} simulations: the galaxy colour bimodality},
	volume = {475},
	issn = {0035-8711},
	shorttitle = {First results from the {IllustrisTNG} simulations},
	url = {https://ui.adsabs.harvard.edu/abs/2018MNRAS.475..624N},
	doi = {10.1093/mnras/stx3040},
	urldate = {2024-03-04},
	journal = {Monthly Notices of the Royal Astronomical Society},
	author = {Nelson, Dylan and Pillepich, Annalisa and Springel, Volker and Weinberger, Rainer and Hernquist, Lars and Pakmor, Rüdiger and Genel, Shy and Torrey, Paul and Vogelsberger, Mark and Kauffmann, Guinevere and Marinacci, Federico and Naiman, Jill},
	month = mar,
	year = {2018},
	note = {ADS Bibcode: 2018MNRAS.475..624N},
	pages = {624--647},
}

@article{pillepich_first_2018,
	title = {First results from the {IllustrisTNG} simulations: the stellar mass content of groups and clusters of galaxies},
	volume = {475},
	issn = {0035-8711},
	shorttitle = {First results from the {IllustrisTNG} simulations},
	url = {https://ui.adsabs.harvard.edu/abs/2018MNRAS.475..648P},
	doi = {10.1093/mnras/stx3112},
	urldate = {2024-03-04},
	journal = {Monthly Notices of the Royal Astronomical Society},
	author = {Pillepich, Annalisa and Nelson, Dylan and Hernquist, Lars and Springel, Volker and Pakmor, Rüdiger and Torrey, Paul and Weinberger, Rainer and Genel, Shy and Naiman, Jill P. and Marinacci, Federico and Vogelsberger, Mark},
	month = mar,
	year = {2018},
	note = {ADS Bibcode: 2018MNRAS.475..648P},
	pages = {648--675},
}

@article{springel_first_2018,
	title = {First results from the {IllustrisTNG} simulations: matter and galaxy clustering},
	volume = {475},
	issn = {0035-8711},
	shorttitle = {First results from the {IllustrisTNG} simulations},
	url = {https://ui.adsabs.harvard.edu/abs/2018MNRAS.475..676S},
	doi = {10.1093/mnras/stx3304},
	urldate = {2024-03-04},
	journal = {Monthly Notices of the Royal Astronomical Society},
	author = {Springel, Volker and Pakmor, Rüdiger and Pillepich, Annalisa and Weinberger, Rainer and Nelson, Dylan and Hernquist, Lars and Vogelsberger, Mark and Genel, Shy and Torrey, Paul and Marinacci, Federico and Naiman, Jill},
	month = mar,
	year = {2018},
	note = {ADS Bibcode: 2018MNRAS.475..676S},
	pages = {676--698},
}

@article{kaviraj_horizon-agn_2017,
	title = {The {Horizon}-{AGN} simulation: evolution of galaxy properties over cosmic time},
	volume = {467},
	issn = {0035-8711},
	shorttitle = {The {Horizon}-{AGN} simulation},
	url = {https://ui.adsabs.harvard.edu/abs/2017MNRAS.467.4739K},
	doi = {10.1093/mnras/stx126},
	urldate = {2024-03-01},
	journal = {Monthly Notices of the Royal Astronomical Society},
	author = {Kaviraj, S. and Laigle, C. and Kimm, T. and Devriendt, J. E. G. and Dubois, Y. and Pichon, C. and Slyz, A. and Chisari, E. and Peirani, S.},
	month = jun,
	year = {2017},
	note = {ADS Bibcode: 2017MNRAS.467.4739K},
	pages = {4739--4752},
}

@article{snyder_automated_2019,
	title = {Automated {Distant} {Galaxy} {Merger} {Classifications} from {Space} {Telescope} {Images} using the {Illustris} {Simulation}},
	volume = {486},
	issn = {0035-8711, 1365-2966},
	url = {http://arxiv.org/abs/1809.02136},
	doi = {10.1093/mnras/stz1059},
	number = {3},
	urldate = {2024-03-01},
	journal = {Monthly Notices of the Royal Astronomical Society},
	author = {Snyder, Gregory F. and Rodriguez-Gomez, Vicente and Lotz, Jennifer M. and Torrey, Paul and Quirk, Amanda C. N. and Hernquist, Lars and Vogelsberger, Mark and Freeman, Peter E.},
	month = jul,
	year = {2019},
	note = {arXiv:1809.02136 [astro-ph]},
	pages = {3702--3720},
}

@article{kaviraj_importance_2014,
	title = {The importance of minor-merger-driven star formation and black hole growth in disc galaxies},
	volume = {440},
	issn = {0035-8711},
	url = {https://ui.adsabs.harvard.edu/abs/2014MNRAS.440.2944K},
	doi = {10.1093/mnras/stu338},
	urldate = {2024-02-15},
	journal = {Monthly Notices of the Royal Astronomical Society},
	author = {Kaviraj, Sugata},
	month = jun,
	year = {2014},
	note = {ADS Bibcode: 2014MNRAS.440.2944K},
	pages = {2944--2952},
}

@article{barrows_census_2023,
	title = {A {Census} of {WISE}-selected {Dual} and {Offset} {AGNs} {Across} the {Sky}: {New} {Constraints} on {Merger}-driven {Triggering} of {Obscured} {AGNs}},
	volume = {951},
	issn = {0004-637X},
	shorttitle = {A {Census} of {WISE}-selected {Dual} and {Offset} {AGNs} {Across} the {Sky}},
	url = {https://ui.adsabs.harvard.edu/abs/2023ApJ...951...92B},
	doi = {10.3847/1538-4357/acd2d3},
	urldate = {2024-02-13},
	journal = {The Astrophysical Journal},
	author = {Barrows, R. Scott and Comerford, Julia M. and Stern, Daniel and Assef, Roberto J.},
	month = jul,
	year = {2023},
	note = {ADS Bibcode: 2023ApJ...951...92B},
	pages = {92},
}

@article{crain_eagle_2015,
	title = {The {EAGLE} simulations of galaxy formation: calibration of subgrid physics and model variations},
	volume = {450},
	issn = {1365-2966, 0035-8711},
	shorttitle = {The {EAGLE} simulations of galaxy formation},
	url = {http://academic.oup.com/mnras/article/450/2/1937/984366/The-EAGLE-simulations-of-galaxy-formation},
	doi = {10.1093/mnras/stv725},
	language = {en},
	number = {2},
	urldate = {2023-12-20},
	journal = {Monthly Notices of the Royal Astronomical Society},
	author = {Crain, Robert A. and Schaye, Joop and Bower, Richard G. and Furlong, Michelle and Schaller, Matthieu and Theuns, Tom and Dalla Vecchia, Claudio and Frenk, Carlos S. and McCarthy, Ian G. and Helly, John C. and Jenkins, Adrian and Rosas-Guevara, Yetli M. and White, Simon D. M. and Trayford, James W.},
	month = jun,
	year = {2015},
	pages = {1937--1961},
}

@article{schaye_eagle_2015,
	title = {The {EAGLE} project: simulating the evolution and assembly of galaxies and their environments},
	volume = {446},
	issn = {1365-2966, 0035-8711},
	shorttitle = {The {EAGLE} project},
	url = {http://academic.oup.com/mnras/article/446/1/521/1316115/The-EAGLE-project-simulating-the-evolution-and},
	doi = {10.1093/mnras/stu2058},
	language = {en},
	number = {1},
	urldate = {2023-12-20},
	journal = {Monthly Notices of the Royal Astronomical Society},
	author = {Schaye, Joop and Crain, Robert A. and Bower, Richard G. and Furlong, Michelle and Schaller, Matthieu and Theuns, Tom and Dalla Vecchia, Claudio and Frenk, Carlos S. and McCarthy, I. G. and Helly, John C. and Jenkins, Adrian and Rosas-Guevara, Y. M. and White, Simon D. M. and Baes, Maarten and Booth, C. M. and Camps, Peter and Navarro, Julio F. and Qu, Yan and Rahmati, Alireza and Sawala, Till and Thomas, Peter A. and Trayford, James},
	month = jan,
	year = {2015},
	pages = {521--554},
}

@article{ellison_galaxy_2008,
	title = {{GALAXY} {PAIRS} {IN} {THE} {SLOAN} {DIGITAL} {SKY} {SURVEY}. {I}. {STAR} {FORMATION}, {ACTIVE} {GALACTIC} {NUCLEUS} {FRACTION}, {AND} {THE} {LUMINOSITY}/{MASS}-{METALLICITY} {RELATION}},
	volume = {135},
	issn = {0004-6256, 1538-3881},
	url = {https://iopscience.iop.org/article/10.1088/0004-6256/135/5/1877},
	doi = {10.1088/0004-6256/135/5/1877},
	language = {en},
	number = {5},
	urldate = {2023-12-20},
	journal = {The Astronomical Journal},
	author = {Ellison, Sara L. and Patton, David R. and Simard, Luc and McConnachie, Alan W.},
	month = may,
	year = {2008},
	pages = {1877--1899},
}

@article{barro_candelsshards_2019,
	title = {The {CANDELS}/{SHARDS} {Multiwavelength} {Catalog} in {GOODS}-{N}: {Photometry}, {Photometric} {Redshifts}, {Stellar} {Masses}, {Emission}-line {Fluxes}, and {Star} {Formation} {Rates}},
	volume = {243},
	issn = {1538-4365},
	shorttitle = {The {CANDELS}/{SHARDS} {Multiwavelength} {Catalog} in {GOODS}-{N}},
	url = {https://iopscience.iop.org/article/10.3847/1538-4365/ab23f2},
	doi = {10.3847/1538-4365/ab23f2},
	language = {en},
	number = {2},
	urldate = {2023-12-20},
	journal = {The Astrophysical Journal Supplement Series},
	author = {Barro, Guillermo and Pérez-González, Pablo G. and Cava, Antonio and Brammer, Gabriel and Pandya, Viraj and Moral, Carmen Eliche and Esquej, Pilar and Domínguez-Sánchez, Helena and Pampliega, Belen Alcalde and Guo, Yicheng and Koekemoer, Anton M. and Trump, Jonathan R. and Ashby, Matthew L. N. and Cardiel, Nicolas and Castellano, Marco and Conselice, Christopher J. and Dickinson, Mark E. and Dolch, Timothy and Donley, Jennifer L. and Briones, Néstor Espino and Faber, Sandra M. and Fazio, Giovanni G. and Ferguson, Henry and Finkelstein, Steve and Fontana, Adriano and Galametz, Audrey and Gardner, Jonathan P. and Gawiser, Eric and Giavalisco, Mauro and Grazian, Andrea and Grogin, Norman A. and Hathi, Nimish P. and Hemmati, Shoubaneh and Hernán-Caballero, Antonio and Kocevski, Dale and Koo, David C. and Kodra, Dritan and Lee, Kyoung-Soo and Lin, Lihwai and Lucas, Ray A. and Mobasher, Bahram and McGrath, Elizabeth J. and Nandra, Kirpal and Nayyeri, Hooshang and Newman, Jeffrey A. and Pforr, Janine and Peth, Michael and Rafelski, Marc and Rodríguez-Munoz, Lucia and Salvato, Mara and Stefanon, Mauro and Wel, Arjen Van Der and Willner, Steven P. and Wiklind, Tommy and Wuyts, Stijn},
	month = jul,
	year = {2019},
	pages = {22},
}

@article{baes_efficient_2011,
	title = {{EFFICIENT} {THREE}-{DIMENSIONAL} {NLTE} {DUST} {RADIATIVE} {TRANSFER} {WITH} {SKIRT}},
	volume = {196},
	issn = {0067-0049, 1538-4365},
	url = {https://iopscience.iop.org/article/10.1088/0067-0049/196/2/22},
	doi = {10.1088/0067-0049/196/2/22},
	language = {en},
	number = {2},
	urldate = {2023-12-20},
	journal = {The Astrophysical Journal Supplement Series},
	author = {Baes, Maarten and Verstappen, Joris and De Looze, Ilse and Fritz, Jacopo and Saftly, Waad and Vidal Pérez, Edgardo and Stalevski, Marko and Valcke, Sander},
	month = oct,
	year = {2011},
	pages = {22},
}

@article{camps_skirt_2015,
	title = {{SKIRT}: an {Advanced} {Dust} {Radiative} {Transfer} {Code} with a {User}-{Friendly} {Architecture}},
	volume = {9},
	issn = {22131337},
	shorttitle = {{SKIRT}},
	url = {http://arxiv.org/abs/1410.1629},
	doi = {10.1016/j.ascom.2014.10.004},
	urldate = {2023-12-20},
	journal = {Astronomy and Computing},
	author = {Camps, Peter and Baes, Maarten},
	month = mar,
	year = {2015},
	note = {arXiv:1410.1629 [astro-ph]},
	pages = {20--33},
}

@article{conroy_propagation_2009,
	title = {{THE} {PROPAGATION} {OF} {UNCERTAINTIES} {IN} {STELLAR} {POPULATION} {SYNTHESIS} {MODELING}. {I}. {THE} {RELEVANCE} {OF} {UNCERTAIN} {ASPECTS} {OF} {STELLAR} {EVOLUTION} {AND} {THE} {INITIAL} {MASS} {FUNCTION} {TO} {THE} {DERIVED} {PHYSICAL} {PROPERTIES} {OF} {GALAXIES}},
	volume = {699},
	issn = {0004-637X, 1538-4357},
	url = {https://iopscience.iop.org/article/10.1088/0004-637X/699/1/486},
	doi = {10.1088/0004-637X/699/1/486},
	language = {en},
	number = {1},
	urldate = {2023-12-20},
	journal = {The Astrophysical Journal},
	author = {Conroy, Charlie and Gunn, James E. and White, Martin},
	month = jul,
	year = {2009},
	pages = {486--506},
}

@article{conroy_propagation_2010,
	title = {{THE} {PROPAGATION} {OF} {UNCERTAINTIES} {IN} {STELLAR} {POPULATION} {SYNTHESIS} {MODELING}. {III}. {MODEL} {CALIBRATION}, {COMPARISON}, {AND} {EVALUATION}},
	volume = {712},
	issn = {0004-637X, 1538-4357},
	url = {https://iopscience.iop.org/article/10.1088/0004-637X/712/2/833},
	doi = {10.1088/0004-637X/712/2/833},
	language = {en},
	number = {2},
	urldate = {2023-12-20},
	journal = {The Astrophysical Journal},
	author = {Conroy, Charlie and Gunn, James E.},
	month = apr,
	year = {2010},
	pages = {833--857},
}

@article{groves_modeling_2008,
	title = {Modeling the {Pan}‐{Spectral} {Energy} {Distribution} of {Starburst} {Galaxies}. {IV}. {The} {Controlling} {Parameters} of the {Starburst} {SED}},
	volume = {176},
	issn = {0067-0049, 1538-4365},
	url = {https://iopscience.iop.org/article/10.1086/528711},
	doi = {10.1086/528711},
	language = {en},
	number = {2},
	urldate = {2023-12-20},
	journal = {The Astrophysical Journal Supplement Series},
	author = {Groves, Brent and Dopita, Michael A. and Sutherland, Ralph S. and Kewley, Lisa J. and Fischera, Jörg and Leitherer, Claus and Brandl, Bernhard and Van Breugel, Wil},
	month = jun,
	year = {2008},
	pages = {438--456},
}

@article{ouchi_large_2009,
	title = {{LARGE} {AREA} {SURVEY} {FOR} z = 7 {GALAXIES} {IN} {SDF} {AND} {GOODS}-{N}: {IMPLICATIONS} {FOR} {GALAXY} {FORMATION} {AND} {COSMIC} {REIONIZATION}*},
	volume = {706},
	issn = {0004-637X, 1538-4357},
	shorttitle = {{LARGE} {AREA} {SURVEY} {FOR} z = 7 {GALAXIES} {IN} {SDF} {AND} {GOODS}-{N}},
	url = {https://iopscience.iop.org/article/10.1088/0004-637X/706/2/1136},
	doi = {10.1088/0004-637X/706/2/1136},
	language = {en},
	number = {2},
	urldate = {2023-12-20},
	journal = {The Astrophysical Journal},
	author = {Ouchi, Masami and Mobasher, Bahram and Shimasaku, Kazuhiro and Ferguson, Henry C. and Fall, S. Michael and Ono, Yoshiaki and Kashikawa, Nobunari and Morokuma, Tomoki and Nakajima, Kimihiko and Okamura, Sadanori and Dickinson, Mark and Giavalisco, Mauro and Ohta, Kouji},
	month = dec,
	year = {2009},
	pages = {1136--1151},
}

@article{mclure_luminosity_2009,
	title = {The luminosity function, halo masses and stellar masses of luminous {Lyman}-break galaxies at redshifts 5 {\textless} z {\textless} 6},
	volume = {395},
	issn = {00358711, 13652966},
	url = {https://academic.oup.com/mnras/article-lookup/doi/10.1111/j.1365-2966.2009.14677.x},
	doi = {10.1111/j.1365-2966.2009.14677.x},
	language = {en},
	number = {4},
	urldate = {2023-12-20},
	journal = {Monthly Notices of the Royal Astronomical Society},
	author = {McLure, R. J. and Cirasuolo, M. and Dunlop, J. S. and Foucaud, S. and Almaini, O.},
	month = jun,
	year = {2009},
	pages = {2196--2209},
}

@article{j_bouwens_alma_2016,
	title = {{ALMA} {SPECTROSCOPIC} {SURVEY} {IN} {THE} {HUBBLE} {ULTRA} {DEEP} {FIELD}: {THE} {INFRARED} {EXCESS} {OF} {UV}-{SELECTED} \textit{z} = 2–10 {GALAXIES} {AS} {A} {FUNCTION} {OF} {UV}-{CONTINUUM} {SLOPE} {AND} {STELLAR} {MASS}},
	volume = {833},
	issn = {1538-4357},
	shorttitle = {{ALMA} {SPECTROSCOPIC} {SURVEY} {IN} {THE} {HUBBLE} {ULTRA} {DEEP} {FIELD}},
	url = {https://iopscience.iop.org/article/10.3847/1538-4357/833/1/72},
	doi = {10.3847/1538-4357/833/1/72},
	language = {en},
	number = {1},
	urldate = {2023-12-20},
	journal = {The Astrophysical Journal},
	author = {J. Bouwens, Rychard and Aravena, Manuel and Decarli, Roberto and Walter, Fabian and Da Cunha, Elisabete and Labbé, Ivo and E. Bauer, Franz and Bertoldi, Frank and Carilli, Chris and Chapman, Scott and Daddi, Emanuele and Hodge, Jacqueline and J. Ivison, Rob and Karim, Alex and Le Fevre, Olivier and Magnelli, Benjamin and Ota, Kazuaki and Riechers, Dominik and R. Smail, Ian and Van Der Werf, Paul and Weiss, Axel and Cox, Pierre and Elbaz, David and Gonzalez-Lopez, Jorge and Infante, Leopoldo and Oesch, Pascal and Wagg, Jeff and Wilkins, Steve},
	month = dec,
	year = {2016},
	pages = {72},
}

@article{finkelstein_observational_2016,
	title = {Observational {Searches} for {Star}-{Forming} {Galaxies} at \textit{z} {\textgreater} 6},
	volume = {33},
	issn = {1323-3580, 1448-6083},
	url = {https://www.cambridge.org/core/product/identifier/S1323358016000266/type/journal_article},
	doi = {10.1017/pasa.2016.26},
	language = {en},
	urldate = {2023-12-20},
	journal = {Publications of the Astronomical Society of Australia},
	author = {Finkelstein, Steven L.},
	year = {2016},
	pages = {e037},
}

@article{oesch_dearth_2018,
	title = {The {Dearth} of \textit{z} ∼ 10 {Galaxies} in {All} \textit{{HST}} {Legacy} {Fields}—{The} {Rapid} {Evolution} of the {Galaxy} {Population} in the {First} 500 {Myr}},
	volume = {855},
	issn = {1538-4357},
	url = {https://iopscience.iop.org/article/10.3847/1538-4357/aab03f},
	doi = {10.3847/1538-4357/aab03f},
	language = {en},
	number = {2},
	urldate = {2023-12-20},
	journal = {The Astrophysical Journal},
	author = {Oesch, P. A. and Bouwens, R. J. and Illingworth, G. D. and Labbé, I. and Stefanon, M.},
	month = mar,
	year = {2018},
	pages = {105},
}

@article{draine_dust_2007,
	title = {Dust {Masses}, {PAH} {Abundances}, and {Starlight} {Intensities} in the {SINGS} {Galaxy} {Sample}},
	volume = {663},
	issn = {0004-637X, 1538-4357},
	url = {https://iopscience.iop.org/article/10.1086/518306},
	doi = {10.1086/518306},
	language = {en},
	number = {2},
	urldate = {2023-12-20},
	journal = {The Astrophysical Journal},
	author = {Draine, B. T. and Dale, D. A. and Bendo, G. and Gordon, K. D. and Smith, J. D. T. and Armus, L. and Engelbracht, C. W. and Helou, G. and Kennicutt, Jr., R. C. and Li, A. and Roussel, H. and Walter, F. and Calzetti, D. and Moustakas, J. and Murphy, E. J. and Rieke, G. H. and Bot, C. and Hollenbach, D. J. and Sheth, K. and Teplitz, H. I.},
	month = jul,
	year = {2007},
	pages = {866--894},
}

@article{schartmann_towards_2005,
	title = {Towards a physical model of dust tori in {Active} {Galactic} {Nuclei}: {Radiative} transfer calculations for a hydrostatic torus model},
	volume = {437},
	issn = {0004-6361, 1432-0746},
	shorttitle = {Towards a physical model of dust tori in {Active} {Galactic} {Nuclei}},
	url = {http://www.aanda.org/10.1051/0004-6361:20042363},
	doi = {10.1051/0004-6361:20042363},
	language = {en},
	number = {3},
	urldate = {2023-12-20},
	journal = {Astronomy \& Astrophysics},
	author = {Schartmann, M. and Meisenheimer, K. and Camenzind, M. and Wolf, S. and Henning, Th.},
	month = jul,
	year = {2005},
	pages = {861--881},
}

@article{bottrell_combined_2022,
	title = {The combined and respective roles of imaging and stellar kinematics in identifying galaxy merger remnants},
	volume = {511},
	issn = {0035-8711, 1365-2966},
	url = {https://academic.oup.com/mnras/article/511/1/100/6482852},
	doi = {10.1093/mnras/stab3717},
	language = {en},
	number = {1},
	urldate = {2023-12-20},
	journal = {Monthly Notices of the Royal Astronomical Society},
	author = {Bottrell, Connor and Hani, Maan H and Teimoorinia, Hossen and Patton, David R and Ellison, Sara L},
	month = feb,
	year = {2022},
	pages = {100--119},
}

@misc{arrieta_explainable_2019,
	title = {Explainable {Artificial} {Intelligence} ({XAI}): {Concepts}, {Taxonomies}, {Opportunities} and {Challenges} toward {Responsible} {AI}},
	shorttitle = {Explainable {Artificial} {Intelligence} ({XAI})},
	url = {http://arxiv.org/abs/1910.10045},
	doi = {10.48550/arXiv.1910.10045},
	urldate = {2023-12-10},
	publisher = {arXiv},
	author = {Arrieta, Alejandro Barredo and Díaz-Rodríguez, Natalia and Del Ser, Javier and Bennetot, Adrien and Tabik, Siham and Barbado, Alberto and García, Salvador and Gil-López, Sergio and Molina, Daniel and Benjamins, Richard and Chatila, Raja and Herrera, Francisco},
	month = dec,
	year = {2019},
	note = {arXiv:1910.10045 [cs]},
}

@article{selvaraju_grad-cam_2020,
	title = {Grad-{CAM}: {Visual} {Explanations} from {Deep} {Networks} via {Gradient}-based {Localization}},
	volume = {128},
	issn = {0920-5691, 1573-1405},
	shorttitle = {Grad-{CAM}},
	url = {http://arxiv.org/abs/1610.02391},
	doi = {10.1007/s11263-019-01228-7},
	number = {2},
	urldate = {2023-12-10},
	journal = {International Journal of Computer Vision},
	author = {Selvaraju, Ramprasaath R. and Cogswell, Michael and Das, Abhishek and Vedantam, Ramakrishna and Parikh, Devi and Batra, Dhruv},
	month = feb,
	year = {2020},
	note = {arXiv:1610.02391 [cs]},
	pages = {336--359},
}

@article{nelson_first_2019,
	title = {First results from the {TNG50} simulation: galactic outflows driven by supernovae and black hole feedback},
	volume = {490},
	issn = {0035-8711},
	shorttitle = {First results from the {TNG50} simulation},
	url = {https://ui.adsabs.harvard.edu/abs/2019MNRAS.490.3234N},
	doi = {10.1093/mnras/stz2306},
	urldate = {2023-12-09},
	journal = {Monthly Notices of the Royal Astronomical Society},
	author = {Nelson, Dylan and Pillepich, Annalisa and Springel, Volker and Pakmor, Rüdiger and Weinberger, Rainer and Genel, Shy and Torrey, Paul and Vogelsberger, Mark and Marinacci, Federico and Hernquist, Lars},
	month = dec,
	year = {2019},
	note = {ADS Bibcode: 2019MNRAS.490.3234N},
	pages = {3234--3261},
}

@article{ackermann_using_2018,
	title = {Using transfer learning to detect galaxy mergers},
	volume = {479},
	issn = {0035-8711, 1365-2966},
	url = {https://academic.oup.com/mnras/article/479/1/415/5017494},
	doi = {10.1093/mnras/sty1398},
	language = {en},
	number = {1},
	urldate = {2023-08-03},
	journal = {Monthly Notices of the Royal Astronomical Society},
	author = {Ackermann, Sandro and Schawinski, Kevin and Zhang, Ce and Weigel, Anna K and Turp, M Dennis},
	month = sep,
	year = {2018},
	pages = {415--425},
}

@article{pearson_identifying_2019,
	title = {Identifying galaxy mergers in observations and simulations with deep learning},
	volume = {626},
	copyright = {© ESO 2019},
	issn = {0004-6361, 1432-0746},
	url = {https://www.aanda.org/articles/aa/abs/2019/06/aa35355-19/aa35355-19.html},
	doi = {10.1051/0004-6361/201935355},
	language = {en},
	urldate = {2023-08-03},
	journal = {Astronomy \& Astrophysics},
	publisher = {EDP Sciences},
	author = {Pearson, W. J. and Wang, L. and Trayford, J. W. and Petrillo, C. E. and Tak, F. F. S. van der},
	month = jun,
	year = {2019},
	pages = {A49},
}

@misc{desmons_galaxy_2023,
	title = {Galaxy {And} {Mass} {Assembly} ({GAMA}): {Comparing} {Visually} and {Spectroscopically} {Identified} {Galaxy} {Merger} {Samples}},
	shorttitle = {Galaxy {And} {Mass} {Assembly} ({GAMA})},
	url = {http://arxiv.org/abs/2305.17894},
	doi = {10.1093/mnras/stad1639},
	urldate = {2023-06-08},
	author = {Desmons, Alice and Brough, Sarah and Martínez-Lombilla, Cristina and De Propris, Roberto and Holwerda, Benne and Sánchez, Ángel R. López},
	month = may,
	year = {2023},
	note = {arXiv:2305.17894 [astro-ph]},
}

@article{lotz_galaxy_2008,
	title = {Galaxy merger morphologies and time-scales from simulations of equal-mass gas-rich disc mergers},
	volume = {391},
	issn = {0035-8711},
	url = {https://ui.adsabs.harvard.edu/abs/2008MNRAS.391.1137L},
	doi = {10.1111/j.1365-2966.2008.14004.x},
	urldate = {2022-04-20},
	journal = {Monthly Notices of the Royal Astronomical Society},
	author = {Lotz, Jennifer M. and Jonsson, Patrik and Cox, T. J. and Primack, Joel R.},
	month = dec,
	year = {2008},
	note = {ADS Bibcode: 2008MNRAS.391.1137L},
	pages = {1137--1162},
}

@article{krist_20_2011,
	title = {20 years of {Hubble} {Space} {Telescope} optical modeling using {Tiny} {Tim}},
	volume = {8127},
	url = {https://ui.adsabs.harvard.edu/abs/2011SPIE.8127E..0JK},
	doi = {10.1117/12.892762},
	urldate = {2022-04-19},
	author = {Krist, John E. and Hook, Richard N. and Stoehr, Felix},
	month = oct,
	year = {2011},
	note = {Conference Name: Optical Modeling and Performance Predictions V
ADS Bibcode: 2011SPIE.8127E..0JK},
	pages = {81270J},
}

@article{mantha_major_2018,
	title = {Major merging history in {CANDELS}. {I}. {Evolution} of the incidence of massive galaxy-galaxy pairs from z = 3 to z ∼ 0},
	volume = {475},
	issn = {0035-8711},
	url = {https://ui.adsabs.harvard.edu/abs/2018MNRAS.475.1549M},
	doi = {10.1093/mnras/stx3260},
	urldate = {2022-04-19},
	journal = {Monthly Notices of the Royal Astronomical Society},
	author = {Mantha, Kameswara Bharadwaj and McIntosh, Daniel H. and Brennan, Ryan and Ferguson, Henry C. and Kodra, Dritan and Newman, Jeffrey A. and Rafelski, Marc and Somerville, Rachel S. and Conselice, Christopher J. and Cook, Joshua S. and Hathi, Nimish P. and Koo, David C. and Lotz, Jennifer M. and Simmons, Brooke D. and Straughn, Amber N. and Snyder, Gregory F. and Wuyts, Stijn and Bell, Eric F. and Dekel, Avishai and Kartaltepe, Jeyhan and Kocevski, Dale D. and Koekemoer, Anton M. and Lee, Seong-Kook and Lucas, Ray A. and Pacifici, Camilla and Peth, Michael A. and Barro, Guillermo and Dahlen, Tomas and Finkelstein, Steven L. and Fontana, Adriano and Galametz, Audrey and Grogin, Norman A. and Guo, Yicheng and Mobasher, Bahram and Nayyeri, Hooshang and Pérez-González, Pablo G. and Pforr, Janine and Santini, Paola and Stefanon, Mauro and Wiklind, Tommy},
	month = apr,
	year = {2018},
	note = {ADS Bibcode: 2018MNRAS.475.1549M},
	pages = {1549--1573},
}

@article{pillepich_first_2019,
	title = {First results from the {TNG50} simulation: the evolution of stellar and gaseous discs across cosmic time},
	volume = {490},
	issn = {0035-8711},
	shorttitle = {First results from the {TNG50} simulation},
	url = {https://ui.adsabs.harvard.edu/abs/2019MNRAS.490.3196P},
	doi = {10.1093/mnras/stz2338},
	urldate = {2022-04-19},
	journal = {Monthly Notices of the Royal Astronomical Society},
	author = {Pillepich, Annalisa and Nelson, Dylan and Springel, Volker and Pakmor, Rüdiger and Torrey, Paul and Weinberger, Rainer and Vogelsberger, Mark and Marinacci, Federico and Genel, Shy and van der Wel, Arjen and Hernquist, Lars},
	month = dec,
	year = {2019},
	note = {ADS Bibcode: 2019MNRAS.490.3196P},
	pages = {3196--3233},
}

@article{rodriguez-gomez_merger_2015,
	title = {The merger rate of galaxies in the {Illustris} simulation: a comparison with observations and semi-empirical models},
	volume = {449},
	issn = {0035-8711},
	shorttitle = {The merger rate of galaxies in the {Illustris} simulation},
	url = {https://ui.adsabs.harvard.edu/abs/2015MNRAS.449...49R},
	doi = {10.1093/mnras/stv264},
	urldate = {2022-04-19},
	journal = {Monthly Notices of the Royal Astronomical Society},
	author = {Rodriguez-Gomez, Vicente and Genel, Shy and Vogelsberger, Mark and Sijacki, Debora and Pillepich, Annalisa and Sales, Laura V. and Torrey, Paul and Snyder, Greg and Nelson, Dylan and Springel, Volker and Ma, Chung-Pei and Hernquist, Lars},
	month = may,
	year = {2015},
	note = {ADS Bibcode: 2015MNRAS.449...49R},
	pages = {49--64},
}

@article{nevin_accurate_2019,
	title = {Accurate {Identification} of {Galaxy} {Mergers} with {Imaging}},
	volume = {872},
	issn = {0004-637X},
	url = {https://ui.adsabs.harvard.edu/abs/2019ApJ...872...76N},
	doi = {10.3847/1538-4357/aafd34},
	urldate = {2022-04-19},
	journal = {The Astrophysical Journal},
	author = {Nevin, R. and Blecha, L. and Comerford, J. and Greene, J.},
	month = feb,
	year = {2019},
	note = {ADS Bibcode: 2019ApJ...872...76N},
	pages = {76},
}

@article{conselice_evolution_2014,
	title = {The {Evolution} of {Galaxy} {Structure} {Over} {Cosmic} {Time}},
	volume = {52},
	issn = {0066-4146},
	url = {https://ui.adsabs.harvard.edu/abs/2014ARA&A..52..291C/abstract},
	doi = {10.1146/annurev-astro-081913-040037},
	language = {en},
	urldate = {2022-04-19},
	journal = {Annual Review of Astronomy and Astrophysics},
	author = {Conselice, Christopher J.},
	month = aug,
	year = {2014},
	pages = {291--337},
}

@article{guo_candels_2013,
	title = {{CANDELS} {Multi}-wavelength {Catalogs}: {Source} {Detection} and {Photometry} in the {GOODS}-{South} {Field}},
	volume = {207},
	issn = {0067-0049},
	shorttitle = {{CANDELS} {Multi}-wavelength {Catalogs}},
	url = {https://ui.adsabs.harvard.edu/abs/2013ApJS..207...24G},
	doi = {10.1088/0067-0049/207/2/24},
	urldate = {2022-04-19},
	journal = {The Astrophysical Journal Supplement Series},
	author = {Guo, Yicheng and Ferguson, Henry C. and Giavalisco, Mauro and Barro, Guillermo and Willner, S. P. and Ashby, Matthew L. N. and Dahlen, Tomas and Donley, Jennifer L. and Faber, Sandra M. and Fontana, Adriano and Galametz, Audrey and Grazian, Andrea and Huang, Kuang-Han and Kocevski, Dale D. and Koekemoer, Anton M. and Koo, David C. and McGrath, Elizabeth J. and Peth, Michael and Salvato, Mara and Wuyts, Stijn and Castellano, Marco and Cooray, Asantha R. and Dickinson, Mark E. and Dunlop, James S. and Fazio, G. G. and Gardner, Jonathan P. and Gawiser, Eric and Grogin, Norman A. and Hathi, Nimish P. and Hsu, Li-Ting and Lee, Kyoung-Soo and Lucas, Ray A. and Mobasher, Bahram and Nandra, Kirpal and Newman, Jeffery A. and van der Wel, Arjen},
	month = aug,
	year = {2013},
	note = {ADS Bibcode: 2013ApJS..207...24G},
	pages = {24},
}

@article{vogelsberger_high-redshift_2020,
	title = {High-redshift {JWST} predictions from {IllustrisTNG}: dust modelling and galaxy luminosity functions},
	volume = {492},
	issn = {0035-8711},
	shorttitle = {High-redshift {JWST} predictions from {IllustrisTNG}},
	url = {https://ui.adsabs.harvard.edu/abs/2020MNRAS.492.5167V},
	doi = {10.1093/mnras/staa137},
	urldate = {2022-04-19},
	journal = {Monthly Notices of the Royal Astronomical Society},
	author = {Vogelsberger, Mark and Nelson, Dylan and Pillepich, Annalisa and Shen, Xuejian and Marinacci, Federico and Springel, Volker and Pakmor, Rüdiger and Tacchella, Sandro and Weinberger, Rainer and Torrey, Paul and Hernquist, Lars},
	month = mar,
	year = {2020},
	note = {ADS Bibcode: 2020MNRAS.492.5167V},
	pages = {5167--5201},
}

@article{shen_high-redshift_2020,
	title = {High-redshift {JWST} predictions from {IllustrisTNG}: {II}. {Galaxy} line and continuum spectral indices and dust attenuation curves},
	volume = {495},
	issn = {0035-8711},
	shorttitle = {High-redshift {JWST} predictions from {IllustrisTNG}},
	url = {https://ui.adsabs.harvard.edu/abs/2020MNRAS.495.4747S},
	doi = {10.1093/mnras/staa1423},
	urldate = {2022-04-19},
	journal = {Monthly Notices of the Royal Astronomical Society},
	author = {Shen, Xuejian and Vogelsberger, Mark and Nelson, Dylan and Pillepich, Annalisa and Tacchella, Sandro and Marinacci, Federico and Torrey, Paul and Hernquist, Lars and Springel, Volker},
	month = jul,
	year = {2020},
	note = {ADS Bibcode: 2020MNRAS.495.4747S},
	pages = {4747--4768},
}

@article{ferreira_galaxy_2020,
	title = {Galaxy {Merger} {Rates} up to z ∼ 3 {Using} a {Bayesian} {Deep} {Learning} {Model}: {A} {Major}-merger {Classifier} {Using} {IllustrisTNG} {Simulation} {Data}},
	volume = {895},
	issn = {0004-637X},
	shorttitle = {Galaxy {Merger} {Rates} up to z ∼ 3 {Using} a {Bayesian} {Deep} {Learning} {Model}},
	url = {https://ui.adsabs.harvard.edu/abs/2020ApJ...895..115F},
	doi = {10.3847/1538-4357/ab8f9b},
	urldate = {2022-04-19},
	journal = {The Astrophysical Journal},
	author = {Ferreira, Leonardo and Conselice, Christopher J. and Duncan, Kenneth and Cheng, Ting-Yun and Griffiths, Alex and Whitney, Amy},
	month = jun,
	year = {2020},
	note = {ADS Bibcode: 2020ApJ...895..115F},
	pages = {115},
}

@article{cavanagh_morphological_2021,
	title = {Morphological classification of galaxies with deep learning: comparing 3-way and 4-way {CNNs}},
	volume = {506},
	issn = {0035-8711},
	shorttitle = {Morphological classification of galaxies with deep learning},
	url = {https://ui.adsabs.harvard.edu/abs/2021MNRAS.506..659C},
	doi = {10.1093/mnras/stab1552},
	urldate = {2022-04-19},
	journal = {Monthly Notices of the Royal Astronomical Society},
	author = {Cavanagh, Mitchell K. and Bekki, Kenji and Groves, Brent A.},
	month = sep,
	year = {2021},
	note = {ADS Bibcode: 2021MNRAS.506..659C},
	pages = {659--676},
}

@article{ciprijanovic_deepmerge_2021,
	title = {{DeepMerge} - {II}. {Building} robust deep learning algorithms for merging galaxy identification across domains},
	volume = {506},
	issn = {0035-8711},
	url = {https://ui.adsabs.harvard.edu/abs/2021MNRAS.506..677C},
	doi = {10.1093/mnras/stab1677},
	urldate = {2022-04-19},
	journal = {Monthly Notices of the Royal Astronomical Society},
	author = {Ćiprijanović, A. and Kafkes, D. and Downey, K. and Jenkins, S. and Perdue, G. N. and Madireddy, S. and Johnston, T. and Snyder, G. F. and Nord, B.},
	month = sep,
	year = {2021},
	note = {ADS Bibcode: 2021MNRAS.506..677C},
	pages = {677--691},
}
\bibliographystyle{aasjournal}

\appendix

\section{Key Performance Metrics of Greyscale CNN} \label{sec:app}
\FloatBarrier
\textbf{We trained the model on two separate versions of a greyscale CNN, as stated in Section~\ref{sec:impactSFR}.
One version was given a single channel input that summed all three filters. 
The other was given a single F160W image.
We choose F160W as our single filter to omit rest-frame UV light.
We test 1) if a single filter can provide enough information for low-mass merger detection at $z\sim1$, and 2) if a single channel CNN's latent space still highlights sSFR.}

\textbf{We use the same hyperparameters as the three-channel model for the summed version, as it still has information from all three filters.
However, we do rerun Optuna for the F160W only version, as its dataset is fundamentally different than the three-channel model and thus requires a different set of hyperparameters.
For the F160W only model we use an initial learning rate of $5.346\times10^{-7}$, exponential learning rate decay of 0.90, weight decay of 2.38$\times10^{-8}$, dropout of 0.55, and a batch size of 16.
We apply the same cosine annealing learning rate scheduler as the three-channel model.}

\begin{figure}
    \centering
    \includegraphics[width=0.49\linewidth]{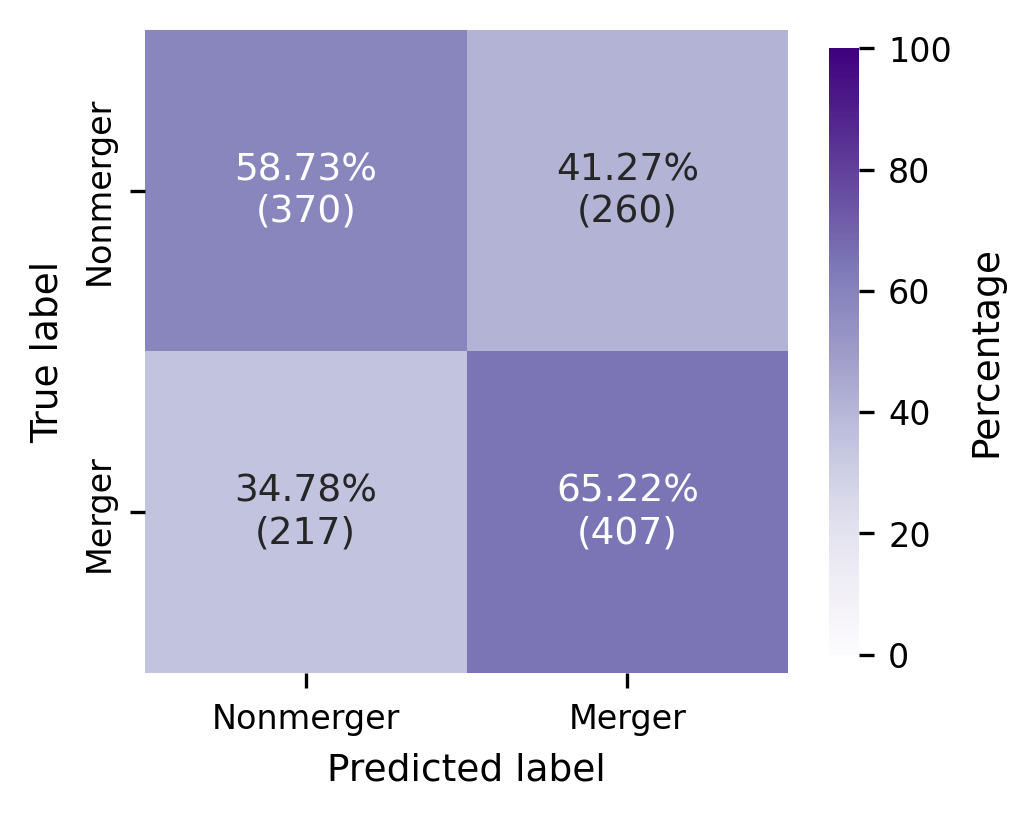}
    \includegraphics[width=0.49\linewidth]{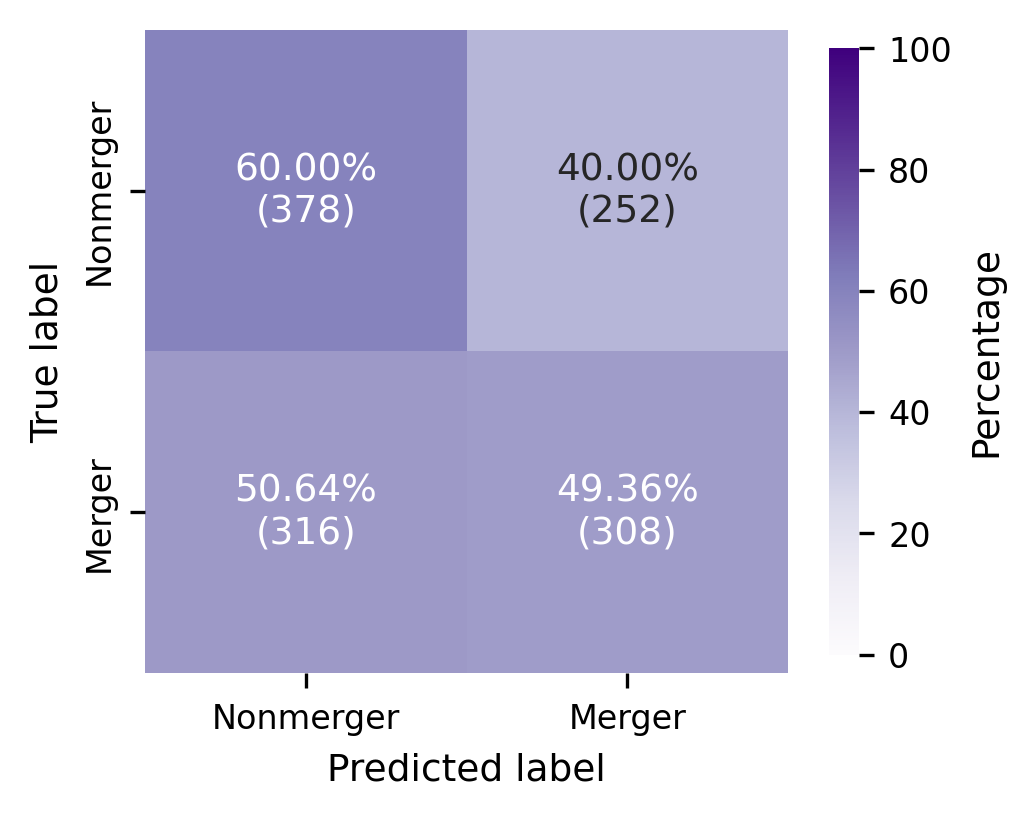}
    \caption{\textbf{Same as Figure~\ref{fig:confusion} with the greyscale data. \emph{Left:} Summed greyscale model. \emph{Right:} F160W only model. Note that both models have lower overall accuracies, and the performance across the two classes is imbalanced. The summed model is more successful at identifying mergers, while, conversely, the F160W only model is more successful at identifying nonmergers, seen by the quadrants on the diagonal.}}
    \label{fig:greyCM}
\end{figure}

\begin{figure}
    \centering
    \includegraphics[width=0.49\linewidth]{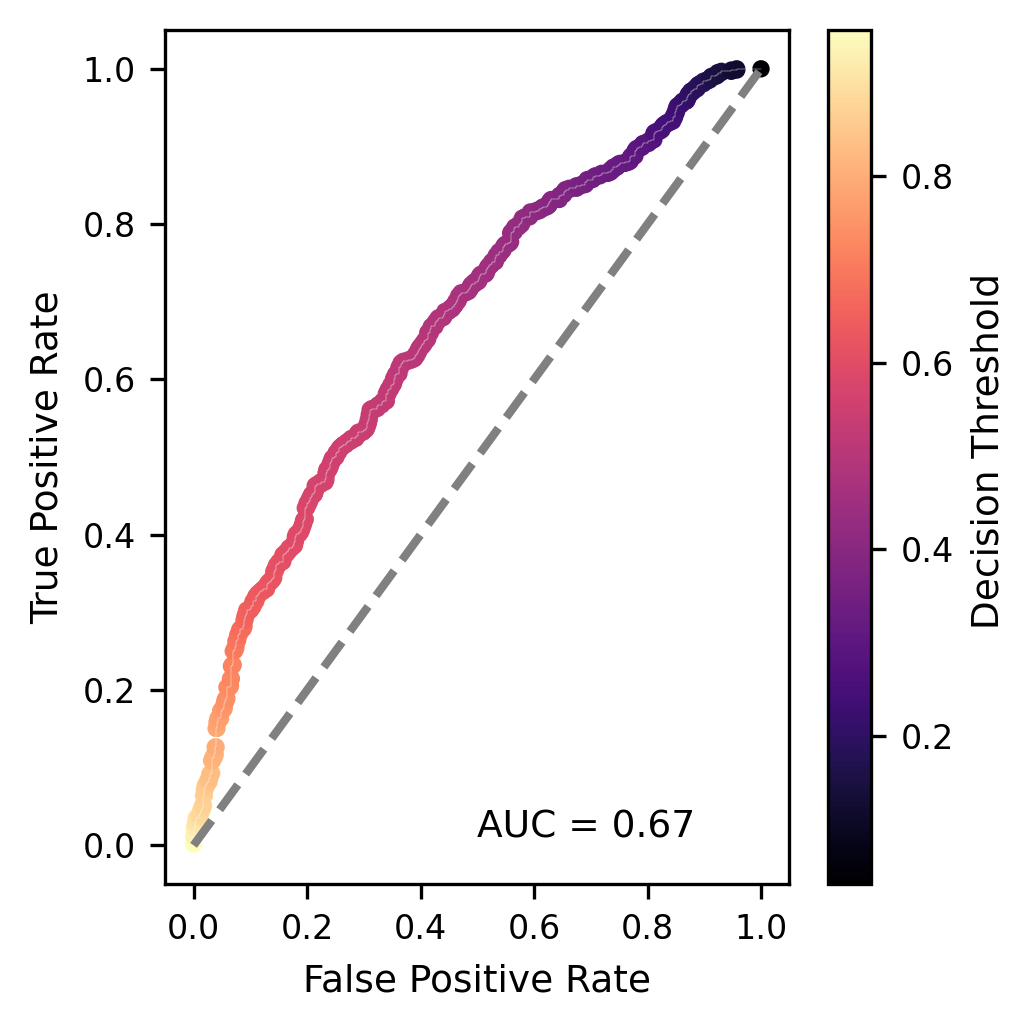}
    \includegraphics[width=0.49\linewidth]{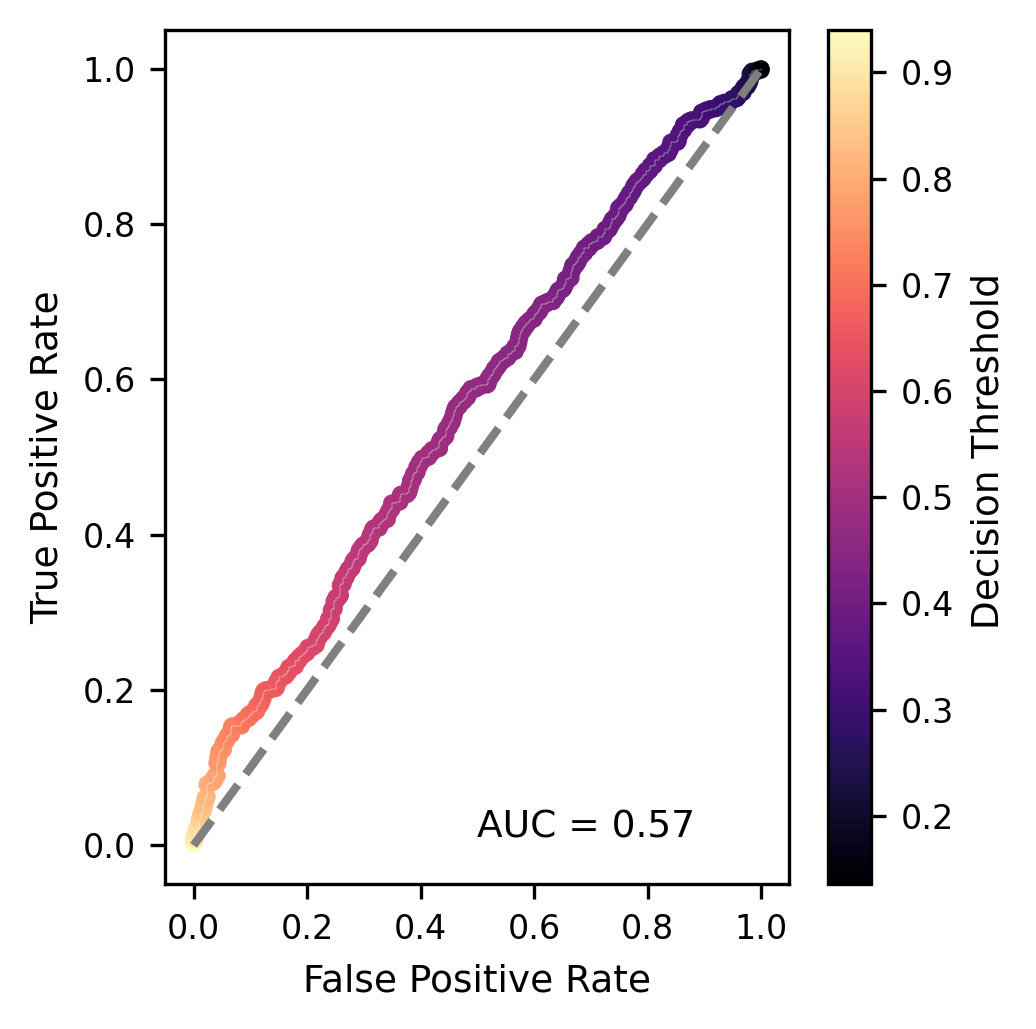}
    \caption{\textbf{Same as Figure~\ref{fig:roc} for our two greyscale models. \emph{Left:} Summed greyscale model. \emph{Right:} F160W only model. For the summed version, the AUC is only 0.02 lower than for the three-channel model, but 0.12 less for our F160W only model (which is essentially randomly guessing).}}
    \label{fig:greyROC}
\end{figure}

\textbf{Here we show some metrics of those two models to emphasize that the multi-band model outperformed the greyscale model.
\revtwo{Neither greyscale model provided a meaningful computational advantage compared to the multi-band model.}
We use our most accurate random seed, Seed 626, for all figures below, while in Table~\ref{tab:pcbegrey} we present means and standard deviations for all metrics calculated across the three models trained with different random seeds.
The summed models performed within a few percentage points lower to the three-channel model in overall accuracy (62.01\% compared to 63.56\%) and other performance metrics (top row of Table~\ref{tab:pcbegrey}). 
However, when we examine the confusion matrix of Seed 626 in Figure~\ref{fig:greyCM} (left), we see that the model does not perform equally well on both classes. 
The bottom right quadrant has 65.22\% of mergers identified, but the top left has only 58.73\% of nonmergers identified.}
\textbf{The F160W only models perform extremely poorly by all metrics. The accuracy of 54.70\% (bottom row of Table~\ref{tab:pcbegrey}) shows that the model is essentially randomly guessing, and this is confirmed in the Seed 626 ROC plot (right on Figure~\ref{fig:greyROC}), with the curve being barely above the 1-1 line.}

\begin{figure}
    \centering
    \includegraphics[width=0.49\linewidth]{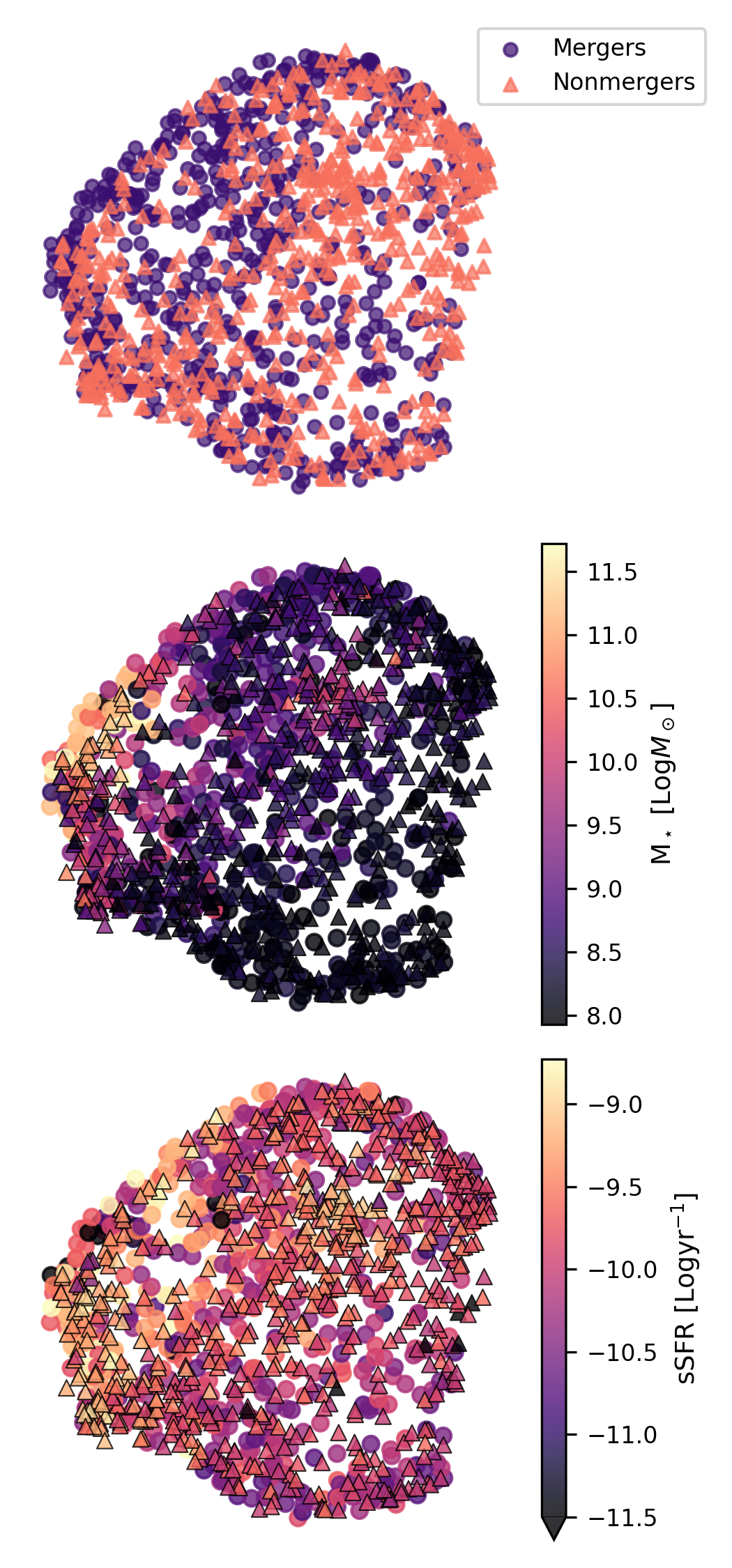}
    \includegraphics[width=0.49\linewidth]{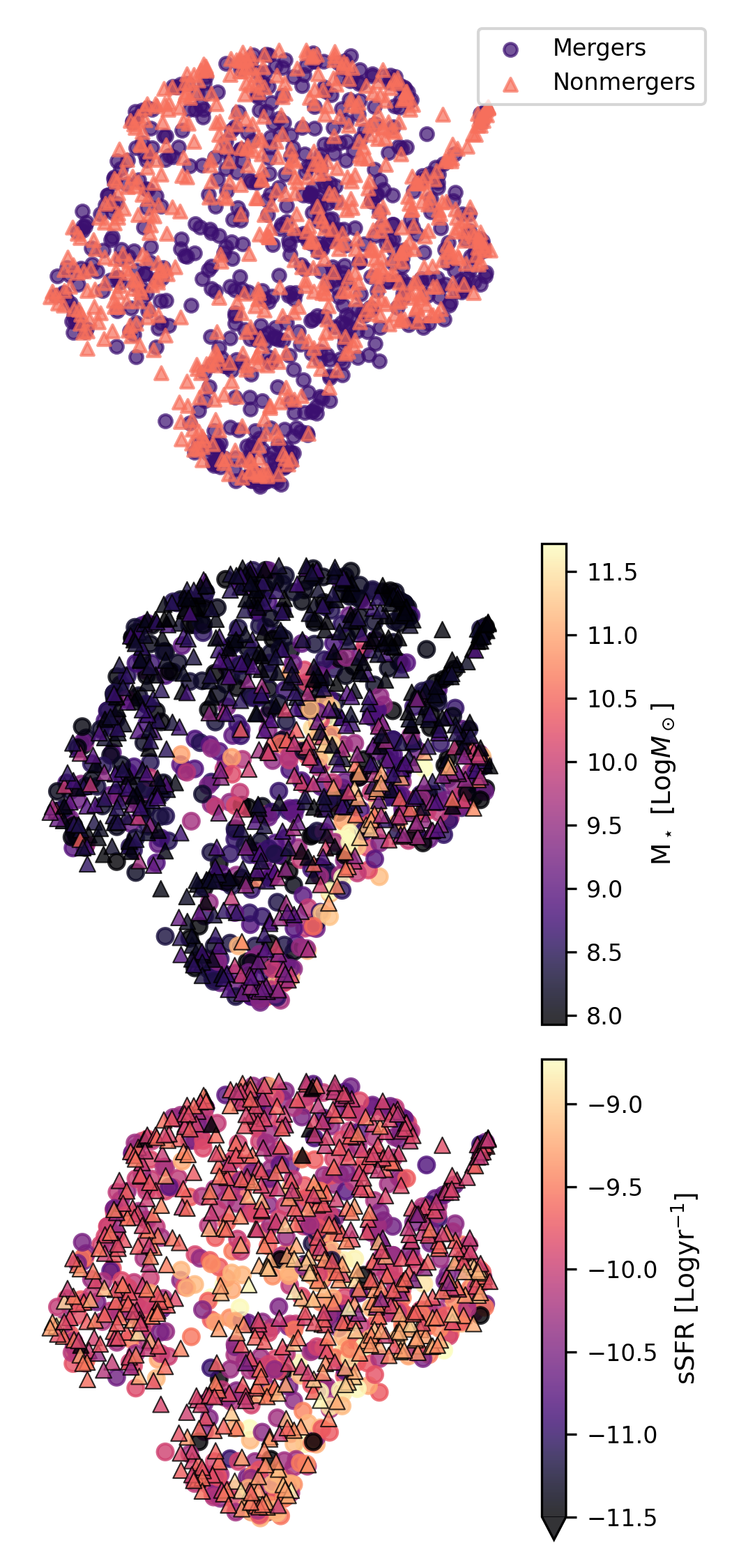}
    \caption{\textbf{Same as Figure~\ref{fig:UMAP} for the two greyscale models. \emph{Left:} Summed greyscale model. \emph{Right:} F160W only model. The trends in the summed model are identical to the trends seen in the three-channel model. This implies that the combined image did not solve the star formation rate bias. However, the F160W only model shows a more random distribution both in stellar mass and sSFR.}}
    \label{fig:greyUMAP}
\end{figure}

\textbf{Figure~\ref{fig:greyUMAP} shows the UMAPs for the Seed 626 greyscale models. In the left row we can see that the summed model was still able to detect stellar mass and sSFR similarly to the three-channel model. 
We see gradients increasing in both from right to left. 
Therefore, the summed filters, as expected, contained similar information to feeding the filters in as separate channels, and the sSFR information was still used.
However, the F160W only model (right column), was not able to learn properly and does not group galaxies strongly by either physical quantity. 
The distribution is far more random, and combined with the ROC cruve confirms that a single channel is not enough information for merger detection at $M_\star>10^8M_\odot$ and $z\sim1$.}

\begin{table}[h!]
    \centering
    \begin{tabular}{c|*{6}c}
    \multicolumn{6}{c}{Performance Metrics} \\
    \midrule
        Model & Accuracy & Purity & Completeness & Brier Score & ECE & AUC\\
        \midrule
        Summed Flux &
         62.04$\pm{1.35}$\% &
         61.29$\pm{0.81}$\% &
         65.92$\pm{2.15}$\% & 
         0.23$\pm{0.01}$ & 
         0.03$\pm{0.01}$ & 
         0.67$\pm{0.01}$\\

        F160W Only &
         53.99$\pm{1.01}$\% &
         53.70$\pm{1.59}$\% &
         49.36$\pm{4.91}$\% & 
         0.26$\pm{0.01}$ & 
         0.10$\pm{0.02}$ & 
         0.55$\pm{0.02}$\\
          
    \end{tabular}
    \caption{\textbf{Same as Table~\ref{tab:pcbe} for our greyscale models-- Accuracy, Purity, Completeness, Brier Score, ECE, and AUC. The values shown are the mean and standard deviation of the three random seeds.}}
    \label{tab:pcbegrey}
\end{table}

\end{document}